\documentclass[12pt]{article}
\usepackage{graphicx}
\usepackage{amsmath}
\usepackage{amssymb}
\usepackage{ulem}
\usepackage{enumerate}
\usepackage{cite}
\usepackage{latexsym}
\begin{document}
%\begin{flushright}
%BHU-PHYS-CAS Preprint\\
%arXiv: 1005.5067 [hep-th]
%\end{flushright}
\begin{center}
{\bf {\large{Nilpotent Symmetries of a Modified Massive Abelian 3-Form \\ Theory: Augmented Superfield Approach}}}

\vskip 3.4cm

{\sf  A. K. Rao$^{(a)}$, R. P. Malik$^{(a,b)}$}\\
$^{(a)}$ {\it Physics Department, Institute of Science,}\\
{\it Banaras Hindu University, Varanasi - 221 005, (U.P.), India}\\

\vskip 0.1cm

$^{(b)}$ {\it DST Centre for Interdisciplinary Mathematical Sciences,}\\
{\it Institute of Science, Banaras Hindu University, Varanasi - 221 005, India}\\
{\small {\sf {e-mails: amit.akrao@gmail.com; rpmalik1995@gmail.com}}}
\end{center}

\vskip 3.0 cm

\noindent
{\bf Abstract:}
We derive the off-shell nilpotent and absolutely anticommuting (anti-)BRST symmetry transformations for
any arbitrary D-dimensional  St{$\ddot u$}ckelberg-modified
{\it massive} Abelian 3-form theory within the framework of augmented version of superfield approach (AVSA) to 
Becchi-Rouet-Stora-Tyutin (BRST) formalism where, in addition to the horizontality condition (HC), we exploit 
the theoretical strength of the gauge invariant restriction (GIR) to deduce the proper transformations for the 
gauge, associated (anti-)ghost fields, auxiliary fields, St{$\ddot u$}ckelberg compensating field, etc. In fact, it is an elegant and delicate combination  
of HC and GIR (within the ambit of AVSA) that is {\it crucial} for {\it all} our  discussions and derivations.
One of the highlights of our present endeavor is the deduction  of a {\it new} set of  (anti-)BRST invariant 
Curci-Ferrari (CF)-type restrictions which are {\it not}
found in the {\it massless} version of our present theory where {\it only} the HC plays an important role in the derivations of {\it all} the (anti-)BRST
transformations and a very {\it specific} set of CF-type restrictions. 
The {\it alternative} ways of the derivation of the {\it full} set of the {\it latter}, from various theoretical 
considerations, are also interesting results of our present investigation.



\newpage

\section {Introduction}
The key ideas behind the (super)string theories have been at the forefront of research in the realm of theoretical high energy physics
because these theories are the most promising candidates to provide a precise theoretical description of the 
theory of quantum gravity (see, e.g. [1-5] for details). Furthermore, as far as the idea of {\it complete}
unification of {\it all} the fundamental interactions of nature is concerned, the  modern developments in (super)string theories provide
a glimpse of hope to accomplish this goal (which is {\it unthinkable} within the framework of 
quantum field theories which are {\it not} based on the ideas of (super)strings). In addition to the 
above {\it positive} aspects, these fundamental developments have led to 
many {\it other} branches of research activities in theoretical high energy physics [1-5]. 
For instance, the studies  of noncommutative geometry and related field theories, higher $p$-form ($p = 2, 3, ...$) 
gauge theories, AdS/CFT correspondence, higher spin gauge theories, 
specific kinds of supersymmetric Yang-Mills theories, etc., owe their origin to the key developments in the realm of (super)string theories.
The central theme of our present investigation is connected with the study of the Becchi-Rouet-Stora-Tyutin (BRST)
{\it quantized} version of the {\it massive} Abelian 3-form gauge theory which, it goes without saying,  is inspired by the 
theoretical developments in (super)string theories where the higher $p$-form ($p = 2, 3, ...$) gauge fields appear in their 
quantum excitations. Our present study goes {\it beyond} the realm of the standard model of particle physics
(see, e.g. [6-8] and references therein) which is based on the quantum field theory of the non-Abelian 1-form  Yang-Mills gauge theory.

The BRST-quantized theories are endowed with the nilpotent and absolutely  anticommuting (anti-)BRST transformations 
corresponding to a given {\it classical} local gauge symmetry transformation for a {\it classical} gauge theory.
The usual superfield approach (USFA) to BRST formalism (see, e.g. [9-11] and references therein), based on the geometrical aspects
of mathematics, addresses  the issues related to the BRST-quantized version of the gauge theory where the celebrated horizontality condition (HC)
plays a decisive role in the derivation of $(i)$ the (anti-)BRST symmetries for the 1-form gauge and associated (anti-)ghost fields, and
$(ii)$ the Curci-Ferrari (CF) condition (see, e.g. [12]) for the non-Abelian 1-form gauge theory 
({\it without} any interaction with {\it matter} fields). The key ideas behind USFA have been systematically augmented so as to 
derive the (anti-)BRST transformations for the gauge, (anti-)ghost and {\it matter} fields {\it together}
along with the CF-type condition. This approach has been christened by us as the augmented version of superfield
approach (AVSA) to BRST formalism (see, e.g. [13-15] for details) where, in addition to the HC, 
we have exploited the gauge invariant restriction (GIR). It has been found that there is {\it no} conflict between HC and GIR
and they compliment  each-other in a coherent manner.   
In fact, the results of the HC are utilized in the application of GIR that is imposed on the superfields 
that are defined on a suitably chosen supermanifold [see, Eqs. (20), (26) below].

In our present investigation, we apply the theoretical beauty and strength of
AVSA in the context of  St{$\ddot u$}ckelberg-modified [16] version of the {\it massive} Abelian 3-form gauge theory 
in any arbitrary D-dimension of spacetime to derive $(i)$ the {\it proper} (i.e. off-shell nilpotent and absolutely anticommuting)
BRST and anti-BRST symmetry transformations, $(ii)$ the precise sets of (anti-BRST invariant CF-type restrictions,
and $(iii)$ the coupled (but equivalent) (anti-)BRST invariant Lagrangian densities. 
One of the upshots of our whole discussion is the derivation of the precise sets of (anti-)BRST invariant CF-type restrictions from $(i)$
the  requirement  of the absolute anticommutativity of the (anti-)BRST symmetry transformations, $(ii)$ the 
BRST and anti-BRST invariance of the appropriate Lagrangian densities of our theory, and $(iii)$ the Euler-Lagrange (EL) equations 
of motion (EoM) that emerge out from the coupled (but equivalent) (anti-)BRST invariant Lagrangian densities.

In our present endeavor, we exploit the theoretical potential and power of AVSA to 
obtain the {\it full} set of (anti-)BRST symmetry transformations for the St{$\ddot u$}ckelberg-modified D-dimensional
{\it massive} Abelian 3-form gauge theory and show the existence of a {\it new} set of CF-type restrictions which are {\it not} found [17, 18]
in  the {\it massless} version of our present theory where {\it only}
HC plays a decisive role. We establish the existence of {\it all} the CF-type restrictions
of our theory by invoking an elegant combination of HC [cf. Eq. (8)] and a gauge invariant 
restriction [cf. Eq. (26)] within the framework of AVSA.
We also derive the (anti-)BRST invariant Lagrangian densities in the D-dimensional {\it ordinary} space for the 
St{$\ddot u$}ckelberg-modified {\it massive} Abelian theory and establish their (anti-)BRST invariance in an explicit fashion.
The highlight of {\it this} exercise is the derivation of {\it all} the CF-type restrictions of our theory on the 
beautiful idea of symmetry consideration {\it alone}.

The theoretical contents of our present paper are organized as follows. In Sec. 2, we discuss the bare essentials
of the St{$\ddot u$}ckelberg-formalism [16] in the case of a {\it massive} Abelian 3-form theory in any arbitrary D-dimension 
of spacetime. In this section, we also capture some of the  essential features of our earlier works [17, 18] where we have 
derived the nilpotent (anti-)BRST symmetries for the gauge, associated (anti-)ghost and auxiliary fields  from the 
horizontality condition [i.e. $\tilde d\, \tilde A^{(3)} 
= d\, A^{(3)}$] [cf. Eq. (8) below] so that we can obtain the explicit form of
 $\tilde A^{(3)}_{(h)}$ [cf. Eq. (20) below] where the subscript $(h)$ denotes the 
explicit expression for the {\it super} Abelian 3-form {\it after} the application of HC.
Our Sec. 3 deals with the derivation of the (anti-)BRST symmetry transformations for the St{$\ddot u$}ckelberg compensating
field, associated (anti-)ghost fields and auxiliary fields.  The subject matter of Sec. 4 concerns itself with the 
derivation of the coupled (but equivalent) (anti-) BRST invariant Lagrangian densities and we comment on their proper transformations
under the (anti-)BRST symmetry transformations. 
In Sec. 5, we perform thread-bare analysis of the applications of the (anti-)BRST symmetry transformations on the 
coupled (but equivalent) Lagrangian densities and show the existence of the (anti-)BRST invariant CF-type restrictions. 
Finally, in Sec. 6, we summarize our key results, highlight new observations and point out a few future directions for further investigation(s).

In our Appendices A and B, we provide some explicit computations that have been used in the main body of our text.
In particular, in Appendix A, we discuss the {\it explicit} use of HC in the derivations of the 
(anti-)BRST symmetry transformations and (non-)trivial CF-type restrictions which have {\it not} 
been performed in our earlier works [17, 18]. Our Appendix C captures, through a figure, the existence and emergence of
CF-type restrictions on our theory where we concisely provide an explanation for such an existence.

\vskip 0.5cm
\noindent
{\it {Convention and Notations}}:
We adopt the convention of the left-derivative w.r.t. {\it all} the {\it fermionic} fields in {\it all} our computations. Furthermore, 
our D-dimensional background  spacetime manifold is {\it flat} with the metric tensor ${\eta}_{\mu \nu } =$ diag (+1, -1, -1,...) so that the dot product 
between two non-null vectors $P_\mu$ and $Q_\mu$ (on this flat  D-dimensional spacetime manifold) is defined by $P \cdot Q 
= \eta_{\mu \nu }\,P^\mu\, Q^\nu \equiv P_0\,Q_0 - P_i\,Q_i$ where the Greek indices $\mu,\nu,\lambda,... = 0, 1, 2, ... D-1$
denote the time and space directions {\it and} the Latin indices $i, j, k,... = 1, 2,... D-1$ stand for the space directions {\it only}. 
We also adopt the convention of the derivatives: $[{\partial A_{\mu\nu\lambda}}/{\partial A_{\rho\sigma\kappa}}] = 
\frac {1}{3!} \big [\delta_\mu ^\rho \,(\delta_\nu ^\sigma \, \delta _\lambda ^\kappa - \delta _\lambda ^\sigma\, \delta _\nu ^\kappa) + 
\delta_\nu ^\rho \, (\delta_\lambda ^\sigma \,\delta _\mu ^\kappa - \delta _\mu ^\sigma\, \delta _\lambda ^\kappa) + 
\delta_\lambda ^\rho \, (\delta_\mu^\sigma \,\delta _\nu ^\kappa - \delta _\nu ^\sigma\, \delta _\mu ^\kappa)  \big], \; 
(\partial\, B_{\mu\nu}/ \partial\,B_{\alpha\beta}) = \frac{1}{2 !}\, [\delta_\mu^\alpha \delta_\nu^\beta 
- \delta_\nu^\alpha \delta_\mu^\beta ]$, etc., 
for the {\it basic} third-ranked and second-ranked tensor fields.
Throughout the whole body of our text, we denote the (anti-)BRST symmetry transformations by $s_{(a)b}$. 
 The generic superfield $\Phi(x, \theta, \bar\theta)$ is defined on the (D, 2)-dimensional 
supermanifold which is characterized by the superspace coordinates $Z^M = (x^\mu, \theta, \bar\theta)$ where the D-dimensional {\it bosonic} 
coordinates are $x^\mu$ (with $\mu = 0, 1, 2, ... D-1$) and the  Grassmannian variables  $(\theta, \bar\theta)$ obey the 
relationships: $\theta^2 = \bar\theta^2 = 0, \, \theta\, \bar\theta + \bar\theta \, \theta = 0$. The {\it latter}  anticommute 
with all the {\it fermionic} (super)fields  {\it  but} commute with all the {\it bosonic} (super)fields that are present in our theory 
(which is discussed within the framework of AVSA to BRST formalism).

\section{Preliminaries: Celebrated St{$\ddot u$}ckelberg Formalism and Geometrical Horizontality Condition}

This section is divided into two parts. In subsection 2.1, we discuss the key features of the St{$\ddot u$}ckelberg technique
of the compensating fields to derive the {\it modified} Lagrangian density [cf. Eq. (4) below] for the {\it massive} Abelian 3-form theory which 
respects the gauge transformations. Our subsection 2.2 is devoted to the discussion on the horizontality condition (HC)
which yields the (anti-)BRST transformations $[s_{(a)b}]$ for the {\it massless} case  [17, 18] of the Abelian 3-form gauge theory
in any arbitrary D-dimension of spacetime (cf. Appendix A below for details). 
These are essential {\it inputs} for the derivation of the {\it complete} set of off-shell nilpotent and absolutely anticommuting (anti-)BRST
transformations for our {\it entire}  theory which incorporates the gauge fields, associated (anti-)ghost fields, auxiliary fields 
{\it and} the St{$\ddot u$}ckelberg-compensating field [and its associated (anti-ghost fields].

\subsection{St{$\ddot u$}ckelberg Technique: Lagrangian Formulation}

We begin with the Lagrangian density $[{\cal L}_{(0)}]$ for the {\it totally} antisymmetric ($A_{\mu\nu\lambda} 
= -\, A_{\nu\mu\lambda} = -\, A_{\mu\lambda\nu} = -\, A_{\lambda\nu\mu}$) tensor field $(A_{\mu\nu\lambda})$
with the rest mass $m$ (see, e.g. [16-18]) in any arbitrary D-dimension of spacetime (with $\mu,\, \nu,\, \lambda,\, \zeta, .... 
= 0, 1, 2,..., D-1$), namely; 
\begin{eqnarray} 
{\cal L}_{(0)} = \frac {1}{24}\,H^{\mu \nu \lambda \zeta }\,H_{\mu \nu \lambda \zeta} - \frac {m^2}{6} 
\, A_{\mu\nu\lambda }\,A^{\mu\nu\lambda },
\end{eqnarray}
where the field strength tensor $H_{\mu \nu \lambda \zeta} = \partial_\mu\, A_{\nu\lambda\zeta} - \partial_\nu\,
 A_{\lambda \zeta \mu } + \partial_\lambda\, A_{\zeta \mu \nu }  - \partial_\zeta \,A_{\mu \nu \lambda }$
is derived from the 4-form $H^{(4)} = d\, A^{(3)}$ where $A^{(3)} = [(d\,x^\mu \wedge d\,x^\nu \wedge d\, x^\lambda)/3!]\,A_{\mu\nu\lambda}$
defines the {\it totally} antisymmetric tensor field $A_{\mu\nu\lambda}$. Here the operator $d= d\, x^\mu\, \partial_\mu$ (with $d^2 = 0)$
is the exterior derivative (see, e.g. [19-22) for details). The explicit form of $H^{(4)}$ is as follows: 
\begin{eqnarray}
H^{(4)} = d\, A^{(3)} = \frac{1}{4!}\,\Big( d\,x^\mu \wedge d\,x^\nu \wedge d\, x^\lambda \wedge d\, x^\zeta \Big)\, H_{\mu \nu \lambda \zeta}. 
\end{eqnarray}
The Euler-Lagrange (EL) equation of motion (EoM) from the Lagrangian density (1) is: $\partial_\mu\,H^{\mu \nu \lambda \zeta }
 + m^2\, A^{\nu\lambda \zeta } = 0$ which implies the subsidiary conditions: $\partial_\nu\,A^{\nu \lambda \zeta } = 
\partial_\lambda\, A^{\nu \lambda \zeta } = \partial_\zeta\, A^{\nu \lambda \zeta } = 0$ due to the {\it totally}
antisymmetric nature of $H_{\mu \nu \lambda \zeta}$ [cf. Eq. (2)] when $m^2 \neq 0$.
With the  above subsidiary conditions as {\it inputs}, we obtain the Klein-Gordon equation of motion: 
$(\Box + m^2)\,A_{\mu\nu\lambda} = 0 $ which shows that the Lagrangian density (2) describes a relativistic field 
$A_{\mu \nu \lambda }$ with the rest mass equal to $m$. The Lagrangian density (2) is endowed with the second-class constraints in the 
terminology of Dirac's prescription for the classification scheme of constraints [23, 24]. To restore the gauge symmetry transformations
that are generated by the first-class constraints, we apply the following theoretical trick of the St{$\ddot u$}ckelberg formalism
of compensating fields [16]. 
\begin{eqnarray}
A_{\mu\nu \lambda } \quad  \longrightarrow \quad A_{\mu\nu \lambda} \;  \mp  \; \frac{1}{m}\, (\partial_\mu\, \Phi_{\nu\lambda} 
+ \partial_\nu\, \Phi_{\lambda \mu } +  \partial_\lambda\, \Phi_{\mu\nu}),
\end{eqnarray}
where the compensating antisymmetric $(\Phi_{\mu\nu} =-\, \Phi_{\nu\mu})$ tensor field  $(\Phi_{\mu\nu})$ has the {\it same} mass dimension as 
$A_{\mu\nu\lambda}$ field in any arbitrary dimension of spacetime in the natural units where $\hbar  = c = 1$.
Adopting the notation $\Sigma_{\mu \nu \lambda } = \partial_\mu\, \Phi_{\nu\lambda} + \partial_\nu\, \Phi_{\lambda \mu } + 
\partial_\lambda\, \Phi_{\mu\nu}$, we obtain the St{$\ddot u$}ckelberg-modified Lagrangian density [${\cal L}_{(S)}$]
(from ${\cal L}_{(0)} $) as
\begin{eqnarray}
{\cal L}_{(0)} \longrightarrow {\cal L}_{(S)} = \frac {1}{24}\,H^{\mu \nu \lambda \zeta }\,H_{\mu \nu \lambda \zeta} - \frac {m^2}{6} 
\, A^{\mu\nu\lambda }\,A_{\mu\nu\lambda } - \frac{1}{6}\, \Sigma^{\mu \nu \lambda }\, \Sigma_{\mu \nu \lambda }
\pm \frac{m}{3}\,A^{\mu\nu\lambda} \, \Sigma_{\mu \nu \lambda },
\end{eqnarray}
which respects $[\delta_g {\cal L}_{(s)} = 0]$ the following gauge symmetry transformations $(\delta_g)$
\begin{eqnarray}
&&\delta_g\, A_{\mu\nu\lambda} = \partial_\mu\, \Lambda_{\nu\lambda} + \partial_\nu\, \Lambda_{\lambda\mu} 
+ \partial_{\lambda} \Lambda_{\mu\nu}, \qquad \;\,
\delta_g\, \Phi_{\mu\nu} = \pm \, m\, \Lambda_{\mu\nu} - (\partial_\mu\,\Lambda_\nu - \partial_\nu\,\Lambda_\mu),\nonumber\\
&&\delta_g\, \Sigma_{\mu\nu\lambda} = \pm \, m\,(\partial_\mu\, \Lambda_{\nu\lambda} + \partial_\nu\, \Lambda_{\lambda\mu} 
+ \partial_{\lambda} \Lambda_{\mu\nu}), \qquad
\delta_g\, H_{\mu \nu \lambda \zeta } = 0,
\end{eqnarray}
where $(\Lambda_{\mu\nu} = -\, \Lambda_{\nu\mu})$ is the antisymmetric gauge transformation parameter. The Lorentz vector 
$(\Lambda_\mu)$ gauge transformation parameter appears in the theory due to the observation that there is a stage-one 
reducibility in the St{$\ddot u$}ckelberg-redefinition (3) where $\Phi_{\mu\nu}\longrightarrow \Phi_{\mu\nu} 
\pm (\partial_\mu\, \Lambda_\nu - \partial_\nu\, \Lambda_\mu)$ is permitted due to the antisymmetric 
$(\Phi_{\mu\nu} = -\, \Phi_{\nu\mu} )$ nature of the compensating field $\Phi_{\mu\nu} $
that is present in the redefinition [cf. Eq. (3)].

We end this subsection with the following remarks. First of all, we note that, in the  St{$\ddot u$}ckelberg formalism,
it is the exterior derivative $d = d\, x^\mu\, \partial_\mu$ (with $\mu = 0, 1, 2, ..., D-1$, and $d^2 = 0$)
that plays an important role because the redefinition (3) can be expressed in the differential form-language [19-22] as follows
\begin{eqnarray}
A^{(3)} \longrightarrow A^{(3)} \mp \frac{1}{m}\, d\, \Phi^{(2)},
\end{eqnarray}
where the {\it bosonic} 2-form  $\Phi^{(2)} = [(d\,x^\mu \wedge d\, x^\nu)/ 2!]\Phi_{\mu\nu}$ establishes that $\Phi_{\mu\nu}$ is an 
antisymmetric $(\Phi_{\mu\nu} = -\, \Phi_{\nu\mu} )$ compensating field. Second, the Lorentz vector $\Lambda_\mu (x)$ is 
a gauge transformations parameter in the theory due to the stage-one reducibility in the theory.
Third, there is existence of stage-two reducibility, too,  in our theory because the Lorentz vector $\Lambda_\mu$
can be replaced by $\Lambda_\mu\longrightarrow \Lambda_\mu \pm \partial_\mu\, \Lambda$ without spoiling 
the symmetry transformations where $\Lambda(x)$ is a Lorentz scalar gauge transformation parameter.
Fourth, within the framework of BRST formalism, the {\it bosonic} set of transformation parameters 
($\Lambda_{\mu\nu}, \, \Lambda_\mu, \, \Lambda$) will be replaced by the {\it fermionic} (anti-)ghost fields.
Fifth, we observe that the following elegant and useful combination of the $A_{\mu\nu\lambda}$ and $\Phi_{\mu\nu}$
fields remains invariant under the  gauge symmetry transformations $(\delta_g)$:
\begin{eqnarray}
\delta_g\, \Big[A_{\mu\nu\lambda} \mp \frac{1}{m}\, \Sigma_{\mu\nu\lambda}\Big] = 0 \quad \Longleftrightarrow \quad
\delta_g\, \Big[A^{(3)} \mp \frac{1}{m}\, d\, \Phi^{(2)}\Big] = 0.
\end{eqnarray}
We shall see {\it later} that this observation plays an important role in the superfield formalism
(cf. Sec. 3 below). Sixth, the gauge transformations (5) are generated by the first-class constraints (see, e.g. [23-26]). 
The {\it latter} emerge due to the presence of the St{$\ddot u$}ckelberg compensating field $(\Phi_{\mu\nu})$
which is responsible for the conversion of the second-class constraints of the {\it original} Lagrangian density [${\cal L}_{(0)}$]
into the first-class constraints [25, 26]. 
Finally, it is straightforward to note that, due to (6), the 4-form
$H^{(4)} = d\, A^{(3)}$ remains invariant because of  the nilpotency $(d^2 = 0)$ of the exterior derivative $(d)$.
In other words, we have the gauge invariance ($\delta_g \, H_{\mu \nu \lambda \zeta } = 0$) of the field-strength tensor.

\subsection{Horizontality Condition: Superfield Approach}

We have seen that the 4-form $H^{(4)} = d\,A^{(3)}$ (with $d^2 = 0$) defines the field strength (curvature) tensor 
$H_{\mu \nu \lambda \zeta} = \partial_\mu\, A_{\nu\lambda\zeta} - \partial_\nu\,
 A_{\lambda \zeta \mu } + \partial_\lambda\, A_{\zeta \mu \nu }  - \partial_\zeta \,A_{\mu \nu \lambda }$
 which remains invariant ($\delta_g H_{\mu\nu\lambda\zeta} = 0$) under the gauge symmetry transformations (5).
 Hence, it is a {\it physical} quantity in the theory and its generalization on the (D, 2)-dimensional 
 supermanifold must remain independent\footnote{The Grassmannian variables ($\theta, \,  \bar\theta$) are {\it only} a set  of  
mathematical artifacts in superfield approach and they have nothing to do with the {\it physical} quantities which are gauge [i.e. (anti-)BRST] invariant.}  
of the Grassmannian variables $(\theta, \bar\theta)$ that characterize {\it this}
supermanifold along with the D-dimensional {\it bosonic} coordinates $x^\mu (\mu = 0, 1, 2...D - 1)$. In other words, 
the following horizontality condition (HC), namely;
\begin{eqnarray}
\tilde d\, \tilde A^{(3)}  = d\, A^{(3)}, \qquad \tilde d = d x^\mu \partial_\mu + d\theta\,\partial_\theta 
+ d \bar\theta\,\partial_{\bar\theta}, \qquad d = d\, x^\mu\, \partial_\mu,
  \end{eqnarray} 
leads to the derivation of the (anti-)BRST symmetry transformations for the 
gauge field $(A_{\mu\nu\lambda})$, bosonic vector field $\phi_\mu$ and a set of bosonic 
as well as fermionic (anti-)ghost fields (see e.g. [17] for details) that are associated with the gauge 
transformations (5). However, before we proceed further, let us define the 
explicit form of the {\it super} 3-form, namely;
\begin{eqnarray}
\tilde A ^{(3)} = \frac{(dZ^M \wedge dZ^N \wedge dZ^K)}{3!}\, \,\tilde A_{MNK} (x, \theta, \bar\theta),
\end{eqnarray}  
where the superspace coordinates $Z^M = (x^\mu, \theta, \bar\theta)$ characterize the (D, 2)-dimensional supermanifold for a 
given D-dimensional {\it ordinary} Abelian 3-form gauge theory that is discussed within the framework of 
superfield approach to BRST formalism. In an explicit form, the above equation 
can be expressed, in terms of various kinds of {\it bosonic} as well as {\it fermionic} (D, 2)-dimensional superfields, as follows:  
\begin{eqnarray*}
 \tilde A ^{(3)}
& =&   \frac{1}{3!}\, (dx^\mu \wedge dx^\nu \wedge dx^\lambda )\,
\tilde A_{\mu\nu\lambda } (x, \theta, \bar\theta)\, + 
\frac{1}{2} \, (dx^\mu \wedge dx^\nu \wedge d\theta) \, \tilde
A_{\mu\nu\theta} (x, \theta, \bar\theta)\nonumber\\ 
\end{eqnarray*}
\begin{eqnarray}
& + & \frac{1}{2}\,  (dx^\mu \wedge dx^\nu \wedge d \bar\theta)\, 
\tilde A_{\mu\nu\bar\theta}(x, \theta, \bar\theta) + \frac{1}{3!} \, (d\theta \wedge d\theta
\wedge d\theta)\,  \tilde A_{\theta\theta\theta}(x, \theta, \bar\theta) \nonumber\\
 & + & (dx^\mu \wedge d \theta \wedge d\bar\theta)\,  \tilde A_{\mu\theta\bar\theta}(x, \theta, \bar\theta) + \frac{1}{2}\, 
(dx^\mu \wedge d\theta \wedge d \theta) \, \tilde A_{\mu\theta\theta}(x, \theta, \bar\theta) \nonumber\\
& + &  \frac{1}{2} \, (dx^\mu \wedge d\bar\theta \wedge d \bar\theta) \, \tilde A_{\mu\bar\theta\bar\theta}(x, \theta, \bar\theta)
+ \frac{1}{2} \, (d\theta \wedge d\theta \wedge d \bar\theta) \, \tilde A_{\theta\theta\bar\theta}(x, \theta, \bar\theta)\nonumber\\
 &+& \frac{1}{2}\,  (d\theta \wedge d\bar\theta \wedge d \bar\theta) \, \tilde A_{\theta\bar\theta\bar\theta}(x, \theta, \bar\theta)
+  \frac{1}{3!} \, (d \bar\theta \wedge d\bar\theta \wedge d \bar\theta)\,  \tilde A_{\bar\theta\bar\theta\bar\theta}(x, \theta, \bar\theta).
\end{eqnarray}
Here we identify the superfields on the r.h.s. as follows: 
\begin{eqnarray}
&& \tilde A_{\mu\nu\lambda } (x,\theta,\bar\theta) = {\cal A}_{\mu\nu\lambda } (x, \theta, \bar \theta),\quad 
\tilde A_{\mu\nu\theta} (x,\theta,\bar\theta) =  {\bar {\cal F}}_{\mu\nu} (x, \theta, \bar\theta),\nonumber\\
&& \tilde A_{\mu\nu\bar\theta} (x,\theta,\bar\theta) 
= {{\cal F}_{\mu\nu}} (x,\theta,\bar\theta),\quad \tilde A_{\mu\theta\bar\theta} (x,\theta,\bar\theta) = \Phi_\mu
(x,\theta,\bar\theta), \nonumber\\
&& \frac{1}{3!} \tilde A_{\theta\theta \theta} (x,\theta,\bar\theta)
 = {{\bar {\cal F}}}_2 (x,\theta,\bar\theta),\quad  \frac{1}{3!}
\tilde A_{\bar\theta\bar\theta\bar\theta} (x,\theta,\bar\theta) = { {\cal F}}_2  (x,\theta,\bar\theta),\nonumber\\
&& \frac{1}{2}  \tilde A_{\theta\bar\theta\bar\theta} (x,\theta,\bar\theta) = {\cal F}_1
(x,\theta,\bar\theta), \quad \frac{1}{2} \tilde A_{\theta\theta\bar\theta} (x,\theta,\bar\theta) = {
{\bar {\cal F}}}_1 (x,\theta,\bar\theta),  \nonumber\\
&& \frac{1}{2} \tilde A_{\mu\bar\theta\bar\theta} (x,\theta,\bar\theta)
 = \beta_\mu (x, \theta, \bar\theta), \quad \frac{1}{2} \tilde A_{\mu\theta\theta} (x,\theta,\bar\theta) 
=  {\bar \beta}_\mu (x, \theta, \bar\theta). 
\end{eqnarray}
It is clear from the above identifications that the {\it  four bosonic}  superfields are: 
$\tilde {\cal A}_{\mu\nu\lambda } (x, \theta, \bar \theta), \\ \; {\Phi}_{\mu} (x, \theta, \bar \theta), \;  {\beta}_{\mu} (x, \theta, \bar \theta)$ 
and $ \bar {\beta}_{\mu} (x, \theta, \bar \theta)$ where the {\it latter}  two superfields are the 
generalizations of the {\it bosonic} ghost (i.e. ghost-for-ghost) fields which carry the 
ghost numbers $(+ 2)$ and $(- 2)$, respectively. On the contrary, the superfields ${\cal A}_{\mu\nu\lambda } (x, \theta, \bar \theta)$
and ${\Phi}_{\mu} (x, \theta, \bar \theta)$ carry {\it no} ghost number (i.e. zero ghost number)
and they are the generalizations of the gauge field $A_{\mu\nu\lambda} (x)$ and a vector field $\phi_\mu (x)$
which appears in the CF-type restriction. The {\it fermionic} superfields, it is obvious, in our theory are: 
${\tilde {\bar {\cal F}}}_{\mu\nu} (x, \theta, \bar\theta), \; {\tilde {\cal F}_{\mu\nu}} (x,\theta,\bar\theta)$
which are nothing but the generalizations of the (anti-)ghost fields 
$(\bar C_{\mu\nu})C_{\mu\nu}$ in our theory with the ghost numbers $(-1)+1$, respectively.
The presence\footnote{It is the theoretical strength of superfield approach to BRST formalism that the 
presence of the Grassmannian components in the superfields lead to the determination of the ghost numbers for the 
{\it corresponding} ordinary fields. Based on this logic, for instance, we see that the ghost-for-the ghost-for-the ghost fields $(\bar C_2)C_2$ carry the 
ghost numbers $(-3)+3$, respectively. }
 of {\it three} Grassmannian  variables: $\frac{1}{3!} \tilde A_{\theta\theta \theta} (x,\theta,\bar\theta)
 = {\tilde {\bar {\cal F}}}_2 (x,\theta,\bar\theta), \; \frac{1}{3!}
A_{\bar\theta\bar\theta\bar\theta} (x,\theta,\bar\theta) = {\tilde {\cal F}}_2  (x,\theta,\bar\theta)$
ensures  that these {\it fermionic} superfields are the generalizations of ordinary fermionic (anti-)ghost 
fields $(\bar C_2)C_2$ which are ghost-for-ghost-for-ghost fields with the ghost numbers $(-3)+3$ respectively. 
The auxiliary (anti-)ghost fields $(\bar C_1)C_1$ have their generalizations [cf. Eq. (12) below] on the (D, 2)-dimensional supermanifold as:
$\frac{1}{2} \tilde A_{\theta\bar\theta\bar\theta} (x,\theta,\bar\theta) = \tilde {\cal F}_1
(x,\theta,\bar\theta), \; \frac{1}{2} \tilde A_{\theta\theta\bar\theta} (x,\theta,\bar\theta) = {\tilde 
{\bar {\cal F}}}_1 (x,\theta,\bar\theta)$  with the ghost numbers $(+1)-1$, respectively
(see, e.g. [17] for details).

The superfields [that have been defined on the (D, 2)-dual supermanifold] can be expanded {\it along} the
Grassmannian directions $(\theta, \bar\theta)$ by the Taylor expansion about $(\theta = 0, \, \bar\theta = 0)$.
The resulting super expansions are (see, e.g. [9-11, 17, 18] for details)
\begin{eqnarray*}
{\cal A}_{\mu\nu\lambda } (x, \theta, \bar \theta) &=& A_{\mu\nu\lambda}(x) + \theta\, \bar R_{\mu\nu\lambda} (x)
+ \bar\theta\, R_{\mu\nu\lambda}(x) + i\, \theta\, \bar\theta\, S_{\mu\nu\lambda}(x), \nonumber\\
\end{eqnarray*}
\begin{eqnarray}
{\cal F}_{\mu\nu} (x, \theta \, \bar \theta) &=&  C_{\mu\nu} + \theta\, \bar B_{\mu\nu}^{(1)} (x)
+ \bar\theta \, B_{\mu\nu}^{(1)} (x) + i\, \theta \, \bar\theta \, s_{\mu\nu}(x), \nonumber\\
{\cal {\bar F}}_{\mu\nu} (x, \theta \, \bar \theta) &=& \bar C_{\mu\nu} + \theta\,  \bar B_{\mu\nu}^{(2)} (x)
+ \bar\theta \, B_{\mu\nu}^{(2)} (x) + i\, \theta \, \bar\theta \, \bar s_{\mu\nu}(x), \nonumber\\
\tilde \beta_{\mu}(x, \theta, \bar \theta) &=&  \beta_\mu (x) + \theta \bar f_\mu ^{(1)} (x) 
+ \bar\theta\, f_\mu ^{(1)} (x) +  i\, \theta \, \bar\theta \, b_{\mu} (x),\nonumber\\
\tilde {\bar \beta}_{\mu}(x, \theta, \bar \theta) &=& \bar \beta_\mu (x) + \theta \bar f_\mu ^{(2)} (x) 
+ \bar\theta\, f_\mu ^{(2)} (x) +  i\, \theta \, \bar\theta \, \bar b_{\mu} (x),\nonumber\\
\Phi_{\mu}(x, \theta, \bar \theta) &=& \phi_\mu (x) + \theta\, \bar f_\mu ^{(3)} (x) 
+ \bar\theta\, f_\mu ^{(3)} (x) +  i\, \theta \, \bar\theta \, b^{(3)}_{\mu} (x),\nonumber\\
{\cal F}_1 (x, \theta, \bar\theta) &=& C_1 (x) + \theta\;\bar b_1^{(1)} (x)
+ \bar \theta\; b_1^{(1)} (x) + i\;\theta\;\bar\theta\; s_1 (x), \nonumber\\
{{\bar {\cal F}}}_1 (x, \theta, \bar\theta) &=& \bar C_1 (x) + \theta\;\bar b_1^{(2)} (x)
+ \bar \theta\; b_1^{(2)} (x) + i\;\theta\;\bar\theta\; \bar s_1 (x), \nonumber\\
{\cal F}_2 (x, \theta, \bar\theta) &=& C_2 (x) + \theta\;\bar b_2^{(1)} (x)
+ \bar \theta\; b_2^{(1)} (x) + i\;\theta\;\bar\theta\; s_2 (x), \nonumber\\
{{\bar {\cal F}}}_2 (x, \theta, \bar\theta) &=& \bar C_2 (x) + \theta\;\bar b_2^{(2)} (x)
+ \bar \theta\; b_2^{(2)} (x) + i\;\theta\;\bar\theta\; \bar s_2 (x), 
\end{eqnarray}
where, because of the {\it fermionic} $(\theta^2 = \bar\theta^2 = 0, \, \theta\,\bar\theta + \bar\theta\,\theta = 0)$ nature 
of the Grassmannian variables $(\theta, \bar\theta)$, we have the {\it fermionic} as well as {\it bosonic} 
secondary fields on the r.h.s. of the above super expansions. 
For instance, in (12), the set of secondary fields $[S_{\mu\nu\lambda} (x), B_{\mu\nu}^{(1)} (x), 
\bar B_{\mu\nu}^{(1)} (x), B_{\mu\nu}^{(2)} (x), \bar B_{\mu\nu}^{(2)} (x), b_\mu (x), \bar b_\mu (x), b_\mu ^{(3)}(x),
b_1 ^{(1)}(x), \bar b_1 ^{(1)}(x),  b_2 ^{(1)}(x), \bar  b_2 ^{(1)}(x),\\ b_2 ^{(2)}(x),\, \bar b_2 ^{(2)}(x)]$
are {\it bosonic} and the set $[R_{\mu\nu\lambda} (x), \; \bar R_{\mu\nu\lambda} (x), \; s_{\mu\nu} (x), \bar s_{\mu\nu} (x),\; 
f_\mu ^{(1)}(x),\; \bar f_\mu ^{(1)}(x),\\ \, f_\mu ^{(2)}(x), \, \bar f_\mu ^{(2)} (x), \, f_\mu ^{(3)} (x), \,\bar f_\mu ^{(3)} (x), \, s_1 (x), \, 
\bar s_1 (x), \,  s_2 (x),\, \bar s_2 (x)]$ constitutes the collection of {\it fermionic}  secondary fields. These secondary fields have to 
be determined in terms of the basic and auxiliary fields of the theory where the HC [cf. Eq. (8)] plays an important role. 
The {\it latter} theoretical trick  lead to the following (cf. Appendix A below for details): 
\begin{eqnarray}
&& R_{\mu\nu\lambda}
= \partial_\mu C_{\nu\lambda} + \partial_\nu C_{\lambda\mu} + \partial_\lambda C_{\mu\nu}, \quad \bar R_{\mu\nu\lambda}
= \partial_\mu \bar C_{\nu\lambda} + \partial_\nu \bar C_{\lambda\mu} + \partial_\lambda \bar C_{\mu\nu},\quad  \bar b_1^{(2)} + b_2^{(2)} = 0,
\nonumber\\
&&  S_{\mu\nu\lambda} = + i \, [\partial_\mu \bar B_{\nu\lambda}^{(1)} + \partial_\nu \bar B_{\lambda\mu}^{(1)}
+ \partial_\lambda \bar B_{\mu\nu}^{(1)}] \equiv  - i \, [\partial_\mu B_{\nu\lambda}^{(2)} + \partial_\nu B_{\lambda\mu}^{(2)}
+ \partial_\lambda B_{\mu\nu}^{(2)}], \quad  b_\mu = i \, \partial_\mu \bar b_2^{(1)}, \nonumber\\
&& \bar s_{\mu\nu} = + i \, (\partial_\mu \bar f_\nu^{(3)} - \partial_\nu \bar f_\mu^{(3)}) \equiv
- i \, (\partial_\mu  f_\nu^{(2)} - \partial_\nu  f_\mu^{(2)}), 
\quad s_{\mu\nu} = i \, (\partial_\mu \bar f_\nu^{(1)} - \partial_\nu \bar f_\mu^{(1)}) \equiv \nonumber\\
&& - i \, (\partial_\mu  f_\nu^{(3)} - \partial_\nu  f_\mu^{(3)}), \quad B^{(1)}_{\mu\nu} = \partial_\mu \beta_\nu 
- \partial_\nu \beta_\mu, \quad \bar B_{\mu\nu}^{(2)} = \partial_\mu  \bar \beta_\nu - 
\partial_\nu \bar \beta_\mu,\quad \bar f_\mu^{(2)} = \partial_\mu \bar C_2, \nonumber\\
&&\bar b_\mu = - i \, \partial_\mu b_2^{(2)}, \quad b_\mu^{(3)} = - i \, \partial_\mu b_1^{(2)},\quad f_\mu^{(1)} = \partial_\mu C_2,\quad 
 \bar b_1^{(1)} +  b_1^{(2)} = 0,\quad  b_1^{(1)} + \bar b_2^{(1)} = 0, \nonumber\\
&& s_1 = 0, \quad \bar s_1 = 0,\quad s_2 = 0, \quad \bar s_2 = 0, \quad b_2^{(1)} = 0,  \quad \bar b_2^{(2)} = 0. 
\end{eqnarray}
In the above equation, there are {\it three}  equivalences. These turn out to be {\it true} due to the 
following CF-type restrictions of our theory, namely; 
\begin{eqnarray}
\bar B_{\mu\nu}^{(1)} + B_{\mu\nu}^{(2)} = \partial_\mu \phi_\nu - \partial_\nu \phi_\mu, \quad
 \bar f_\mu^{(1)} +  f_\mu^{(3)} = \partial_\mu  C_1,\quad  f_\mu^{(2)} + \bar f_\mu^{(3)} = \partial_\mu \bar C_1.  
\end{eqnarray}
The above CF-type restrictions have been obtained from the HC [cf. Eq. (8)] where we have set the coefficients of the 
following {\it super} 4-form differentials equal to zero, namely;  
\begin{eqnarray}
(dx^\mu \wedge d x^\nu \wedge d\theta \wedge d \bar\theta), \quad 
(dx^\mu \wedge d\theta \wedge d \bar \theta \wedge d \bar\theta), \quad 
(dx^\mu \wedge d\theta \wedge d\theta \wedge d \bar\theta),
\end{eqnarray}
which emerge out from the computation of the super 4-form $\tilde H^{(4)} = \tilde d\, \tilde A ^{(3)}$ [i.e. the l.h.s. of Eq. (8)]. 
It is very interesting to point out that r.h.s. and l.h.s. of Eq. (8) match because of the equality of the 
coefficients of the differentials ($dx^\mu \wedge d x^\nu \wedge d x^\lambda  \wedge d x^\xi$) due to the expressions 
for $R_{\mu\nu\lambda}, \bar R_{\mu\nu\lambda}$ and $S_{\mu\nu\lambda}$ [cf. Eq. (13)]. 
Let us identify the secondary fields: $\bar B_{\mu\nu}^{(1)} = \bar B_{\mu\nu}, \, B_{\mu\nu}^{(2)} = B_{\mu\nu}, \,
 b_2 ^{(2)} = B_2, \, \bar b_1 ^{(2)}  = - \, B_2, \, b_1^{(1)} = -\, B, \, \bar b_2^{(1)} = B,\, 
\bar b_1^{(1)} = B_1, \, b_1^{(2)}  = -\, B_1,\, f_{\mu}^{(3)} = f_\mu, \, \bar f_{\mu}^{(1)} = \bar F_\mu, \, f_{\mu}^{(2)} = F_\mu, \,
\bar f_{\mu}^{(3)} =  \bar f_\mu.$
With these identifications and their substitution in Eq. (12),
we obtain the following  (cf. Appendix A for details)
\begin{eqnarray}
\tilde {\cal A}^{(h)}_{\mu\nu\lambda} (x, \theta, \bar\theta) &=& A_{\mu\nu\lambda}
(x) + \theta \,[s_{ab} A_{\mu\nu\lambda} (x)] + \bar\theta \,[s_b A_{\mu\nu\lambda} (x)] +
 \theta  \,\bar\theta \, [s_b s_{ab} A_{\mu\nu\lambda} (x)], \nonumber\\ 
 {\cal F}^{(h)}_{\mu\nu} (x, \theta, \bar\theta) &=& C_{\mu\nu}
(x) + \theta \, [s_{ab} C_{\mu\nu} (x)] + \bar\theta \, [s_b C_{\mu\nu} (x)]
+ \theta \, \bar\theta \, [s_b s_{ab} C_{\mu\nu} (x)], \nonumber\\
 {\bar {\cal F}}^{(h)}_{\mu\nu} (x, \theta, \bar\theta) &=& \bar C_{\mu\nu} (x) +
\theta \, [s_{ab} \bar C_{\mu\nu} (x)] + \bar\theta \, [s_b \bar C_{\mu\nu} (x)] + 
\theta \, \bar\theta \, [s_b s_{ab} \bar C_{\mu\nu} (x)], \nonumber\\
\tilde \beta^{(h)}_\mu
(x, \theta, \bar\theta ) &=& \beta_\mu (x) + \theta \;[s_{ab} \beta_\mu (x)] +
\bar\theta\; [s_b \beta_\mu (x)] +  \theta\; \bar\theta\; [s_b s_{ab} \beta_\mu (x)],
\nonumber\\
 \tilde {\bar \beta}^{(h)}_\mu (x, \theta, \bar\theta) &=& 
\bar\beta_\mu (x) + \theta \;[s_{ab} \bar\beta_\mu (x)] 
+ \bar \theta\; [s_b \bar\beta_\mu (x)] 
+ \theta\;\bar\theta\; [s_b s_{ab} \bar\beta_\mu (x)], \nonumber\\ 
\Phi^{(h)}_\mu (x, \theta, \bar\theta) &=& \phi_\mu (x) + \theta \;[s_{ab} \phi_\mu (x)] +
\bar\theta\; [s_b \phi_\mu (x)] + \theta \;\bar\theta\; [s_b s_{ab} \phi_\mu (x)],
\nonumber\\ 
{\cal F}^{(h)}_1 (x, \theta, \bar\theta) &=& C_1 (x) + \theta \;(s_{ab} C_1 (x))
+ \bar \theta\; [s_b C_1 (x)] + \theta \;\bar\theta\; [s_b s_{ab} C_1 (x)], \nonumber\\
 {\bar {\cal F}}^{(h)}_1 (x, \theta, \bar\theta) &=& \bar C_1 (x) + \theta\; (s_{ab}
\bar C_1 (x))
+ \bar \theta\; [s_b \bar C_1 (x)] + \theta\;\bar\theta\; [s_b s_{ab} \bar C_1 (x)], \nonumber\\
{\cal F}^{(h)}_2 (x, \theta, \bar\theta) &=& C_2 (x) + \theta\;[s_{ab} C_2 (x)]
+ \bar \theta\; [s_b C_2 (x)] + \theta\;\bar\theta\; [s_b s_{ab} C_2 (x)], \nonumber\\
{\bar {\cal F}}^{(h)}_2 (x, \theta, \bar\theta) &=& \bar C_2 (x) 
+ \theta\;[s_{ab} \bar C_2 (x)]
+ \bar \theta\; [s_b \bar C_2 (x)] 
+ \theta\;\bar\theta\; [s_b s_{ab} \bar C_2 (x)], 
\end{eqnarray}
where the superscript $(h)$ denotes that the above superfields have been obtained after the substitution 
of the secondary fields (13) into the super expansions (12) which have been obtained after the application of the 
HC. In the above equation (16), the explicit form of the (anti-)BRST symmetry transformations $s_{(a)b}$ are as follows: 
\begin{eqnarray}
&& s_{ab} A_{\mu\nu\lambda} = \partial_\mu \bar C_{\nu\lambda} 
+ \partial_\nu \bar C_{\lambda\mu} + \partial_\lambda \bar C_{\mu\nu},
\qquad s_{ab} \bar C_{\mu\nu} = \partial_\mu \bar \beta_\nu - \partial_\nu \bar\beta_\mu, 
 \nonumber\\
&& s_{ab} \bar \beta_\mu = \partial_\mu \bar C_2, \qquad s_{ab} \bar C_2 = 0, \qquad 
s_{ab} \bar B_{\mu\nu} = 0, \qquad
s_{ab} C_1 =  B_1,  \nonumber\\
&& s_{ab} \bar C_1 = - B_2, \qquad s_{ab}  B = 0, \qquad s_{ab} C_2 =  B, 
\quad s_{ab}  \beta_\mu = \bar F_\mu,
\quad s_{ab} \bar F_\mu = 0,  \nonumber\\
&& s_{ab} \bar f_\mu = 0, \quad s_{ab}  F_\mu = - \partial_\mu B_2, \quad 
s_{ab}  f_\mu =  \partial_\mu B_1,
\quad s_{ab} B_{\mu\nu} = -\, (\partial_\mu  F_\nu - \partial_\nu  F_\mu), \nonumber\\
&& s_{ab}  C_{\mu\nu} = \bar B_{\mu\nu}, \qquad  s_{ab} B_1 = 0, \qquad s_{ab}  B_2 = 0,
\qquad s_{ab} \phi_\mu = \bar f_\mu,
\end{eqnarray}
\begin{eqnarray}
&& s_b A_{\mu\nu\lambda} = \partial_\mu C_{\nu\lambda} + \partial_\nu C_{\lambda\mu} + \partial_\lambda C_{\mu\nu},
\qquad s_b C_{\mu\nu} = \partial_\mu \beta_\nu - \partial_\nu \beta_\mu,  \nonumber\\
&& s_b \beta_\mu = \partial_\mu C_2, \quad s_b C_2 = 0, \quad s_b B_{\mu\nu} = 0, \quad
s_b C_1 = - B,  \nonumber\\
&& s_b \bar C_1 = -\, B_1, \quad s_b B_1 = 0, \quad s_b \bar C_2 = B_2, \quad s_b \bar \beta_\mu = F_\mu,
\quad s_b \,  F_\mu = 0, \nonumber\\
&& s_b f_\mu = 0, \quad s_b \bar F_\mu = - \partial_\mu  B, \quad 
s_b \bar f_\mu = -\, \partial_\mu B_1,
\quad s_b \bar B_{\mu\nu} = -\, (\partial_\mu \bar F_\nu - \partial_\nu \bar F_\mu), \nonumber\\
&& s_b \bar C_{\mu\nu} = B_{\mu\nu}, \qquad  s_b  B = 0, \qquad s_b  B_2 = 0, \qquad s_b \phi_\mu = f_\mu.
\end{eqnarray}
It is an elementary exercise  to check that the above (anti-)BRST symmetry transformations $s_{(a)b}$
are off-shell nilpotent $[s_{(a)b}^2 = 0]$ of order two. On the other hand, we note that the absolute anticommutativity property  
$(s_b\, s_{ab} + s_{ab}\, s_b = 0)$ of the (anti-)BRST symmetry transformations for the following fields, namely;
\begin{eqnarray}
\{s_b, s_{ab} \} \;A_{\mu\nu\eta} = 0, \;\qquad 
\{s_b, s_{ab} \} \; C_{\mu\nu} = 0, \;\qquad
\{s_b, s_{ab} \}\; \bar C_{\mu\nu} = 0,
\end{eqnarray}
is  satisfied if and only if the CF-type restrictions: 
$ B_{\mu\nu} + \bar B_{\mu\nu} = \partial_\mu \phi_\nu - \partial_\nu \phi_\mu,\, 
f_\mu +  \bar F_\mu = \partial_\mu C_1,\, 
\bar f_\mu +  F_\mu = \partial_\mu \bar C_1$ are satisfied [which are nothing but Eq. (14) 
but expressed in the {\it new} notations]. It is interesting to point out that the absolute anticommutativity 
property $(s_b\, s_{ab} + s_{ab}\, s_b = 0)$ is automatically  satisfied for the {\it rest} of the fields of our theory. 
A close look at the above CF-type restrictions establishes that the fermionic  pair $(F_\mu, \, \bar F_\mu)$ of fields
 carry the ghost numbers ($-1, +1$), respectively.

We end this section with the {\it final} remark that we have obtained the super Abelian 3-form $\tilde A_{(h)} ^{(3)}$,
after the application of the HC [cf. Appendix A for details], as follows:  
\begin{eqnarray}
 \tilde A ^{(3)} _{(h)}
& =&   \frac{1}{3!}\; (dx^\mu \wedge dx^\nu \wedge dx^\lambda )\; \tilde {A}_{\mu\nu\lambda }^{(h)} (x, \theta, \bar\theta)\; + \;
\frac{1}{2} (dx^\mu \wedge dx^\nu \wedge d\theta) \;
\bar F_{\mu\nu}^{(h)} (x, \theta, \bar\theta)\nonumber\\ 
& + & \frac{1}{2} (dx^\mu \wedge dx^\nu \wedge d \bar\theta)\; 
F_{\mu\nu}^{(h)} (x, \theta, \bar\theta) + (d\theta \wedge d\theta
\wedge d\theta) \; {\cal {\bar F}}_2 ^{(h)} (x, \theta, \bar\theta) \nonumber\\
 & + & (dx^\mu \wedge d \theta \wedge d\bar\theta) \; \Phi_{\mu} ^{(h)} (x, \theta, \bar\theta) + 
(dx^\mu \wedge d\theta \wedge d \theta) \; {\tilde{\bar \beta}} _\mu  ^{(h)} (x, \theta, \bar\theta) \nonumber\\
& + &  (dx^\mu \wedge d\bar\theta \wedge d \bar\theta) \; {\tilde{\beta}} _\mu  ^{(h)} (x, \theta, \bar\theta)
+ (d\theta \wedge d\theta \wedge d \bar\theta) \; {\cal {\bar F}}_1 ^{(h)} (x, \theta, \bar\theta)\nonumber\\
 &+& (d\theta \wedge d\bar\theta \wedge d \bar\theta) \;{\cal {F}}_1 ^{(h)}(x, \theta, \bar\theta)
+   (d \bar\theta \wedge d\bar\theta \wedge d \bar\theta) \; {\cal {F}}_2 ^{(h)}(x, \theta, \bar\theta).
\end{eqnarray}
In the above, we have the superfields 
after the application of HC and their explicit forms are listed in Eq. (16) where the {\it exact} expressions for the (anti-)BRST 
symmetry transformations $s_{(a)b}$ are quoted in equations (17) and (18), respectively. In the next section, we shall exploit 
the explicit forms of (16) and (20) for the application of AVSA.

\section {(Anti-)BRST Symmetries for the St{$\ddot u$}ckelberg  and Associated (Anti-)Ghost Fields: AVSA}

In this section, we apply the key idea of AVSA to BRST formalism and derive, first of all, 
the (anti-)BRST symmetry transformations for $\Sigma_{\mu\nu\lambda}$. In this connection, 
we note that the gauge invariance [cf. Eq. (7)] implies the following on the (D, 2)-dimensional supermanifold 
\begin{eqnarray}
\tilde {\cal A}^{(h)}_{\mu\nu\lambda}\, (x, \theta, \bar\theta) \,\mp \,\frac {1}{m} \, \tilde \Sigma_{\mu\nu\lambda}\, (x, \theta, \bar\theta)  
= A_{\mu\nu\lambda} (x)\, \mp \, \frac {1}{m} \,\Sigma_{\mu\nu\lambda}\, (x),
\end{eqnarray} 
where the expression for $\tilde {\cal A}^{(h)}_{\mu\nu\lambda} (x, \theta, \bar\theta)$ has been quoted in Eq. (16)
and we have the following super expansion  of $\tilde \Sigma_{\mu\nu\lambda}\, (x, \theta, \bar\theta)$
\begin{eqnarray}
\tilde \Sigma_{\mu\nu\lambda}\, (x, \theta, \bar\theta) = \Sigma_{\mu\nu\lambda}\, (x) +\theta\, \bar P_{\mu\nu\lambda} (x)
+ \bar\theta \, P_{\mu\nu\lambda} (x) + i\, \theta\, \bar\theta\, Q_{\mu\nu\lambda} (x), 
\end{eqnarray} 
where (due to the fermionic nature of $\theta, \bar\theta$) the pair $(P_{\mu\nu\lambda}, \; \bar P_{\mu\nu\lambda})$ 
are the {\it fermionic}  secondary fields and 
$Q_{\mu\nu\lambda} (x)$ is the {\it bosonic} secondary field which have to be determined  in terms of the {\it basic} and 
{\it auxiliary} fields of our D-dimensional {\it massive} Abelian 3-form gauge theory. In fact, the substitution of the explicit 
expression for 
\begin{eqnarray}
\tilde {\cal A}^{(h)}_{\mu\nu\lambda}\, (x, \theta, \bar\theta) & = & {A}_{\mu\nu\lambda}\, (x) + \theta \,
(\partial_\mu \bar C_{\nu\lambda} + \partial_\nu \bar C_{\lambda\mu} + \partial_\lambda \bar C_{\mu\nu})
+ \bar\theta\, (\partial_\mu C_{\nu\lambda} + \partial_\nu C_{\lambda\mu} + \partial_\lambda C_{\mu\nu})\nonumber\\
& + & \theta\, \bar\theta\,(\partial_\mu B_{\nu\lambda} + \partial_\nu B_{\lambda\mu}
+ \partial_\lambda  B_{\mu\nu})\nonumber\\
 & \equiv  & {A}_{\mu\nu\lambda}\, (x) + \theta \,
(\partial_\mu \bar C_{\nu\lambda} + \partial_\nu \bar C_{\lambda\mu} + \partial_\lambda \bar C_{\mu\nu})
+ \bar\theta\, (\partial_\mu C_{\nu\lambda} + \partial_\nu C_{\lambda\mu} + \partial_\lambda C_{\mu\nu})\nonumber\\
& + & \,\theta\, \bar\theta\,[-(\,\partial_\mu \bar B_{\nu\lambda} + \partial_\nu \bar B_{\lambda\mu}
+ \partial_\lambda  \bar B_{\mu\nu})],   
\end{eqnarray}
in Eq. (21) leads to the following: 
\begin{eqnarray}
P_{\mu\nu\lambda} (x) & =  &  \pm\, m\,(\partial_\mu C_{\nu\lambda} + \partial_\nu C_{\lambda\mu} + \partial_\lambda C_{\mu\nu})
\equiv s_b \, \Sigma_{\mu\nu\lambda}, \nonumber\\
\bar P_{\mu\nu\lambda} (x) & = & \pm\, m\,(\partial_\mu \bar C_{\nu\lambda} + \partial_\nu \bar C_{\lambda\mu} + \partial_\lambda \bar C_{\mu\nu})
\equiv s_{ab} \, \Sigma_{\mu\nu\lambda}, \nonumber\\
Q_{\mu\nu\lambda} (x) & = & \pm\,m\, (\partial_\mu B_{\nu\lambda} + \partial_\nu B_{\lambda\mu} + \partial_\lambda B_{\mu\nu})
\equiv s_b\, s_{ab} \, \Sigma_{\mu\nu\lambda},\nonumber\\
& \equiv &  \mp\,m\,  (\partial_\mu \bar B_{\nu\lambda} + \partial_\nu \bar B_{\lambda\mu} + \partial_\lambda \bar B_{\mu\nu})
\equiv -\,  s_{ab}\, s_b \, \Sigma_{\mu\nu\lambda}.
\end{eqnarray}
The above forms of the secondary fields imply that we have obtained 
\begin{eqnarray}
\tilde \Sigma_{\mu\nu\lambda}^{(g)}\, (x, \theta, \bar\theta) & = &  \Sigma_{\mu\nu\lambda}\, (x) +\theta\, (s_{ab}\,\Sigma_{\mu\nu\lambda})
+ \bar\theta \, (s_{b}\,\Sigma_{\mu\nu\lambda}) + \theta\, \bar\theta\, (s_b\, s_{ab} \, \Sigma_{\mu\nu\lambda}),\nonumber\\
& \equiv  &  \Sigma_{\mu\nu\lambda}\, (x) +\theta\, (s_{ab}\,\Sigma_{\mu\nu\lambda})
+ \bar\theta \, (s_{b}\,\Sigma_{\mu\nu\lambda}) + \theta\, \bar\theta\, (-\, s_{ab}\, s_{b} \, \Sigma_{\mu\nu\lambda}),  
\end{eqnarray}
where the superscript $(g)$ denotes the superfield [i.e. $\tilde\Sigma_{\mu\nu\lambda}^{(g)}\, (x, \theta, \bar\theta)$] 
has been obtained after the application of GIR that has been quoted in (21).

At this juncture, we are in the position to exploit the gauge invariant condition (7) for the purpose of the 
application of AVSA to BRST formalism. In other words, we have the following GIR in the language of differential forms, namely;
\begin{eqnarray}
\tilde{A}_{(h)}^{(3)} \mp \, \frac{1}{m}\,\tilde d\, \tilde \Phi^{(2)} \; \equiv \;  A^{(3)} \mp \frac{1}{m}\, d\, \Phi^{(2)},
\end{eqnarray}
where the expansion for $\tilde{A}_{(h)}^{(3)}$ has been quoted in (20) and all the superfields with the superscript $(h)$
have been written in our Appendix A. The expression for the super derivative $\tilde d$ has been quoted in (8)
and the explicit form of the super 2-form is as follows
\begin{eqnarray}
\tilde \Phi^{(2)} &=& \frac{1}{2 !}\,(d\,Z^M \wedge d\, Z^N) \, \tilde\Phi_{MN}
\equiv \frac{1}{2 !}\,(d\,x^\mu \wedge d\, x^\nu) \, \tilde\Phi_{\mu\nu}\, (x,\, \theta,\, \bar\theta)
+ (d\, x^\mu \wedge d\, \theta)\, \tilde\Phi_{\mu\theta}\,(x,\, \theta,\, \bar\theta) \nonumber\\
&+& (d\, x^\mu \wedge d\, \bar\theta)\, \tilde\Phi_{\mu\bar\theta}\,(x,\, \theta,\, \bar\theta)
+ (d\,\theta \wedge d\, \bar\theta)\, \tilde\Phi_{\theta\bar\theta}\,(x,\, \theta,\, \bar\theta)
+ \frac{1}{2 !}\,(d\, \theta \wedge d\, \theta)\, \tilde\Phi_{\theta\theta}\,(x,\, \theta,\, \bar\theta) \nonumber\\
&+& \frac{1}{2 !}\,(d\, \bar\theta \wedge d\, \bar\theta)\, \tilde\Phi_{\bar\theta\bar\theta}\,(x,\, \theta,\, \bar\theta),
\end{eqnarray}
where we identify the fermionic as well as bosonic superfields on the r.h.s. as\footnote{As pointed out earlier, 
the presence of the Grassmannian components of the superfields decides the nature of the corresponding {\it ordinary} fields and their
ghost numbers.}:
\begin{eqnarray}
\tilde\Phi_{\mu\theta} = \tilde{\bar F}_\mu\, ( x,\, \theta,\, \bar\theta), \quad 
\tilde\Phi_{\mu\bar\theta} = \tilde F_\mu\,( x,\, \theta,\, \bar\theta), \quad
\tilde\Phi_{\theta\bar\theta} = \tilde\Phi\,( x,\, \theta,\, \bar\theta),\nonumber\\
\frac{1}{2\, !}\, \tilde\Phi_{\theta\theta} = \tilde {\bar\beta}\,(x,\, \theta,\, \bar\theta),\quad
\frac{1}{2\, !}\, \tilde\Phi_{\bar\theta\bar\theta} = \tilde {\beta}\,(x,\, \theta,\, \bar\theta).
\end{eqnarray}
It is evident that the {\it fermionic} superfields $(\tilde F_\mu, \, \tilde{\bar F}_\mu)$ have the ghost numbers $(+1, -1)$
and the {\it bosonic} superfields $(\tilde\beta,\, \tilde{\bar \beta})$ are characterized by the ghost numbers $(+2, -2)$, respectively.
On the contrary, the scalar superfield $\tilde\Phi\, ( x,\, \theta,\, \bar\theta)$ has the {\it zero} ghost number.
It is straightforward to note that, we have the following
\begin{eqnarray}
\mp \, \frac{1}{m}\, \tilde d \, \tilde \Phi^{(2)} \equiv  &\mp&\, \frac{1}{m}\,\Big[ \frac{1}{3\, !} \, 
(d\,x^\mu \wedge d\, x^\nu \wedge d\, x^\lambda)\, \{\partial_\mu\, \tilde\Phi_{\nu\lambda} + \partial_\nu\, \tilde\Phi_{\lambda\mu}
+ \partial_\lambda\, \tilde\Phi_{\mu\nu} \}  \nonumber\\
&+&  \frac{1}{2 !}\,(d\,x^\mu \wedge d\, x^\nu \wedge d\, \theta) 
\Big\{\partial_\theta \, \tilde\Phi_{\mu\nu}  + (\partial_\mu\, \tilde{\bar F}_\nu - \partial_\nu\, \tilde{\bar F}_\mu)\Big\} \nonumber\\
 &+& \frac{1}{2 !}\,(d\,x^\mu \wedge d\, x^\nu \wedge d\, \bar\theta) 
\Big\{\partial_{\bar\theta} \, \tilde{\bar\Phi}_{\mu\nu}  + (\partial_\mu\,  \tilde F_\nu - \partial_\nu\, \tilde F_\mu)\Big\} \nonumber\\
&+&  (d\, \theta \wedge d\, \theta \wedge d\, \theta)\, \partial_\theta\, \tilde{\bar\beta} 
+ (d\, x^\mu \wedge d\, \theta \wedge d\, \bar\theta)\, \{\partial_\mu\, \tilde\Phi + \partial_{\bar\theta} \tilde{\bar F}_\mu 
+ \partial_\theta\,\tilde F_\mu \}   \nonumber\\
 &+&  (d\, x^\mu \wedge d\, \theta \wedge d\, \theta)\, \{\partial_\mu\, \tilde{\bar\beta} 
+ \partial_\theta\, \tilde{\bar F}_\mu\}
+ (d\, x^\mu \wedge d\, \bar\theta \wedge d\, \bar\theta)\, \{\partial_\mu\, \tilde{\beta} 
+ \partial_{\bar\theta}\, \tilde{F}_\mu\}  \nonumber\\
&+& (d\, \theta \wedge d\, \theta \wedge d\, \bar\theta)\, \{\partial_{\bar\theta}\,\tilde{\bar\beta} + 
\partial_\theta\, \tilde\Phi\} 
+ (d\, \theta \wedge d\, \bar\theta \wedge d\, \bar\theta)\, \{ 
\partial_{\bar\theta}\, \tilde\Phi + \partial_{\theta}\,\tilde{\beta} \} \nonumber\\
&+&  (d\, \bar\theta \wedge d\, \bar\theta \wedge d\, \bar\theta)\, \partial_{\bar\theta}\, \tilde{\beta}\Big].
\end{eqnarray}
We would like to point out that, using the explicit expressions  of (20) and (29), we can compute the l.h.s. of (28) and equate it with the r.h.s.

A close and careful look at (26) demonstrates that all the super 3-differentials with the Grassmannian variables on the l.h.s.
will {\it not} have their counterparts on the r.h.s. As a consequence, {\it their} coefficients will be equal to zero.
This requirement leads to the following
\begin{eqnarray}
&&\bar{\cal F}_2^{(h)} \mp\, \frac {1}{m}\, \partial_\theta\, \tilde{\bar\beta} = 0, \qquad \qquad  \quad 
{\cal F}_2^{(h)} \mp\, \frac {1}{m}\, \partial_{\bar\theta}\, \tilde{\beta} = 0, \nonumber\\
&&\bar{\cal F}_1^{(h)} \mp \, \frac{1}{m}\, (\partial_{\bar\theta}\, \tilde{\bar\beta} + \partial_\theta\, \tilde\Phi) = 0, \quad
{\cal F}_1^{(h)} \mp \, \frac{1}{m}\, (\partial_{\theta}\, \tilde{\beta} + \partial_{\bar\theta}\, \tilde\Phi) = 0, \nonumber\\
&&\tilde{\bar\beta}_\mu^{(h)} \mp \frac{1}{m}\, (\partial_\mu\, \tilde{\bar\beta} + \partial_\theta\, \tilde{\bar F}_\mu) = 0, \quad
\tilde{\beta}_\mu^{(h)} \mp \frac{1}{m}\, (\partial_\mu\, \tilde{\beta} + \partial_{\bar\theta}\, \tilde F_\mu) = 0, \nonumber\\
&&\tilde\Phi_\mu^{(h)} \mp \frac{1}{m}\, (\partial_\mu\, \tilde\Phi + \partial_{\bar\theta} \tilde{\bar F}_\mu + \partial_\theta\, \tilde F_\mu) = 0, \quad
\bar F_{\mu\nu}^{(h)} \mp\, \frac{1}{m}\, (\partial_\theta\, \tilde\Phi_{\mu\nu} + \partial_\mu\, 
\tilde{\bar F}_\nu - \partial_\nu \, \tilde{\bar F}_\mu = 0), \nonumber\\
&&F_{\mu\nu}^{(h)} \mp\, \frac{1}{m}\, (\partial_{\bar\theta}\, \tilde\Phi_{\mu\nu} + \partial_\mu\, \tilde F_\nu - \partial_\nu \, \tilde F_\mu) = 0.
\end{eqnarray}
It goes without saying that only the coefficient of $(d\,x^\mu \wedge d\, x^\nu \wedge d\, x^\lambda)$ will be, 
finally, equated from the l.h.s. and r.h.s. It has been already found that both the sides are {\it equal} provided 
we substitute the expressions for the secondary fields\footnote{It is gratifying to  state that we have derived the exact expressions
for {\it all} the secondary fields in terms of the basic and auxiliary fields of our theory and some of the (non-)trivial CF-type
restrictions in equations (32), (34), (35) and (38) of our present endeavor.} in the super expansions of the following superfields which are 
present in (27) and (28), namely;
\begin{eqnarray}
&&\tilde\beta\,(x, \theta, \bar\theta) = \beta + \theta \, \bar g_1 + \bar \theta \, g_1 + i\, \theta \,\bar\theta \, h_1, \nonumber\\
&&\tilde{\bar\beta}\,(x, \theta, \bar\theta) = \bar\beta + \theta \, \bar g_2 + \bar \theta \, g_2 + i\, \theta \,\bar\theta \, h_2, \nonumber\\
&&\tilde\Phi\,(x, \theta, \bar\theta) =\phi + \theta\, \bar k + \bar\theta\, k + i\, \theta\, \bar\theta \, h, \nonumber\\
&&\tilde F_\mu\,(x, \theta, \bar\theta) = C_\mu + \theta\,\bar p_{\mu}^{(1)} + \bar\theta\, p_\mu^{(1)} + i\, \theta\, \bar\theta\, q_\mu, \nonumber\\
&&\tilde {\bar F}_\mu\,(x, \theta, \bar\theta) = \bar C_\mu + \theta\,\bar p_{\mu}^{(2)} 
+ \bar\theta\, p_\mu^{(2)} + i\, \theta\, \bar\theta\, \bar q_\mu,\nonumber\\
&& \tilde\Phi_{\mu\nu}\, (x, \theta, \bar\theta) = \Phi_{\mu\nu} + \theta\, \bar R_{\mu\nu} + \bar\theta\, R_{\mu\nu}
+ i\, \theta\,\bar\theta\, S_{\mu\nu},
\end{eqnarray}
where the fermionic ($\theta^2 = \bar\theta^2 = 0, \, \theta\,\bar\theta + \bar\theta\, \theta = 0$) nature of the Grassmannian variables
($\theta, \, \bar\theta$) ensures that the set of secondary fields ($k,\, \bar k, \, g_1, \, \bar g_1,\, g_2, \, 
\bar g_2, \, q_\mu, \, \bar q_\mu, \,  R_{\mu\nu},\,\bar R_{\mu\nu}$)
are fermionic in nature and the fields $(h,\, h_1, \, h_2,\, p_\mu^{(1)}, \, \bar p_\mu^{(1)}, \,p^{(2)}_\mu, \, \bar p_\mu^{(2)},\, S_{\mu\nu})$
form a {\it bosonic} set of secondary fields which are to be determined in terms of the {\it basic} and {\it auxiliary}
 fields of our theory. In the above context, 
it can be noted that the {\it first four} entries in equation (30) lead to the following expressions for the secondary fields
\begin{eqnarray}
&&g_1 = \pm \, m\, C_2, \;\; \bar g_2 = \pm \, m\, \bar C_2,\; \; h_1 = \pm\,i\,  m \, B, \;\; h_2 = \mp\, i\, m\, B_2 \nonumber\\
&&k + \bar g_1 = \pm \, m\, C_1, \qquad g_2 + \bar k = \pm \, m\, \bar C_1, \qquad h = \pm\, i\, m\, B_1, 
\end{eqnarray}
which demonstrate that we have obtained the explicit expressions for some of the secondary fields in terms of 
the auxiliary and basic fields of our theory. Furthermore, with identifications: 
$g_2 = F, \;\;\bar k = \bar f, \;\; k = f, \;\; \bar g_1 = \bar F$, we have also derived the new set of CF-type restrictions: 
$f + \bar F = \pm\, m\, C_1$ and $\bar f + F = \pm\, m\, \bar C_1 $ which are over and above the CF-type 
restrictions (14). Ultimately, we have derived the super expansions of the {\it three} bosonic superfields
[cf. Eqs (23), (28)] which are the generalization of the (anti-)ghost fields $(\beta,\, \bar\beta)$ with ghost numbers $(+2, -\,2)$
and the scalar field $\phi\, (x)$ with ghost number zero. These super expansions can be explicitly written as 
\begin{eqnarray}
\tilde \beta^{(g)}\, (x, \theta, \bar\theta) &=& \beta + \theta\, (\bar F) + \bar\theta\, (\pm\, m\, C_2) + \theta\,\bar\theta\, (\mp\, m\, B)\nonumber\\
&\equiv& \beta + \theta\, (s_{ab}\, \beta) + \bar\theta\, (s_b\, \beta) + \theta\, \bar\theta\, (s_b\, s_{ab}\,\beta),\nonumber\\
\tilde {\bar\beta}^{(g)}\, (x, \theta, \bar\theta) &=& \bar\beta + \theta\, (\pm\, m\, \bar C_2) 
+ \bar\theta\, (F) + \theta\,\bar\theta\, (\pm\, m\, B_2)\nonumber\\
&\equiv& \bar\beta + \theta\, (s_{ab}\, \bar\beta) + \bar\theta\, (s_b\, \bar\beta) + \theta\, \bar\theta\, (s_b\, s_{ab}\,\bar\beta),\nonumber\\
\tilde\Phi^{(g)}\,(x, \theta, \bar\theta) &=& \phi + \theta\, (\bar f) + \bar\theta\, (f) + \theta\,\bar\theta \, (\mp\, m\, B_1) \nonumber\\
&\equiv& \phi + \theta\, (s_{ab}\, \phi) + \bar\theta\, (s_b\, \phi) + \theta\, \bar\theta\, (s_b\, s_{ab}\,\phi),
\end{eqnarray}
where the superscript $(g)$ on the superfields denotes the fact that the superfields on the l.h.s. have been derived after the 
application of the GIR in (26). It goes without saying that we have already obtained the (anti-)BRST symmetry transformations:
\begin{eqnarray}
&&s_b\, \beta = \pm\, m\, C_2, \quad s_b\, \phi = f, \quad  s_b\, f = 0,\quad  s_b\, \bar\beta = F,  \nonumber\\ 
&&s_b\, \bar F = \mp\, m\, B, \quad  s_b\, C_2 = 0, \quad  s_b\, F = 0, \quad  s_b\, \bar f = \mp\, m\, B_1, \nonumber\\
&&s_{ab}\,\beta = \bar F, \quad \quad  s_{ab}\,\phi = \bar f, \quad  s_{ab}\,\bar f = 0, 
\qquad s_{ab}\,\bar\beta = \pm\, m\, \bar C_2,  \nonumber\\
&& s_{ab}\, \bar F = 0, \quad \quad s_{ab}\,\bar C_2 = 0, \;\; s_{ab}\, F = \mp\, m\, B_2, \;\; s_{ab}\, f = \pm\, m\, B_1,
\end{eqnarray}
where the off-shell nilpotency property has been taken into account for the (anti-)BRST symmetry transformations for some of the 
fields. In addition, we have derived $s_b\, \bar F = \mp\, m\, B$ and $s_{ab}\, F = \mp\, m\, B_2$ from the requirement of the 
(anti-)BRST invariance of the CF-type restrictions: $f +\bar F = \pm\, m\, C_1$ and $\bar f + F = \pm\, m\, \bar C_1$.
As a side remarks, it is interesting to point out that the CF-type restriction: $f + \bar F =\pm \, m\, C_1$ has been derived 
from equating the coefficient of the super differential $(d \theta \wedge d \bar\theta \wedge d \bar\theta )$ equal to zero and similar
exercise has been performed with the super differential $(d \theta \wedge d \theta \wedge d \bar\theta )$ which leads to the derivation of the CF-type 
restriction $\bar f + F =\pm \, m\, \bar C_1$.
At this stage, we equate the coefficient of $(d x^\mu \wedge d \theta \wedge d \theta), \; (d x^\mu \wedge d \bar\theta \wedge d \bar \theta)$
and $(d x^\mu \wedge d \theta \wedge d \bar\theta)$ equal to zero. These amount to taking into account 
the {\it fifth}, {\it sixth} and {\it seventh} entries of equation (30) which lead to the following expression for some of the secondary 
fields and a beautiful relationship: 
\begin{eqnarray}
&&\bar p_\mu^{(2)} = \pm \, m\, \bar\beta_\mu - \partial_\mu\, \bar\beta, \qquad \quad  \bar q_\mu 
= -\, i\, \big [\pm \, m\, F_\mu -\partial_\mu\, F \big ], \nonumber\\
&&p_\mu^{(1)} = \pm \, m\, \beta_\mu - \partial_\mu\, \beta, \qquad  \quad q_\mu 
= +\,  i\, \big [\pm \, m\, \bar F_\mu -\partial_\mu\, \bar F \big ], \nonumber\\
&&\bar p_\mu^{(1)} + p_\mu^{(2)} = \pm \, m \, \phi_\mu - \partial_\mu\, \phi \; \; \Longrightarrow \; \;
\bar B_\mu + B_\mu = \pm\, m\, \phi_\mu - \partial_\mu\, \phi. 
\end{eqnarray}
In the above , we have identified $\bar p_\mu^{(1)} = \bar B_\mu$ and $p_\mu^{(2)} = B_\mu$ which lead to the derivation of
a new CF-type restriction $\bar B_\mu + B_\mu = \pm\, m\, \phi_\mu - \partial_\mu\, \phi$.
This has been obtained from the condition that the coefficient of the super differential $(d x^\mu \wedge d \theta \wedge d \bar \theta)$
should be set equal to zero. Ultimately, we obtain the following super expansions 
\begin{eqnarray}
\tilde F_\mu^{(g)}\, (x,\, \theta, \, \bar\theta) = C_\mu + \theta\, \bar{B}_\mu + \bar\theta \,[\pm \, m\, \beta_\mu - \partial_\mu\, \beta]
+ \theta\bar\theta\, [\mp \, m\, \bar F_\mu -\partial_\mu\, \bar F], \nonumber\\
\tilde {\bar F}_\mu^{(g)}\, (x,\, \theta, \, \bar\theta) = \bar C_\mu + \theta\,[\pm \, m\, \bar\beta_\mu - \partial_\mu\, \bar\beta]
 + \bar\theta \, {B}_\mu
+ \theta\bar\theta\, [\mp \, m\,  F_\mu -\partial_\mu\, F],
\end{eqnarray}
which show that we have already derived: $s_b\, C_\mu = \pm\, m\, \beta_\mu - \partial_\mu\, \beta, \; 
s_b\, (\pm\, m\, \beta_\mu - \partial_\mu\, \beta) = 0,\; s_b\, \bar C_\mu = B_\mu, \; s_b\, B_\mu = 0, \; 
s_{ab}\, C_\mu = \bar B_\mu,\; s_{ab}\, \bar B_\mu = 0, \; s_{ab}\, \bar C_\mu = \pm\, m\, \bar\beta_\mu
 - \partial_\mu\, \bar\beta, \; s_{ab}\, ( \pm\, m\, \bar\beta_\mu -  \partial_\mu\, \bar\beta) = 0$
where the property of the off-shell nilpotency has been taken into account.

The stage is now set to implement the last two conditions of (30). These are
\begin{eqnarray}
\bar F_{\mu\nu}^{(h)} \pm \, \frac{1}{m}\, [\partial_\theta\, \tilde\Phi_{\mu\nu} + \partial_\mu\, \tilde {\bar F}_\nu^{(g)}
 - \partial_\nu\,\, \tilde{\bar F}_\mu^{(g)}] = 0, \nonumber\\
F_{\mu\nu}^{(h)} \pm \, \frac{1}{m}\, [\partial_{\bar\theta}\, \tilde\Phi_{\mu\nu} + \partial_\mu\, \tilde { F}_\nu^{(g)}
 - \partial_\nu\,\, \tilde{F}_\mu^{(g)}] = 0,  
\end{eqnarray}
where the explicit super expansions for the $\bar F_{\mu\nu}^{(h)}$ and $ F_{\mu\nu}^{(h)}$ are listed in 
Appendix A and (36) has to be taken into account for the explicit super expansions of $\tilde{\bar F}_\mu^{(g)}$ and $\tilde{F}_\mu^{(g)}$.
These substitutions in (37) lead to the following
\begin{eqnarray}
\bar R_{\mu\nu} = \pm m\, \bar C_{\mu\nu} - (\partial_\mu\, \bar C_\nu - \partial_\nu\, \bar C_\mu), \qquad 
S_{\mu\nu} = -\, i\, [\pm\, m\, B_{\mu\nu} - (\partial_\mu\, B_\nu - \partial_\nu\, B_\mu)], \nonumber\\
R_{\mu\nu} = \pm m\,  C_{\mu\nu} - (\partial_\mu\,  C_\nu - \partial_\nu\,  C_\mu), \qquad 
S_{\mu\nu} = +\, i\, [\pm\, m\, \bar B_{\mu\nu} - (\partial_\mu\, \bar B_\nu - \partial_\nu\, \bar B_\mu)],
\end{eqnarray}
where {\it both} the expression for $S_{\mu\nu}$ [cf. Eq. (38)] are {\it equal} provided we take into account the CF-type restrictions:
$B_{\mu\nu} + \bar B_{\mu\nu}  = \partial_\mu\, \phi_\nu - \partial_\nu\, \phi_\mu$ and 
$B_\mu + \bar B_\mu = \pm m\, \phi_\mu - \partial_\mu\, \phi$.  We are now in the position to equate the 
coefficients of $(d\, x^\mu \wedge d\, x^\nu \wedge d\, x^\lambda)$ from the l.h.s. and r.h.s. in the GIR that is quoted in (26).
This can be explicitly expressed as 
\begin{eqnarray}
\tilde A_{\mu\nu\lambda}^{(h)}\, (x, \, \theta, \bar\theta)  \mp \, \frac{1}{m}\, \Big[\partial_\mu\, \tilde\Phi_{\nu\lambda}^{(g)}\, (x, \, \theta, \bar\theta)
 + \partial_\nu\, \tilde\Phi_{\lambda\mu}^{(g)}\, (x, \, \theta, \bar\theta)  + \partial_\lambda\,\tilde\Phi_{\mu\nu}^{(g)}
\, (x, \, \theta, \bar\theta) \Big] \nonumber\\
=  A_{\mu\nu\lambda}\,(x)\mp \frac{1}{m}\, \Big[\partial_\mu\, \Phi_{\nu\lambda}\,(x) + \partial_\nu\, \Phi_{\lambda\mu} \,(x)
+ \partial_\lambda\, \Phi_{\mu\nu}\,(x)\Big],
\end{eqnarray}
where the precise expression for $\tilde A_{\mu\nu\lambda}^{(h)}\, (x, \, \theta, \bar\theta)$ is quoted in Appendix A [cf. Eq. (A.16)] and 
the explicit super expansion of $\tilde\Phi_{\mu\nu}^{(g)}\, (x, \, \theta, \bar\theta) $ is as follows:
\begin{eqnarray}
\tilde\Phi_{\mu\nu}^{(g)}\, (x, \, \theta, \bar\theta) &=&  \Phi_{\mu\nu}\, (x) + \theta\, \big[\pm m\, \bar C_{\mu\nu}
 - (\partial_\mu\, \bar C_\nu - \partial_\nu\, \bar C_\mu)\big]\nonumber\\
 &+& \bar\theta\, \big[\pm m\,  C_{\mu\nu} 
- (\partial_\mu\,  C_\nu - \partial_\nu\,  C_\mu)\big] 
 + \theta\,\bar\theta \, \big[\pm\, m\, B_{\mu\nu} - (\partial_\mu\, B_\nu - \partial_\nu\, B_\mu)\big] \nonumber\\
 &\equiv& \Phi_{\mu\nu}\, (x) + \theta\, \big[\pm m\, \bar C_{\mu\nu}
 - (\partial_\mu\, \bar C_\nu - \partial_\nu\, \bar C_\mu)\big] \nonumber\\
&+& \bar\theta\, \big[\pm m\,  
C_{\mu\nu} - (\partial_\mu\,  C_\nu - \partial_\nu\,  C_\mu)\big] 
 + \theta\,\bar\theta\, \big[\mp \, m\, \bar B_{\mu\nu} - (\partial_\mu\, \bar B_\nu - \partial_\nu\, \bar B_\mu)\big].
\end{eqnarray}
In the above, we have taken into account the inputs from (38). It is straightforward to note that (39) is satisfied in 
a precise manner. In other words, we find that the coefficient of $\theta, \; \bar\theta$ and $\theta\bar\theta$ of the l.h.s. are equal to zero
so that only the spacetime dependent {\it ordinary} terms exist which sum up to produce the r.h.s.
Thus, we have computed  {\it all} the off-shell nilpotent (anti-)BRST symmetry transformations for our {\it modified} Abelian 3-form
gauge theory where the {\it mass} and (anti-)BRST symmetry transformations co-exist {\it together}.

\section{Coupled Lagrangian Densities: (Anti-)BRST Symmetry Transformations and CF-Type Restrictions}

In this section, we explicitly write down the coupled (but equivalent) Lagrangian densities for our present
{\it massive} gauge theory by exploiting the standard techniques of BRST formalism. In particular, we demonstrate that the Lagrangian density
${\cal L}_{B}^{(T)}$ respects {\it perfect} BRST symmetry transformations as does ${\cal L}_{\bar B}^{(T)}$ w.r.t. the 
anti-BRST symmetry transformations. 
The CF-type restrictions are also derived from the equations of motion that emerge out from ${\cal L}_{B}^{(T)}$ and ${\cal L}_{\bar B}^{(T)}$.
For the purpose of BRST-quantization scheme, we have to incorporate the gauge-fixing and Faddeev-Popov (FP) ghost terms into 
the {\it total} BRST and anti-BRST invariant Lagrangian densities  ${\cal L}_{B}^{(T)}$ and ${\cal L}_{\bar B}^{(T)}$ which are 
{\it coupled} but {\it equivalent} as far as 
the (anti-)BRST symmetry transformations are concerned [cf. Sec. 5]. The gauge-fixing and FP-ghost terms can be constructed from the 
{\it basic} fields that appear on the r.h.s. of (12) [i.e. $ A_{\mu\nu\lambda}, \; \bar C_{\mu\nu}, \; C_{\mu\nu},
\; \bar\beta_\mu, \; \beta_\mu, \; \phi_\mu, \; \bar C_2, \; C_2, \; \bar C_1, \; C_1$]  as well as on the r.h.s. of 
(31) [i.e. $\Phi_{\mu\nu}, \; \bar C_\mu, \; C_\mu, \; \bar\beta, \; \beta, \; \phi$]. 
Thus, our present section is divided into two parts. In subsection 4.1, we consider the gauge-fixing and FP-ghost terms
for the {\it massless} Abelian 3-form theory where HC plays an important role. In subsection 4.2, we deal with the 
gauge-fixing and FP-ghost terms corresponding to the St$\ddot u$ckelberg compensating field and associated fermionic and bosonic (anti-)ghost 
fields where GIR plays an important and decisive role.
Finally, we write down the {\it total} gauge-fixing and FP-ghost terms for our theory. 
We end this section by  demonstrating  that ${\cal L}_{B}^{(T)}$ and ${\cal L}_{\bar B}^{(T)}$ (i.e. the coupled Lagrangian densities)
respect {\it perfect} BRST and anti-BRST symmetry transformations, respectively.

\subsection{Gauge-Fixing and FP-Ghost Terms for Massless Abelian 3-Form Theory: Basic Fields Present in HC}

In our earlier work [17, 18], for the {\it massless} Abelian 3-form gauge theory, we have 
the following as the gauge-fixing and FP-ghost terms, namely;
\begin{eqnarray}
s_b\,  s_{ab}\,  \Bigl [ \frac{1}{2}
\bar C_2 C_2 - \frac{1}{2} \bar C_1 C_1 - \frac{1}{2} \bar C_{\mu\nu} C^{\mu\nu} 
- \frac{1}{2} \phi_\mu \phi^\mu  + \bar \beta_\mu \beta^\mu  - \frac{1}{6} A_{\mu\nu\lambda} A^{\mu\nu\lambda} \Bigr ], \nonumber\\
-\,  s_{ab}\, s_b\,  \Bigl [ \frac{1}{2}
\bar C_2 C_2 - \frac{1}{2} \bar C_1 C_1 - \frac{1}{2} \bar C_{\mu\nu} C^{\mu\nu} 
- \frac{1}{2} \phi_\mu \phi^\mu  + \bar \beta_\mu \beta^\mu  - \frac{1}{6} A_{\mu\nu\lambda} A^{\mu\nu\lambda} \Bigr ].
\end{eqnarray}
The above expressions lead to the Lagrangian densities [17, 18] which are a part of the 
{\it perfectly} BRST {\it and} anti-BRST invariant coupled (but equivalent) Lagrangian densities 
${\cal L}_{B}^{(HC)} = {\cal L}_{gf}^{(B, H)} + {\cal L}_{FP}^{(B, H)}$ and 
${\cal L}_{\bar B}^{(HC)} = {\cal L}_{gf}^{(\bar B, H)} + {\cal L}_{FP}^{(\bar B, H)} $, respectively.  These
Lagrangian densities, in their explicit forms, are as follows:
\begin{eqnarray*}
{\cal L}_{gf}^{(B, H)} + {\cal L}_{FP}^{(B, H)}  = (\partial_\mu A^{\mu\nu\lambda}) B_{\nu\lambda}  
+ \frac{1}{2} B_{\mu\nu}\,\bar B^{\mu\nu}
+ (\partial_\mu \bar C_{\nu\lambda} + \partial_\nu \bar C_{\lambda\mu} + 
\partial_\lambda \bar C_{\mu\nu}) (\partial^\mu C^{\nu\lambda}) \nonumber\\
 - \, (\partial_\mu \bar \beta_\nu - \partial_\nu \bar \beta_\mu) (\partial^\mu \beta^\nu)
 - \, B B_2 - \frac{1}{2} B_1^2 + (\partial_\mu \bar C^{\mu\nu}) f_\nu
-  \, (\partial_\mu  C^{\mu\nu})  F_\nu\nonumber\\
 -  \, \partial_\mu \bar C_2 \partial^\mu C_2 + \bar f^\mu f_\mu - \bar F^\mu F_\mu - 
(\partial \cdot \beta) B_2 - (\partial \cdot \phi) B_1
+ \, (\partial \cdot \bar \beta) B,
\end{eqnarray*}
\begin{eqnarray}
{\cal L}_{gf}^{(\bar B, H)} + {\cal L}_{FP}^{(\bar B, H)}  =  - (\partial_\mu A^{\mu\nu\lambda}) \bar B_{\nu\lambda} 
+ \frac{1}{2}  B_{\mu\nu} \bar B^{\mu\nu}
+ (\partial_\mu \bar C_{\nu\lambda} + \partial_\nu \bar C_{\lambda\mu} + 
\partial_\lambda \bar C_{\mu\nu}) (\partial^\mu C^{\nu\lambda}) \nonumber\\
 -\,(\partial_\mu \bar \beta_\nu - \partial_\nu \bar \beta_\mu) (\partial^\mu \beta^\nu)
 - \, B B_2 - \frac{1}{2} B_1^2 - (\partial_\mu  \bar C^{\mu\nu})\bar F_\nu
+  (\partial_\mu  C^{\mu\nu}) \bar f_\nu \nonumber\\
 -  \, \partial_\mu \bar C_2 \partial^\mu C_2 + \bar f^\mu f_\mu - \bar F^\mu F_\mu 
- \, (\partial \cdot \beta) B_2 - (\partial \cdot \phi) B_1
+ (\partial \cdot \bar \beta) B.
\end{eqnarray}
We would like to comment on the choice of the combination of the basic fields that have been
incorporated into the square brackets of (41). We note that every term of the square bracket has the 
mass dimension $[M]^{(D-2)}$  {\it individually}  in the natural units $(\hbar  = c = 1)$. Furthermore, the ghost number for all the individual term is 
{\it zero} and all the individual terms are Lorentz scalars as is required by the basic tenets of the construction of the  proper 
coupled (but equivalent) Lagrangian densities  of a theory.
The sign of the individual term has been chosen {\it judiciously} so that the EL-EoMs w.r.t. the  auxiliary fields from (42) yield the 
Curci-Ferrari restrictions: $ B_{\mu\nu} + \bar B_{\mu\nu} = \partial_\mu \phi_\nu - \partial_\nu \phi_\mu,\, 
f_\mu +  \bar F_\mu = \partial_\mu C_1$ and $ \bar f_\mu +  F_\mu = \partial_\mu \bar C_1$.
In addition, we lay emphasis on the fact that the {\it basic} fields, appearing in the square bracket (41), are 
those that have appeared in the definition of $\tilde A^{(3)}$ [cf. Eqs (10) and (12)].
Using the CF-type restrictions: $ B_{\mu\nu} + \bar B_{\mu\nu} = \partial_\mu \phi_\nu - \partial_\nu \phi_\mu,\, 
f^\mu +  \bar F^\mu = \partial^\mu C_1,$ and $ \bar f^\mu +  F^\mu = \partial^\mu \bar C_1$,
we write down the sum of the gauge-fixing and FP-ghost terms, in the {\it massless} sector of the Abelian 3-form theory,  as 
\begin{eqnarray}
{\cal L}_B^{(HC)} &=& {\cal L}_{gf}^{(B, H)} + {\cal L}_{FP}^{(B, H)}  = (\partial_\mu A^{\mu\nu\lambda}) B_{\nu\lambda} 
 - \frac{1}{2} B_{\mu\nu}\, B^{\mu\nu} + \frac{1}{2}\, B^{\mu\nu}\, (\partial_\mu\,\phi_\nu - \partial_\nu\, \phi_\mu)- B B_2 \nonumber\\
&-& \frac{1}{2} B_1^2 
+ (\partial_\mu \bar C_{\nu\lambda} + \partial_\nu \bar C_{\lambda\mu} + 
\partial_\lambda \bar C_{\mu\nu}) (\partial^\mu C^{\nu\lambda}) 
 - (\partial_\mu \bar \beta_\nu - \partial_\nu \bar \beta_\mu) (\partial^\mu \beta^\nu)\nonumber\\ 
 &-& \partial_\mu \bar C_2 \partial^\mu  C_2  - 
(\partial \cdot \beta) \, B_2 - (\partial \cdot \phi)\, B_1
+ (\partial \cdot \bar \beta)\, B  - 2 \,  F^\mu f_\mu  \nonumber\\
 &+& ( \partial_\mu  \bar C^{\mu\nu} + \partial^\nu \bar C_1 ) \, f_\nu - (\partial_\mu  C^{\mu\nu} + \partial^\nu C_1)\, F_\nu,
\end{eqnarray}
\begin{eqnarray}
{\cal L}_{\bar B}^{(HC)} &=& {\cal L}_{gf}^{(\bar B, H)} + {\cal L}_{FP}^{(\bar B, H)}  = -\ (\partial_\mu A^{\mu\nu\lambda}) \bar B_{\nu\lambda} 
 - \frac{1}{2} \bar B_{\mu\nu}\,\bar B^{\mu\nu} + \frac{1}{2}\, \bar B^{\mu\nu}\, (\partial_\mu\,\phi_\nu - \partial_\nu\, \phi_\mu)- B B_2 \nonumber\\
&-& \frac{1}{2} B_1^2 
+ (\partial_\mu \bar C_{\nu\lambda} + \partial_\nu \bar C_{\lambda\mu} + 
\partial_\lambda \bar C_{\mu\nu}) (\partial^\mu C^{\nu\lambda}) 
 - (\partial_\mu \bar \beta_\nu - \partial_\nu \bar \beta_\mu) (\partial^\mu \beta^\nu)\nonumber\\ 
 &-& \partial_\mu \bar C_2 \partial^\mu  C_2  - 
(\partial \cdot \beta)\,  B_2 - (\partial \cdot \phi)\, B_1
+ (\partial \cdot \bar \beta)\, B  + 2 \,  \bar F^\mu \bar f_\mu  \nonumber\\
 &+& ( \partial_\mu  C^{\mu\nu} - \partial^\nu C_1 ) \, \bar f_\nu - (\partial_\mu \bar C^{\mu\nu} - \partial^\nu \bar C_1)\, \bar F_\nu,
\end{eqnarray}
where superscript $(HC)$ on the Lagrangian densities on the l.h.s. denotes the sum of gauge-fixing and Faddeev-Popov ghost 
terms corresponding to the HC that is valid for the {\it massless} Abelian 3-form gauge theory. The superscript $(H)$, on the r.h.s., carries
forward this information.
It should be recalled that the terms such as $(- 2 \,  F^\mu f_\mu )$ and $(+ 2 \,  \bar F^\mu \bar f_\mu )$ have
ghost number equal to {\it zero} because ghost numbers for $(\bar F_\mu, \, F_\mu)$ are equal to $(+1, -1)$, respectively.
Similarly the ghost number is {\it zero} for the terms: $(\partial_\mu \bar C^{\mu\nu} + \partial^\nu \bar C_1)\, \bar F_\nu$
and $-\, [(\partial_\mu  C^{\mu\nu} + \partial^\nu C_1)\, F_\nu] $.
A close look at   ${\cal L}_{B}^{(HC)} $ and ${\cal L}_{\bar B}^{(HC)} $ demonstrate that  ${\cal L}_{B}^{(HC)} $  is expressed in terms of the auxiliary fields
$(B_{\mu\nu}, \, f_\mu, \, F_\mu)$. On the other hand, the Lagrangian density ${\cal L}_{\bar B}^{(HC)} $  is expressed in terms of 
$(\bar B_{\mu\nu}, \, \bar f_\mu, \, \bar F_\mu)$.
It is straightforward to note that the following EL-EoMs emerge out from ${\cal L}_{B}^{(HC)}$ and ${\cal L}_{\bar B}^{(HC)}$
w.r.t. the set of auxiliary fields $(B_{\mu\nu}, \, f_\mu, \, F_\mu)$ 
and their counterpart set of auxiliary fields $(\bar B_{\mu\nu}, \, \bar f_\mu, \, \bar F_\mu)$; namely;
\begin{eqnarray}
 &&B_{\mu\nu} = (\partial^\lambda \, A_{\lambda\mu\nu}) + \frac{1}{2}\,(\partial_\mu\,\phi_\nu - \partial_\nu\, \phi_\mu), \quad
 2\, F_\mu = \partial^\nu\, \bar C_{\nu\mu} + \partial_\mu\, \bar C_1, \nonumber\\
 &&\bar B_{\mu\nu} = -\, (\partial^\lambda \, A_{\lambda\mu\nu}) + \frac{1}{2}\,(\partial_\mu\,\phi_\nu - \partial_\nu\, \phi_\mu), \quad
 2\, \bar f_\mu = -\, \partial^\nu\, \bar C_{\nu\mu} + \partial_\mu\, \bar C_1,\nonumber\\
 &&2\, f_\mu = \partial^\nu\,  C_{\nu\mu} + \partial_\mu\,  C_1, \quad 
 2\, \bar F_\mu = -\, \partial^\nu\,  C_{\nu\mu} + \partial_\mu\,  C_1.
\end{eqnarray}
It is clear from the above equations of motion that we have obtained the CF-type restrictions:
 $B_{\mu\nu} + \bar B_{\mu\nu} = \partial_\mu \phi_\nu - \partial_\nu \phi_\mu,\;  
f_\mu +  \bar F_\mu = \partial_\mu C_1$ and $ \bar f_\mu +  F_\mu = \partial_\mu \bar C_1$.
It should be recalled that these (anti-)BRST invariant restrictions have emerged out in our theory
 because of the HC (i.e. $\tilde d\, \tilde A^{(3)} = d\, A^{(3)}$)
which amounts to the gauge invariance ($\delta_g \, H_{\mu \nu \lambda \zeta } = 0$) of the 
field-strength tensor. Hence, the choice of the combinations of the fields in Eq. (41) are {\it correct} which
lead to the derivation of the gauge-fixing and FP-ghost terms for the free {\it massless} Abelian 3-form gauge theory.

\subsection{Gauge-Fixing and FP-Ghost Terms for the St$\ddot u$ckelberg-Modified Massive Abelian 3-Form Theory: GIR}

In this subsection, we focus on the construction of the gauge-fixing and FP-ghost terms for the {\it basic } fields that 
are present in GIR [cf. Eq. (26)]  which are explicitly written on the r.h.s. of (31) as the {\it first} terms 
(i.e. $\Phi_{\mu\nu}, \, \bar  C_\mu, \, C_\mu, \, \bar\beta, \, \beta,\, \phi$) in all the super expansions. 
The analogue of (41) can be written for the {\it massive} sector [i.e. the St$\ddot u$ckelberg antisymmetric field $(\Phi_{\mu\nu} = \Phi_{\nu\mu})$
and associated gauge-fixing and FP-ghost terms] as follows
\begin{eqnarray}
{\cal L}_B^{(GIR)} \equiv {\cal L}_{gf}^{(B, G)} + {\cal L}_{FP}^{(B, G)} = s_b\, s_{ab}\, \Big[-\, \frac{1}{4}\, \Phi_{\mu\nu}\, \Phi^{\mu\nu}
+ \bar\beta \, \beta -\, \frac{1}{2}\,  \phi^2 -\, \frac{1}{2}\, \bar C_\mu\, C^\mu  \Big], \nonumber\\
{\cal L}_{\bar B}^{(GIR)} \equiv {\cal L}_{gf}^{(\bar B, G)} + {\cal L}_{FP}^{(\bar B, G)} = -\, s_{ab}\,s_b \Big[-\, \frac{1}{4}\, \Phi_{\mu\nu}\, \Phi^{\mu\nu}
+ \bar\beta \, \beta -\, \frac{1}{2}\,  \phi^2 -\, \frac{1}{2}\, \bar C_\mu\, C^\mu  \Big],
\end{eqnarray}
where the superscript $(GIR)$ on the l.h.s. denotes the importance of GIR in our discussions. The superscript $(G)$ on the
gauge-fixing and FP-ghost terms of the Lagrangian densities, on the r. h. s., encodes
this information. 
The explicit computations of the r.h.s. in terms of the (anti-)BRST transformations [$s_{(a)b}$] [cf. Eqs. (B.1), (B.2)] lead to the following
\begin{eqnarray*}
{\cal L}_B^{(GIR)} &=& {\cal L}_{gf}^{B, G} + {\cal L}_{FP}^{B, G} \equiv 
-\, (\partial_\mu \Phi^{\mu\nu})\, B_\nu \mp \frac{m}{2}\, B^{\mu\nu}\, \Phi_{\mu\nu} + \frac{m^2}{2}\, \bar C_{\mu\nu}\, C^{\mu\nu} \nonumber\\
&+& (\partial_\mu \bar C_\nu - \partial_\nu \bar C_\mu)\, \partial^\mu C^\nu \pm m\, (\partial_\mu \bar C^{\mu\nu})\, C_\nu
\pm\, m\, \bar C^\nu\, (\partial^\mu C_{\mu\nu}) \nonumber\\
\end{eqnarray*}
\begin{eqnarray} 
& -& \, \frac{1}{2}\, \Big[\pm \, m\, \bar\beta^\mu - \partial^\mu \bar\beta \Big]\, \Big[ \pm \, m\, \beta_\mu - \partial_\mu \beta \Big]
+ \frac{1}{2}\, B_\mu \bar B^\mu \mp\, \frac{m}{2}\, F^\mu\, C_\mu  \nonumber\\
&-& \frac{1}{2}\, F\, (\partial \cdot C)
\mp \frac{m}{2}\, \bar C_\mu f^\mu - \frac{1}{2}\, (\partial \cdot \bar C)\, f \pm m\, B_1\, \phi + \bar f \, f -\bar F\, F \nonumber\\
&-&\, m^2\, \bar C_2\, C_2 \mp\, m\, B \, 
\bar\beta \pm m\, B_2\, \beta, 
\end{eqnarray}
which is a part of the {\it perfectly} BRST invariant Lagrangian density ${\cal L}_B^{(T)} $ [cf. Appendix B]. On the contrary,
its counterpart gauge-fixing and FP-ghost terms which are a part of the {\it perfectly} anti-BRST 
Lagrangian density ${\cal L}_{\bar B}^{(T)} $ [cf. Appendix B] are:
\begin{eqnarray}
{\cal L}_{\bar B}^{(GIR)} &=& {\cal L}_{gf}^{(\bar B, G)} + {\cal L}_{FP}^{(\bar B, G)} \equiv 
(\partial_\mu \Phi^{\mu\nu})\,  \bar B_\nu \pm \frac{m}{2}\, \bar B^{\mu\nu}\, \Phi_{\mu\nu} + \frac{m^2}{2}\, \bar C_{\mu\nu}\, C^{\mu\nu} \nonumber\\
&+& (\partial_\mu \bar C_\nu - \partial_\nu \bar C_\mu)\, \partial^\mu C^\nu) \pm m\, (\partial_\mu \bar C^{\mu\nu})\, C_\nu
\pm\, m\, \bar C^\nu\, (\partial^\mu C_{\mu\nu}) \nonumber\\
& -& \, \frac{1}{2}\, \Big[\pm \, m\, \bar\beta^\mu - \partial^\mu \bar\beta \Big]\, \Big[ \pm m\, \beta_\mu - \partial_\mu \beta \Big]
+ \frac{1}{2}\, B_\mu \bar B^\mu \pm\, \frac{m}{2}\, \bar f^\mu\, C_\mu  \nonumber\\
&+& \frac{1}{2}\, \bar f\, (\partial \cdot C)
\pm \frac{m}{2}\, \bar C_\mu \bar F^\mu + \frac{1}{2}\, (\partial \cdot \bar C)\, \bar F \pm m\, B_1\, \phi + \bar f \, f -\bar F\, F \nonumber\\
&-&\, m^2\, \bar C_2\, C_2 \mp\, m\, B \bar\beta \pm m\, B_2\, \beta. 
\end{eqnarray}
At this stage, we take the help of the CF-type restrictions:
$B_\mu + \bar B_\mu = \pm m\, \phi_\mu - \partial_\mu\phi, \; f + \bar F = \pm m\, C_1$ and $\bar f + F = \pm \, m\, \bar C_1$
to recast some of the useful and characteristic terms of the above Lagrangian density $({\cal L}_{gf} + {\cal L}_{FP})$ as follows
\begin{eqnarray}
&&\frac{1}{2}\, B^\mu \bar B_\mu = \frac{1}{2}\, B^\mu\, \big[- B_\mu \pm m\, \phi_\mu -\, \partial_\mu\, \phi \big]
\equiv -\, \frac{1}{2}\, B^\mu B_\mu + \frac{1}{2}\,B^\mu \big[ \pm m\, \phi_\mu -\, \partial_\mu\, \phi \big], \nonumber\\
&&\bar f \, f - \bar F\, F = -\, 2\, F\, f \pm m\, F\, C_1 \pm m\, \bar C_1 \, f, \nonumber\\ 
&&\frac{1}{2}\, B^\mu \bar B_\mu = \frac{1}{2}\,\big[- \bar B_\mu \pm m\, \phi_\mu -\, \partial_\mu\, \phi \big]\, \bar  B^\mu
\equiv -\, \frac{1}{2}\,\bar B^\mu \bar B_\mu + \frac{1}{2}\,\bar B^\mu \big[ \pm m\, \phi_\mu -\, \partial_\mu\, \phi \big], \nonumber\\
&&\bar f \, f - \bar F\, F = 2 \, \bar F\, \bar f \mp m\,  C_1 \, \bar f  \pm m\, \bar C_1\, \bar F,
\end{eqnarray}
where the top {\it two} entries are for ${\cal L}_{B}^{(GIR)} $ and bottom {\it two} are for the 
Lagrangian density ${\cal L}_{\bar B}^{(GIR)} $ (cf. Sec. 5 below). It will be noted that
 ${\cal L}_{ B}^{(GIR)} $ has been expressed in terms of the auxiliary fields
$(B_\mu, \, f, \, F)$ and ${\cal L}_{\bar B}^{(GIR)} $ has been written in terms of the 
set $(\bar B_\mu, \, \bar f, \, \bar F)$ of auxiliary fields (cf. Sec. 5 for details).

We are in the  position now to express the total gauge-fixing and FP-ghost terms which are constructed from the 
{\it basic} and {\it auxiliary} fields that are present in the HC and GIR. These terms are nothing but the sum of (44) and (48) as well as the sum 
total of (43) and (47) as (see, e.g., Appendix B for details): 
\begin{eqnarray}
{\cal L}_{ gf}^{(B)} + {\cal L}_{ FP}^{(B)} \equiv {\cal L}_{B}^{(HC)} + {\cal L}_{B}^{(GIR)},  \nonumber\\
{\cal L}_{gf}^{(\bar B)} + {\cal L}_{FP}^{(\bar B)} \equiv {\cal L}_{\bar B}^{(HC)} + {\cal L}_{\bar B}^{(GIR)}.  
\end{eqnarray}
The equations of motions w.r.t. the {\it bosonic} auxiliary fields $B_{\mu\nu}$ and $\bar B_{\mu\nu}$ that 
emerge out from (50) (see, e.g., Appendix B for details)
are as follows:
\begin{eqnarray}
\frac{\partial}{\partial B_{\mu\nu}}\, \Big[  {\cal L}_{ B}^{(HC)} + {\cal L}_{ B}^{(GIR)} \Big] = (\partial^\lambda A_{\lambda\mu\nu})
+ \frac{1}{2}\, (\partial_\mu\phi_\nu - \partial_\nu\phi_\mu) \mp \frac{m}{2}\, \Phi_{\mu\nu} - B_{\mu\nu} = 0, \nonumber\\
\frac{\partial}{\partial \bar B_{\mu\nu}}\, \Big[  {\cal L}_{ \bar B}^{(HC)} + {\cal L}_{ \bar B}^{(GIR)} \Big] = -\, (\partial^\lambda A_{\lambda\mu\nu})
+ \frac{1}{2}\, (\partial_\mu\phi_\nu - \partial_\nu\phi_\mu) \pm \frac{m}{2}\, \Phi_{\mu\nu} - \bar B_{\mu\nu} = 0. 
\end{eqnarray}
These are modified versions of (45). However, they lead to the Curic-Ferrari type restrictions:
$B_{\mu\nu} + \bar B_{\mu\nu} = \partial_\mu \phi_\nu - \partial_\nu \phi_\mu$ as mentioned earlier. 
Similarly, we note that we have the following equations of motion w.r.t. the fermionic 
auxiliary fields\footnote{The equations of motion (51), (52), (54) and (55) can be derived {\it easily} from the 
total Lagrangian densities that have been written in (B.5), (B.6), (B.8) and (B.9) (see Appendix B for details). }
\begin{eqnarray}
\frac{\partial}{\partial f_{\mu}}\, \Big[  {\cal L}_{ B}^{(HC)} + {\cal L}_{ B}^{(GIR)} \Big] &=& -\, (\partial_\nu\, \bar C^{\nu\mu}
+ \partial^\mu\, \bar C_1) + 2\, F^\mu \pm \frac{m}{2}\, \bar C^\mu = 0 \nonumber\\
&\Longrightarrow&   2\, F^\mu  = (\partial_\nu\, \bar C^{\nu\mu} + \partial^\mu\, \bar C_1)  \mp \frac{m}{2}\, \bar C^\mu, \nonumber\\
\frac{\partial}{\partial \bar F_{\mu}}\, \Big[  {\cal L}_{ \bar B}^{(HC)} + {\cal L}_{\bar B}^{(GIR)} \Big] &=&  (\partial_\nu\,  C^{\nu\mu}
- \partial^\mu\,  C_1) + 2\, \bar f^\mu \mp \frac{m}{2}\, \bar C^\mu = 0 \nonumber\\
&\Longrightarrow&   2\, \bar f^\mu  = - \, \partial_\nu\, \bar C^{\nu\mu} + \partial^\mu\, \bar C_1  \pm \frac{m}{2}\, \bar C^\mu, \nonumber\\
\frac{\partial}{\partial F_{\mu}}\, \Big[  {\cal L}_{ B}^{(HC)} + {\cal L}_{ B}^{(GIR)} \Big] &=& ( \partial_\nu\,  C^{\nu\mu}
+ \partial^\mu\,  C_1) \mp \frac{m}{2}\, C^\mu  -   2\, f^\mu = 0 \nonumber\\
&\Longrightarrow&   2\, f^\mu  = (\partial_\nu\, C^{\nu\mu}) + \partial^\mu\,  C_1  \mp \frac{m}{2}\,  C^\mu, \nonumber\\
\frac{\partial}{\partial \bar f_{\mu}}\, \Big[  {\cal L}_{\bar B}^{(HC)} + {\cal L}_{\bar B}^{(GIR)} \Big] &=&  -   2\, \bar F^\mu 
- (\partial_\nu\,  C^{\nu\mu} - \partial^\mu\,  C_1) \mp \frac{m}{2}\, C^\mu  = 0 \nonumber\\
&\Longrightarrow&   2\, \bar F^\mu  = -\, (\partial_\nu\, C^{\nu\mu}) + \partial^\mu\,  C_1  \mp \frac{m}{2}\,  C^\mu.
\end{eqnarray}
A close and  careful observations of (52) imply the following:
\begin{eqnarray}
f_\mu +  \bar F_\mu = \partial_\mu C_1, \qquad \qquad   \bar f_\mu +  F_\mu = \partial_\mu \bar C_1.
\end{eqnarray}
Thus, we observe that $(i)$ the combination of the fields chosen in (46) and their ghost numbers as well as the mass dimension in natural units, 
and $(ii)$ the numerical factors and their signs, etc., are {\it perfectly} alright as they lead to the derivation of the 
(anti-)BRST invariant CF-type restrictions for the EL-EoMs w.r.t. the auxiliary fields.

Finally, before we end this section, we observe the following explicit EoMs w.r.t. the {\it fermionic} auxiliary fields
 from the Lagrangian densities
\begin{eqnarray}
\frac{\partial}{\partial f}\, \Big[  {\cal L}_{ B}^{(HC)} + {\cal L}_{ B}^{(GIR)} \Big]  = 0  \quad \Longrightarrow \quad
2 \, F   =  \pm m\, \bar C_1 - \, \frac{1}{2}\, (\partial \cdot \bar C), \nonumber\\
\frac{\partial}{\partial F}\, \Big[  {\cal L}_{ B}^{(HC)} + {\cal L}_{ B}^{(GIR)} \Big]  = 0 \quad \Longrightarrow \quad
2 \, f   =  \pm m\, C_1 - \, \frac{1}{2}\, (\partial \cdot  C), \nonumber\\
\frac{\partial}{\partial \bar f}\, \Big[  {\cal L}_{\bar B}^{(HC)} + {\cal L}_{\bar B}^{(GIR)} \Big]  = 0\quad \Longrightarrow \quad
2 \, \bar F   =  \pm m\,  C_1 + \, \frac{1}{2}\, (\partial \cdot C), \nonumber\\
\frac{\partial}{\partial \bar F}\, \Big[  {\cal L}_{\bar B}^{(HC)} + {\cal L}_{\bar B}^{(GIR)} \Big]  = 0 \quad \Longrightarrow \quad
2 \, \bar f   =  \pm m\,  \bar C_1 + \, \frac{1}{2}\, (\partial \cdot \bar C).
\end{eqnarray}
It is straightforward to note that the above equations are crucial for the derivation of the CF-type restrictions: 
$ f + \bar F = \pm m\, C_1$ and $\bar f + F = \pm \, m\, \bar C_1$. Last but not least, we are in position to find out the EL-EoMs 
w.r.t. $B_\mu$ and $\bar B_\mu$ fields as:
\begin{eqnarray}
\frac{\partial}{\partial  \bar B_\mu}\, \Big[  {\cal L}_{\bar B}^{(HC)} + {\cal L}_{\bar B}^{(GIR)} \Big] = (\partial_\nu\, \Phi^{\nu\mu})
- \bar B^\mu \pm \frac{m}{2}\,\phi^\mu - \frac{1}{2}\, \partial^\mu\phi = 0, \nonumber\\
\frac{\partial}{\partial   B_\mu}\, \Big[  {\cal L}_{ B}^{(HC)} + {\cal L}_{ B}^{(GIR)} \Big] = -\, (\partial_\nu\, \Phi^{\nu\mu})
-  B^\mu \pm \frac{m}{2}\,\phi^\mu - \frac{1}{2}\, \partial^\mu\phi = 0.
\end{eqnarray}
The above equation implies the following
\begin{eqnarray}
B_\mu + \bar B_\mu = \pm m\, \phi_\mu - \partial_\mu\phi,
\end{eqnarray}
which is nothing but the CF-type restriction. To sum up, we have derived all the CF-type restrictions of our theory by 
taking into account the proper forms of the gauge-fixing and FP-ghost terms for our theory. 
The standard techniques and tricks of the BRST formalism have been at the heart of our derivations of the 
(anti-)BRST invariant CF-type restrictions which are the hallmark of a BRST-{\it quantized} gauge theory [10, 11].
It is very interesting  to point out that, from the mathematical structure of (41) and (46), it is clear that 
${\cal L}_{B}^{(T)}$ and ${\cal L}_{\bar B}^{(T)}$ will be BRST and anti-BRST invariant because of the off-shell nilpotency 
[$s_{(a)b}^2 = 0$] of the (anti-)BRST symmetry transformations  [$s_{(a)b}$]. This is due to the observation that
 $s_{(a)b}\, {\cal L}_{(S)} =0$ where ${\cal L}_{(S)} $ [cf. Eq. (4)] is the {\it classically} gauge invariant 
 St$\ddot u$ckelberg-modified Lagrangian density.
We elaborate all these statements very clearly in the next section.

\section{Symmetry Invariance and CF-Type Restrictions}

Our present section is devoted to the thread-bare analysis of the BRST and anti-BRST symmetry invariance of the Lagrangian densities 
${\cal L}_B^{(T)}$ and ${\cal L}_{\bar B}^{(T)}$ (cf. Appendix B) which are coupled (but equivalent). We demonstrate that the ``equivalence" of these 
Lagrangian densities is due to the existence of a set of CF-type restrictions on our theory [cf. Eq. (69) below]. 
First of all, we note that the  St$\ddot u$ckelberg-modified Lagrangian density ${\cal L}_{(S)}$, {\it common} in the coupled (but equivalent) total
Lagrangian densities ${\cal L}_B^{(T)}$ and ${\cal L}_{\bar B}^{(T)}$, remains (anti-)BRST invariant 
[i.e. $s_{(a)b}\, {\cal L}_{(S)} =0$]. Now we focus on the gauge-fixing and FP-ghost terms of ${\cal L}_B^{(T)}$ and ${\cal L}_{\bar B}^{(T)}$,
respectively, and study their (anti-)BRST invariance(s). To achieve this goal, our present section is divided into {\it three} parts.
In subsection 5.1, we concentrate on the Lagrangian density ${\cal L}_{B}^{(HC)}$ [cf. Eq. (43)]
and Lagrangian density ${\cal L}_{\bar B}^{(HC)}$ [cf. Eq. (44)] and focus on their (anti-)BRST invariance(s). 
Our subsection 5.2 is devoted to the discussion on the (anti-)BRST invariance of Lagrangian density 
${\cal L}_{B}^{(GIR)}$ [cf. Eq. (47)] and Lagrangian density ${\cal L}_{\bar B}^{(GIR)}$ [cf. Eq. (48)].
Finally, in subsection 5.3, we discuss and comment on some of the key aspects of the (anti-)BRST invariance of the {\it total} Lagrangian densities 
${\cal L}_{B}^{(T)}$ and ${\cal L}_{\bar B}^{(T)}$ of our present {\it modified} version of the massive Abelian 3-form theory.

\subsection{(Anti-)BRST Invariance: ${\cal L}_{B}^{(HC)}$ and  ${\cal L}_{\bar B}^{(HC)}$}

First of all, we apply the BRST symmetry transformations $s_b$ [cf. Eq. (B.1)] on the Lagrangian density 
${\cal L}_{B}^{(HC)}$ [cf. Eq. (43)] which leads to following {\it total} spacetime derivative:\\
\begin{eqnarray}
s_b\, {\cal L}_{B}^{(HC)} &=& \partial_\mu\, \Big[ (\partial^\mu\, C^{\nu\lambda} + \partial^\nu\, C^{\lambda\mu}
+ \partial^\lambda\, C^{\mu\nu}) \, B_{\nu\lambda}  + B^{\mu\nu}\, f_\nu - B_2\, \partial^\mu\,C_2  \nonumber\\
&-& B_1\, f^\mu + B\, F^\mu - (\partial^\mu\, \beta^\nu - \partial^\nu\, \beta^\mu)\, F_\nu \Big].
\end{eqnarray}
Similarly, the application of the anti-BRST symmetry transformations $s_{ab}$ 
[cf. Eq. (B.2)] on the Lagrangian densities ${\cal L}_{\bar B}^{(HC)}$ [cf. Eq. (44)] yields the following:
\begin{eqnarray}
s_{ab}\, {\cal L}_{\bar B}^{(HC)} &=& \partial_\mu\, \Big[\bar B^{\mu\nu}\, \bar f_\nu  - (\partial^\mu\, \bar C^{\nu\lambda} 
+ \partial^\nu\, \bar C^{\lambda\mu}+ \partial^\lambda\, \bar C^{\mu\nu}) \, \bar B_{\nu\lambda}  + B \, \partial^\mu\, \bar C_2 \nonumber\\
&-& B_2 \, \bar F^\mu - B_1\, \bar f^\mu - (\partial^\mu\, \bar \beta^\nu - \partial^\nu\, \bar \beta^\mu)\, \bar F_\nu \Big].
\end{eqnarray}
It is clear from our observations in (57) and (58) that  ${\cal L}_{ B}^{(HC)}$ and  ${\cal L}_{\bar B}^{(HC)}$ respect 
{\it perfect} BRST and anti-BRST symmetry transformations because they transform to the total spacetime derivative 
where {\it no} EL-EoMs and/or  CF-type restrictions are invoked for their validity.

To be precise, we have
demonstrated that the Lagrangian density ${\cal L}_{\bar B}^{(HC)}$ is {\it perfectly} anti-BRST invariant because the 
action integral $S_1 = \int d^D x \,{\cal L}_{\bar B}^{(HC)}$ remains invariant
[i.e. $s_{ab}\, S_1 = 0]$ due to the Gauss divergence  theorem where {\it all} the {\it physical} fields vanish-off as $x\rightarrow \pm \infty$.
In exactly similar fashion, we have shown  that ${\cal L}_{B}^{(HC)}$ transforms under the BRST symmetry transformations 
$(s_b)$ such that the action integral  $S_2 = \int d^D x \,{\cal L}_{ B}^{(HC)}$ remains invariant  $(s_{b}\, S_2 = 0)$
under the BRST transformations because of the Gauss divergence theorem.
The {\it latter} symmetry is {\it also} a {\it perfect} symmetry because we have not imposed any kind of {\it external} conditions 
(e.g. EL-EoMs and/or  CF-type restrictions) for its proof. Let us take this opportunity to clarify that we define a {\it perfect} 
symmetry is the {\it one} under which the action integral remains invariant {\it without} any kind of {\it outside} restrictions.

Within the realm of the symmetry considerations, we wish to establish the existence of the (anti-)BRST invariant CF-type restrictions:
$B_{\mu\nu} + \bar B_{\mu\nu} = \partial_\mu\, \phi_\nu - \partial_\nu\, \phi_\mu, \, 
f_\mu + \bar F_\mu = \partial_\mu\,  C_1$ and $\bar f_\mu + F_\mu = \partial_\mu\, \bar C_1$ [cf. Eq. (14)]. 
Towards this goal in mind, first of all, we apply the continuous and nilpotent BRST symmetry transformations $(s_b)$ on the
{\it perfectly} anti-BRST invariant Lagrangian density [${\cal L}_{\bar B}^{(HC)}$] which yields the following:
\begin{eqnarray}
s_b\, {\cal L}_{\bar B}^{(HC)} &=& \partial_\mu\, \Big[ -\, (\partial^\mu\, C^{\nu\lambda} + \partial^\nu\, C^{\lambda\mu}
+ \partial^\lambda\, C^{\mu\nu}) \, B_{\nu\lambda}  - B^{\mu\nu}\, \bar F_\nu - B_2\, \partial^\mu\,C_2   + C^{\mu\nu}\, \partial_\nu\, B_1\nonumber\\
 &+& 2\, A^{\mu\nu\lambda}\, \partial_\nu\, \bar F_\lambda- B_1\, f^\mu + B\, F^\mu + (\partial^\mu\, \beta^\nu 
- \partial^\nu\, \beta^\mu)\, \bar f_\nu  - \bar C^{\mu\nu}\, \partial_\nu\, B\Big] \nonumber\\
 &-& [\bar f^\mu + F^\mu - \partial^\mu\, \bar C_1]\, (\partial_\mu\, B) + \frac{1}{2}\, \Big[ B^{\mu\nu} + \bar B^{\mu\nu}
- (\partial^\mu\, \phi^\nu - \partial^\nu\, \phi^\mu)  \Big]\, (\partial_\mu\, \bar F_\nu - \partial_\nu\, \bar F_\mu)
\nonumber\\
 &-& (\partial^\mu\, \beta^\nu  - \partial^\nu\, \beta^\mu)\, \partial_\mu\, [\bar f_\nu + F_\nu - \partial_\nu\, \bar C_1]
 + [f^\mu + \bar F^\mu - \partial^\mu \, C_1]\, (\partial_\mu\, B_1) \nonumber\\
 &+& (\partial^\mu\, C^{\nu\lambda} + \partial^\nu\, C^{\lambda\mu} + \partial^\lambda\, C^{\mu\nu}) \,
 [B_{\nu\lambda} + \bar B_{\nu\lambda} - (\partial_\nu\, \phi_\lambda - \partial_\lambda\, \phi_\nu)] .
\end{eqnarray}
It is self-evident that, if we impose the (anti-)BRST invariant CF-type restrictions from {\it outside}, the 
{\it perfectly} anti-BRST invariant part [cf. Eq. (58)] of the Lagrangian density [${\cal L}_{\bar B}^{(HC)}$]
respects the BRST symmetry, too. In fact, if we invoke the validity of the CF-type restrictions: 
$B_{\mu\nu} + \bar B_{\mu\nu} = \partial_\mu\, \phi_\nu - \partial_\nu\, \phi_\mu,\, 
f_\mu + \bar F_\mu = \partial_\mu\,  C_1$ and $\bar f_\mu + F_\mu = \partial_\mu\, \bar C_1$, we 
obtain the following explicit transformation for the Lagrangian density ${\cal L}_{\bar B}^{(HC)}$:
\begin{eqnarray}
s_b\, {\cal L}_{\bar B}^{(HC)} &=& \partial_\mu\, \Big[ -\, (\partial^\mu\, C^{\nu\lambda} + \partial^\nu\, C^{\lambda\mu}
+ \partial^\lambda\, C^{\mu\nu}) \, B_{\nu\lambda}  - B^{\mu\nu}\, \bar F_\nu - B_2\, \partial^\mu\,C_2   + C^{\mu\nu}\, \partial_\nu\, B_1\nonumber\\
&+& 2\, A^{\mu\nu\lambda}\, \partial_\nu\, \bar F_\lambda- B_1\, f^\mu + B\, F^\mu + (\partial^\mu\, \beta^\nu 
- \partial^\nu\, \beta^\mu)\, \bar f_\nu  - \bar C^{\mu\nu}\, \partial_\nu\, B \Big]. 
\end{eqnarray}
In exactly similar fashion, when we apply the anti-BRST symmetry transformations $(s_{ab})$ on the 
{\it perfectly} BRST invariant  [cf. Eq. (57)] Lagrangian density ${\cal L}_{ B}^{(HC)}$, we obtain the following explicit transformation 
for the {\it latter} under $s_{ab}\,$ [cf. $(B.2)$]:
\begin{eqnarray}
s_{ab}\, {\cal L}_{ B}^{(HC)} & =& \partial_\mu\, \Big[ (\partial^\mu\, \bar C^{\nu\lambda} + \partial^\nu\, \bar C^{\lambda\mu}
+ \partial^\lambda\, \bar C^{\mu\nu}) \, B_{\nu\lambda}  - \bar  B^{\mu\nu}\, F_\nu + B \, \partial^\mu\,\bar C_2   - C^{\mu\nu}\, \partial_\nu\, B_2\nonumber\\
 &-& 2\, A^{\mu\nu\lambda}\, \partial_\nu\,  F_\lambda- \bar f^\mu\, B_1 - \bar F^\mu \, B_2 + (\partial^\mu\, \bar \beta^\nu 
- \partial^\nu\, \bar \beta^\mu)\, f_\nu  - \bar C^{\mu\nu}\, \partial_\nu\, B_1 \Big] \nonumber\\
 &+& \frac{1}{2}\, B^{\mu\nu} \partial_\mu \, [\bar f_\nu + F_\nu - \partial_\nu \bar C_1]+ \frac{1}{2}\, \Big[ B^{\mu\nu} + \bar B^{\mu\nu}
- (\partial^\mu \phi^\nu - \partial^\nu \phi^\mu)  \Big]\, (\partial_\mu  F_\nu - \partial_\nu F_\mu)
 \nonumber\\
 &-& (\partial^\mu\, \bar \beta^\nu  - \partial^\nu\, \bar  \beta^\mu)\, \partial_\mu\, [ f_\nu + \bar F_\nu - \partial_\nu\, C_1]
 + [f^\mu + \bar F^\mu - \partial^\mu \, C_1]\, (\partial_\mu\, B_2) \nonumber\\
 &-& (\partial^\mu\, \bar C^{\nu\lambda} + \partial^\nu\, \bar C^{\lambda\mu} + \partial^\lambda\, \bar C^{\mu\nu}) \,
 \partial_\mu\, \Big[B_{\nu\lambda} + \bar B_{\nu\lambda} - (\partial_\nu\, \phi_\lambda - \partial_\lambda\, \phi_\nu)\Big].
\end{eqnarray}
A careful and close look at the above equation demonstrates that $s_{ab}\, {\cal L}_{ B}^{(HC)} $ is {\it also}
a {\it total} spacetime derivative like Eq. (60) provided we invoke the validity of the CF-type restrictions:
$B_{\mu\nu} + \bar B_{\mu\nu} = \partial_\mu\, \phi_\nu - \partial_\nu\, \phi_\mu,\, 
f_\mu + \bar F_\mu = \partial_\mu\,  C_1$ and $\bar f_\mu + F_\mu = \partial_\mu\, \bar C_1$ which imply:
\begin{eqnarray}
s_{ab}\, {\cal L}_{ B}^{(HC)} &=& \partial_\mu\, \Big[ (\partial^\mu\, \bar C^{\nu\lambda} + \partial^\nu\, \bar C^{\lambda\mu}
+ \partial^\lambda\, \bar C^{\mu\nu}) \, B_{\nu\lambda}  - \bar  B^{\mu\nu}\, F_\nu + B \, \partial^\mu\,\bar C_2   - C^{\mu\nu}\, \partial_\nu\, B_2\nonumber\\
 &-& 2\, A^{\mu\nu\lambda}\, \partial_\nu\,  F_\lambda- \bar f^\mu\, B_1 - \bar F^\mu \, B_2 + (\partial^\mu\, \bar \beta^\nu 
- \partial^\nu\, \bar \beta^\mu)\, f_\nu  - \bar C^{\mu\nu}\, \partial_\nu\, B_1 \Big].
\end{eqnarray}
We lay emphasis on the fact that {\it both} the Lagrangian densities ${\cal L}_{ B}^{(HC)} $ and ${\cal L}_{\bar B}^{(HC)} $
are {\it equivalent} w.r.t. the (anti-)BRST symmetry transformations as {\it both} of them respect {\it both} of these off-shell
nilpotent symmetry transformations on a submanifold of Hilbert space of quantum fields where the CF-type restrictions (14) are satisfied.
On this submanifold, the BRST and anti-BRST symmetry transformations have their {\it own} identities 
as they absolutely anticommute with each-other. In addition, it is worthwhile to point out that ${\cal L}_{ B}^{(HC)} $
and ${\cal L}_{\bar B}^{(HC)} $ respect on their own {\it perfect} BRST and anti-BRST symmetries, respectively. However, the 
application of the anti-BRST symmetry transformations on ${\cal L}_{ B}^{(HC)} $ and, 
in exactly similar fashion, the application of the BRST symmetry transformation on  ${\cal L}_{\bar B}^{(HC)} $ lead to the 
existence of the (anti-)BRST invariant CF-type restrictions for their {\it equivalence}.

We end this subsection with the remark that, in the derivation of (59) and (61), we have used, at some places, the 
standard theoretical trick that the summation of the symmetric and antisymmetric indices turns out to be {\it zero}. 
This has led to the derivation of the CF-type restrictions on the r.h.s. on (59) and (61).
As far as the symmetry considerations are concerned, this is an {\it alternative} way to show the existence of the 
(anti-)BRST invariant CF-type restrictions on our theory {\it besides} the requirement of the 
absolute anticommutativity (i.e. $s_b\, s_{ab} + s_{ab}\, s_b =0)$ of the (anti-)BRST symmetry transformations [cf. Appendix B].

\subsection{(Anti-)BRST Invariance: ${\cal L}_{\bar B}^{(GIR)}$ and  ${\cal L}_{B}^{(GIR)}$}

In this section, first of all, we show the (anti-)BRST invariance(s) of ${\cal L}_{\bar B}^{(GIR)}$ and  ${\cal L}_{B}^{(GIR)}$,
respectively. For this purpose, by using the bottom {\it two} lines of (49),  the Lagrangian density ${\cal L}_{\bar B}^{(GIR)}$
[cf. Eq. (48)] can be re-expressed as follows: 
\begin{eqnarray}
{\cal L}_{\bar B}^{(GIR)} &=& {\cal L}_{gf}^{(\bar B)} + {\cal L}_{FP}^{(\bar B)} \equiv 
(\partial_\mu \Phi^{\mu\nu})\,  \bar B_\nu \pm \frac{m}{2}\, \bar B^{\mu\nu}\, \Phi_{\mu\nu} + \frac{m^2}{2}\, \bar C_{\mu\nu}\, C^{\mu\nu} \nonumber\\
&+& (\partial_\mu \bar C_\nu - \partial_\nu \bar C_\mu)\, \partial^\mu C^\nu \pm m\, (\partial_\mu \bar C^{\mu\nu})\, C_\nu
\pm\, m\, \bar C^\nu\, (\partial^\mu C_{\mu\nu}) \nonumber\\
& -& \, \frac{1}{2}\, \big(\pm \, m\, \bar\beta^\mu - \partial^\mu \bar\beta \big)\, \big( \pm m\, \beta_\mu - \partial_\mu \beta \big)
- \frac{1}{2}\,  \bar B^\mu\, \bar B_\mu + \frac{1}{2}\,  \bar B^{\mu}\, \big(\pm \, m\phi_\mu - \partial_\mu\, \phi \big)   \nonumber\\
&+& \frac{1}{2}\, \bar f\, (\partial \cdot C)
\pm \frac{m}{2}\, \bar C_\mu \bar F^\mu + \frac{1}{2}\, (\partial \cdot \bar C)\, \bar F \pm m\, B_1\, \phi 
 \pm\, \frac{m}{2}\, \bar f^\mu\, C_\mu \nonumber\\
&-&\, m^2\, \bar C_2\, C_2 \mp\, m\, B \bar\beta \pm m\, B_2\, \beta + 2\, \bar F\, \bar f \pm m\, \bar f \, C_1 \mp\, m\, \bar F\, \bar C_1.
\end{eqnarray}
It should be noted, at this crucial juncture, that the above Lagrangian density has been expressed in terms of the auxiliary fields
($\bar B_\mu, \; \bar F, \; \bar f$) of (48) because the auxiliary fields ($B_\mu, \; F,\; f$) have been replaced by using the CF-type restrictions:
$B_\mu + \bar B_\mu = \pm m\, \phi_\mu - \partial_\mu \, \phi, \; $ 
$ f + \bar F = \pm m\, C_1$ and $\bar f + F = \pm \, m\, \bar C_1$.
We are now in the position to apply the anti-BRST symmetry transformations ($s_{ab}$) of Appendix B [cf. Eq. (B.2)] on the 
{\it above} Lagrangian density which leads to the following explicit expression 
\begin{eqnarray}
s_{ab}\, {\cal L}_{\bar B}^{(GIR)} &=& \partial_\mu\, \Big[-\,(\partial^\mu\, \bar C^\nu -\partial^\nu\,
 \bar C^\mu)\, \bar B_\nu \mp \, m\, \bar B^{\mu\nu}\, \bar C_\nu  
-\, \frac{1}{2} \, \bar B^\mu\, \bar f \nonumber\\
&\pm& m\, C^{\mu\nu}\, (\pm m\, \bar \beta_\nu - \partial_\nu\, \bar \beta)
\pm\, m\, (\partial^\mu\, \bar\beta^\nu - \partial^\nu\,\bar\beta^\mu)\,C_\nu \nonumber\\
&+& \frac{1}{2}\, (\pm\, m\, \bar \beta^\mu - \partial^\mu\, \bar\beta)\, \bar F    \Big].
\end{eqnarray}
As a consequence, it is clear that the action integral corresponding to ${\cal L}_{\bar B}^{(GIR)}$ remains invariant under the 
anti-BRST symmetry transformations [cf. Eq. (B.2)] due to Gauss's divergence theorem because of the fact that all the {\it physical} field
vanish-off as $x\rightarrow  \pm \infty$.

Against the backdrop of our discussions related with  ${\cal L}_{\bar B}^{(GIR)}$ and its anti-BRST invariance, we concentrate on the 
Lagrangian density  ${\cal L}_{B}^{(GIR)}$  [cf. Eq. (47)]. Using the top {\it two} equations of (49) as {\it inputs}, we obtain the 
following form of the Lagrangian density  ${\cal L}_{ B}^{(GIR)}$:
\begin{eqnarray}
{\cal L}_B^{(GIR)} &=& {\cal L}_{gf}^{B} + {\cal L}_{FP}^{B} \equiv 
-\, (\partial_\mu \Phi^{\mu\nu})\, B_\nu \mp \frac{m}{2}\, B^{\mu\nu}\, \Phi_{\mu\nu} + \frac{m^2}{2}\, \bar C_{\mu\nu}\, C^{\mu\nu} \nonumber\\
&+& (\partial_\mu \bar C_\nu - \partial_\nu \bar C_\mu)\, (\partial^\mu C^\nu) \pm m\, (\partial_\mu \bar C^{\mu\nu})\, C_\nu
\pm\, m\, \bar C^\nu\, (\partial^\mu C_{\mu\nu}) \nonumber\\
& -& \, \frac{1}{2}\, \big(\pm \, m\, \bar\beta^\mu - \partial^\mu \bar\beta \big)\, \big( \pm \, m\, \beta_\mu - \partial_\mu \beta \big)
-\, \frac{1}{2}\, B^\mu B_\mu   + \frac{1}{2}\,  B^{\mu}\, \big(\pm \, m\phi_\mu - \partial_\mu\, \phi \big)   \nonumber\\
&-& \frac{1}{2}\, F\, (\partial \cdot C)
\mp \frac{m}{2}\, \bar C_\mu f^\mu - \frac{1}{2}\, (\partial \cdot \bar C)\, f \pm m\, B_1\, \phi 
\mp\, \frac{m}{2}\, F^\mu\, C_\mu   -\, 2\, F\, f  \nonumber\\
 &\pm& m\, F\, C_1 \mp m\, f\, \bar C_1 \,
-\, m^2\, \bar C_2\, C_2 \mp\, m\, B \, 
\bar\beta \pm m\, B_2\, \beta.
\end{eqnarray}
It is to be noted that the above Lagrangian density is a function of the set of auxiliary fields ($B_\mu, \; F,\; f$) 
because of the fact that auxiliary fields  
($\bar B_\mu, \; \bar F, \; \bar f$) of (47) have been replaced by ($B_\mu, \; F,\; f$) using the CF-type restrictions:
$B_\mu + \bar B_\mu = \pm m\, \phi_\mu - \partial_\mu \, \phi, \; $ 
$ f + \bar F = \pm m\, C_1$ and $\bar f + F = \pm \, m\, \bar C_1$.
This is the key difference between (47) and (65). The stage is now set for the application 
of the BRST symmetry transformations $(s_b)$ of the Appendix B [cf. Eq. (B.1)] on the Lagrangian density (65). 
This operation yields the following:
\begin{eqnarray}
s_{b}\, {\cal L}_{B}^{(GIR)} &=& \partial_\mu\, \Big[(\partial^\mu\,  C^\nu -\partial^\nu\,  C^\mu)\, B_\nu \pm \, m\, B^{\mu\nu}\, C_\nu  
-\, \frac{1}{2} \,  B^\mu\,  f \nonumber\\
&\mp& m\,\bar C^{\mu\nu}\, \big(\pm m\,  \beta_\nu - \partial_\nu\, \beta \big)
\mp\, m\, (\partial^\mu\, \beta^\nu - \partial^\nu\,\beta^\mu)\,\bar C_\nu \nonumber\\
&+& \frac{1}{2}\, (\pm\, m\, \beta^\mu - \partial^\mu\, \beta)\, F    \Big].
\end{eqnarray}
As a consequence, we observe that the action integral $(S)$,  corresponding to the Lagrangian density ${\cal L}_{B}^{(GIR)}$,
remains invariant $(s_b\, S = 0)$ under the BRST symmetry transformations $(s_b)$ that have been quoted in our Appendix B.

Within the purview of the symmetry considerations, we can show the existence of the CF-type restrictions on our 
theory by proving the Lagrangian densities ${\cal L}_{B}^{(GIR)}$ and ${\cal L}_{\bar B}^{(GIR)}$to be 
{\it equivalent} w.r.t. the nilpotent (anti-)BRST symmetry transformations. In other words, for this purpose, we have to apply 
$(i)$ the anti-BRST  transformation on the {\it perfectly} BRST invariant [cf. Eqs (63), (64)]
Lagrangian density ${\cal L}_{B}^{(GIR)}$, and $(ii)$ the BRST transformations on the {\it perfectly} 
anti-BRST invariant [cf. Eqs. (65), (66)] Lagrangian density ${\cal L}_{\bar B}^{(GIR)}$. In this connection,
 first of all, we apply the BRST symmetry transformations [cf. Eq. (B.1)] on the Lagrangian density ${\cal L}_{\bar B}^{(GIR)}$
[cf. Eq. (65)] which leads to the following:
\begin{eqnarray}
s_b\, {\cal L}_{\bar B}^{(GIR)} &=&  \partial_\mu\, \Big[\pm \, m\,  C^{\mu\nu}\,  B_\nu  \mp \, m\, 
(\partial^\mu\, \beta^\nu - \partial^\nu\, \beta^\mu)\,\bar C_{\nu} 
\mp m \, \bar C^{\mu\nu}\, (\pm\, m\, \beta_\nu - \partial_\nu\, \beta) \nonumber\\
 &\pm& m\, B^{\mu\nu}\,  C_\nu +\, \Phi^{\mu\nu}\, (\pm \, m\,  f_\nu - \partial_\nu\,  f)  +\, \{\pm\, m\,C^{\mu\nu} 
-\, (\partial^\mu\, C^\nu - \partial^\nu\, C^\mu)\}\, \bar B_\nu \nonumber\\
 &\mp& \frac{m}{2}\, C^\mu\, B_1 + \frac{1}{2}\, B^\mu\, \bar F -\, \frac{1}{2}\, \big\{\pm \, m\,  \beta^\mu - \partial^\mu\, \beta  \big\}\, \bar f
\pm \frac{m}{2}\, B\, \bar C^\mu  \Big]  \nonumber\\
&\mp& \frac{m}{2}\, \big[  B_{\mu\nu} + \bar B_{\mu\nu} - (\partial_\mu\, \phi_\nu  - \partial_\nu\, \phi_\mu) \big]
\, \big[(\partial^\mu\, C^\nu - \partial^\nu\, C^\mu) \mp\, m\, C^{\mu\nu} \big] \nonumber\\
&\mp&\, m\, \Phi^{\mu\nu} \, \partial_\mu\, \Big(f_\nu + \bar F_\nu - \partial_\nu\, C_1\Big)
\mp\, m\, B\, \Big(\bar f + F \mp \ m \bar C_1\Big) \pm m\, B_1\,\Big(f + \bar F \mp m\, C_1\Big)  \nonumber\\
&-& \big[ \pm \ m\, C^{\mu\nu} - (\partial^\mu\, C^\nu - \partial^\nu\, C^\mu)\big]\, 
\partial_\mu \, [B_\nu + \bar B_\nu \mp\, m\, \phi_\nu + \partial_\nu\, \phi] \nonumber\\
&-&\, \frac{1}{2}\, \big(\pm\, m\, \beta^\mu - \partial^\mu\, \beta \big)\, \big[\pm\, m\, (\bar f_\mu + F_\mu -\partial_\mu\, \bar C_1)
 - \partial_\mu\, (\bar f + F \mp\, m \bar C_1)\big] \nonumber\\
 &+& \frac{1}{2}\, B^\mu\, \big[ \pm\, m\, \big\{ \bar F_\mu + f_\mu -\, \partial_\mu\, C_1  \big\} 
 - \partial_\mu \big\{ \bar F + f  \mp\, m\, C_1  \big\} \big].
\end{eqnarray}
Thus, on the submanifold of the Hilbert space of quantum fields [defined by (anti-)BRST invariant CF-type restrictions],
we note that ${\cal L}_{\bar B}^{(GIR)}$ {\it also} respects [cf. Eq. (67)] the BRST symmetry transformations [cf. Eq. (B.1)]
thereby showing the {\it equivalence} of ${\cal L}_{\bar B}^{(GIR)}$ {\it with} ${\cal L}_{ B}^{(GIR)}$ w.r.t. the 
(anti-)BRST symmetry transformations of our theory.

To establish the {\it equivalence} of ${\cal L}_{B}^{(GIR)}$ {\it with} ${\cal L}_{\bar B}^{(GIR)}$ and existence of the
(anti-)BRST invariant CF-type type restrictions, we now apply the anti-BRST symmetry transformations [cf. Eq. (B.2)] on 
${\cal L}_{ B}^{(GIR)}$  which leads to the following observation, namely;
\begin{eqnarray}
s_{ab}\, {\cal L}_{B}^{(GIR)} &=&  \partial_\mu\, \Big[\pm \, m\, (\partial^\mu\, \bar\beta^\nu - \partial^\nu\, \bar\beta^\mu)\,C_{\nu} 
\mp\, m\, \bar C^{\mu\nu}\, \bar B_\nu \pm m \, C^{\mu\nu}\, (\pm\, m\, \bar\beta_\nu - \partial_\nu\, \bar\beta) \nonumber\\
 &\mp& m\,\bar B^{\mu\nu}\, \bar C_\nu -\, \Phi^{\mu\nu}\, (\pm \, m\, \bar f_\nu - \partial_\nu\, \bar f)  -\, \{\pm\, m\, \bar C^{\mu\nu} 
-\, (\partial^\mu\, \bar C^\nu - \partial^\nu\, \bar C^\mu)\}\, B_\nu \nonumber\\
 &\pm& \frac{m}{2}\, C^\mu\, B_2 + \frac{1}{2}\,\bar B^\mu\, F -\, \frac{1}{2}\, \big\{\pm \, m\, \bar \beta^\mu - \partial^\mu\, \bar\beta  \big\}  \, f
\pm \frac{m}{2}\, \bar C^\mu\, B_1  \Big]  \nonumber\\
&\pm& m\, B_2\, [ f + \bar F \mp m\, C_1] \pm \, m\, B_1\, [ \bar f +  F \mp m\, \bar C_1] \nonumber\\
&\pm & \frac{m}{2}\,  \big[ (\partial^\mu \, \bar C^\nu - \partial^\nu\, \bar C^\mu) \mp \, m\, \bar C^{\mu\nu}  \big]
\big[B_{\mu\nu} + \bar B_{\mu\nu} - (\partial_\mu\, \phi_\nu  - \partial_\nu\, \phi_\mu) \big]\,\nonumber\\
 &+ & \{ \pm\, m\, \bar C^{\mu\nu} - (\partial^\mu\, \bar C^\nu - \partial^\nu\, \bar C^\mu)  \}\, \partial_\mu\, [B_\nu + 
\bar B_\nu \mp\, m\, \phi_\nu + \partial_\nu\, \phi] \nonumber\\
&+&  \frac{1}{2}\,\bar B^\mu\,  \big [\pm\, m\, (\bar f_{\mu} + F_{\mu} - \partial_\mu\, \bar C_1)  
- \partial_\mu (F + \bar f \mp\, m\, \bar C_1) \big ]\nonumber\\
&-& \frac{1}{2}\, \big(\pm\, m\, \bar \beta^\mu - \partial^\mu \, \bar \beta \big)\, \big[\pm\, m\, (f_\mu 
+ \bar F_\mu -\, \partial_\mu\, C_1) - \partial_\mu \, (f + \bar F \mp\, m\, C_1)\big] \nonumber\\
&-& \frac{1}{2}\, (\pm\, m\, \bar f^\mu - \partial^\mu \, \bar f)\, [B_\mu + \bar B_\mu \mp \, m\, \phi_\mu + \partial_\mu\, \phi] \nonumber\\
&\pm& m\, \Phi^{\mu\nu}\, \big(\partial_\mu\, [\bar f_\nu + F_\nu - \partial_\nu\, \bar C_1]\big).
\end{eqnarray}
The noteworthy point at this stage is the fact that {\it all} the {\it six} CF-type restrictions have appeared  on the 
r.h.s. of the above transformation and (67). If we invoke the validity of these (anti-)BRST CF-type restrictions, namely;
\begin{eqnarray}
&&B_{\mu\nu} + \bar B_{\mu\nu} - (\partial_\mu \phi_\nu - \partial_\nu \phi_\mu) = 0, \qquad   f_\mu +  \bar F_\mu - \partial_\mu C_1  = 0, \; \nonumber\\
&&\bar f_\mu +  F_\mu -  \partial_\mu \bar C_1 = 0,\qquad  \qquad \qquad  \quad f + \bar F  \mp m\, C_1 = 0, \nonumber\\
&& B_\mu + \bar B_\mu  \mp m\, \phi_\mu + \partial_\mu \, \phi = 0,  \qquad \quad  \;\;    \bar f + F  \mp \, m\, \bar C_1 = 0,
\end{eqnarray}
we observe that, under the anti-BRST symmetry transformations [cf. Eq. (B.2)], the {\it perfectly} BRST invariant 
Lagrangian density ${\cal L}_B^{(GIR)}$ transforms to:
\begin{eqnarray}
s_{ab}\, {\cal L}_{B}^{(GIR)} =   &=&  \partial_\mu\, \Big[\pm \, m\, (\partial^\mu\, \bar\beta^\nu - \partial^\nu\, \bar\beta^\mu)\,C_{\nu} 
\mp\, m\, \bar C^{\mu\nu}\, \bar B_\nu \pm m \, C^{\mu\nu}\, (\pm\, m\, \bar\beta_\nu - \partial_\nu\, \bar\beta) \nonumber\\
 &\mp& m\,\bar B^{\mu\nu}\, \bar C_\nu -\, \Phi^{\mu\nu}\, (\pm \, m\, \bar f_\nu - \partial_\nu\, \bar f)  -\, \{\pm\, m\, \bar C^{\mu\nu} 
-\, (\partial^\mu\, \bar C^\nu - \partial^\nu\, \bar C^\mu)\}\, B_\nu \nonumber\\
 &\pm& \frac{m}{2}\, C^\mu\, B_2 + \frac{1}{2}\,\bar B^\mu\, F -\, \frac{1}{2}\, \big\{\pm \, m\, \bar \beta^\mu - \partial^\mu\, \bar\beta  \big\}  \, f
\pm \frac{m}{2}\, \bar C^\mu\, B_1  \Big].
\end{eqnarray}
Thus, we note that, on a specific submanifold of the Hilbert space of quantum fields, the coupled Lagrangian density 
$ {\cal L}_{ B}^{(GIR)}$ and  ${\cal L}_{\bar B}^{(GIR)}$ are {\it equivalent} w.r.t. the (anti-)BRST symmetry transformations
 [cf. Eqs. (B.1),  (B.2)].
The above {\it specific} submanifold is defined in terms of the field equations (69) corresponding to the 
(anti-)BRST invariant CF-type restrictions where the BRST and anti-BRST symmetry transformations absolutely anticommute 
($s_b\, s_{ab} + s_{ab}\, s_b = 0$) with each-other. Exactly similar kind of statement can be made for our observation in (67)
where the imposition of (69) on the r.h.s. leads to the observation that ${\cal L}_{\bar B}^{(GIR)} $ 
transforms to a total spacetime derivative under the off-shell nilpotent BRST symmetry transformations   $s_b$ of our 
present {\it massive} Abelian 3-form theory.

\subsection{(Anti-)BRST Invariance: ${\cal L}_{ B}^{(T)}$ and  ${\cal L}_{\bar B}^{(T)}$}

In this subsection, we have to take into account the {\it total} BRST and anti-BRST symmetry transformations of Appendix B and apply
them on the {\it total} Lagrangian densities of our theory which incorporate the St$\ddot u$ckelberg-modified Lagrangian density 
${\cal L}_{(S)}$ [cf. Eq. (4)] and {\it total} gauge-fixing and FP-ghost terms {\it together}. 
In this connection, we note that the application of the continuous and infinitesimal  BRST symmetry transformations (B.1) on 
${\cal L}_{ B}^{(T)}$ [cf. Eq. (B.5)] leads to the following
\begin{eqnarray}
s_b\, {\cal L}_{B}^{(T)} &=& \partial_\mu\, \Big[ (\partial^\mu\, C^{\nu\lambda} + \partial^\nu\, C^{\lambda\mu}
+ \partial^\lambda\, C^{\mu\nu}) \, B_{\nu\lambda}  + B^{\mu\nu}\, f_\nu - B_2\, \partial^\mu\,C_2  \nonumber\\
&-& B_1\, f^\mu + B\, F^\mu - (\partial^\mu\, \beta^\nu - \partial^\nu\, \beta^\mu)\, F_\nu 
+ \frac{1}{2}\, (\pm\, m\, \beta^\mu - \partial^\mu\, \beta)\, F \nonumber\\
 &+& \,(\partial^\mu\,  C^\nu -\partial^\nu\,  C^\mu)\, B_\nu \pm \, m\, B^{\mu\nu}\, C_\nu  
-\, \frac{1}{2} \,  B^\mu\,  f \nonumber\\
&\mp& m\,\bar C^{\mu\nu}\, \big(\pm m\,  \beta_\nu - \partial_\nu\, \beta \big)
\mp\, m\, (\partial^\mu\, \beta^\nu - \partial^\nu\,\beta^\mu)\,\bar C_\nu  \Big],
\end{eqnarray}
which demonstrates that the action integral $S_1 = \int d^D\, x \, {\cal L}_{ B}^{(T)}$ remains invariant under the 
BRST symmetry transformations (B.1) for the {\it physical} fields which vanish-off as $x\rightarrow \pm \infty$. 
In exactly similar fashion, we observe that the application of the infinitesimal and continuous anti-BRST symmetry transformations (B.2) 
on ${\cal L}_{\bar B}^{(T)}$ [cf. Eqs. (B.8), (B.9)] leads to the following {\it total} spacetime derivative:
\begin{eqnarray}
s_{ab}\, {\cal L}_{\bar B}^{(T)} &=& \partial_\mu\, \Big[\bar B^{\mu\nu}\, \bar f_\nu  - (\partial^\mu\, \bar C^{\nu\lambda} 
+ \partial^\nu\, \bar C^{\lambda\mu}+ \partial^\lambda\, \bar C^{\mu\nu}) \, \bar B_{\nu\lambda}  + B \, \partial^\mu\, \bar C_2 \nonumber\\
&-& B_2 \, \bar F^\mu - B_1\, \bar f^\mu - (\partial^\mu\, \bar \beta^\nu - \partial^\nu\, \bar \beta^\mu)\, \bar F_\nu
+ \frac{1}{2}\, (\pm\, m\, \bar \beta^\mu - \partial^\mu\, \bar\beta)\, \bar F   \nonumber\\
 &-&(\partial^\mu\, \bar C^\nu -\partial^\nu\, \bar C^\mu)\, \bar B_\nu \mp \, m\, \bar B^{\mu\nu}\, \bar C_\nu  
-\, \frac{1}{2} \, \bar B^\mu\, \bar f \nonumber\\ 
&\pm& m\, C^{\mu\nu}\, (\pm m\, \bar \beta_\nu - \partial_\nu\, \bar \beta)
\pm\, m\, (\partial^\mu\, \bar\beta^\nu - \partial^\nu\,\bar\beta^\mu)\,C_\nu\Big].
\end{eqnarray}
The above observation establishes the fact that the action integrals $S_1 = \int d^D\, x \, {\cal L}_{B}^{(T)}$  
and $S_2 = \int d^D\, x \, {\cal L}_{\bar B}^{(T)}$ remain invariant under the BRST and
anti-BRST symmetry transformations (B.1) and (B.2), respectively,  for the {\it physical} fields that vanish-off as $x\rightarrow \pm \infty$.

To prove explicitly the ``equivalence" of the Lagrangian densities ${\cal L}_{B}^{(T)}$ and ${\cal L}_{\bar B}^{(T)}$
on symmetry grounds, we have to apply the anti-BRST symmetry transformations $(s_{ab})$ on ${\cal L}_{ B}^{(T)}$ and 
BRST symmetry transformations $(s_b)$ on the Lagrangian density ${\cal L}_{\bar B}^{(T)}$.
Towards this goal in mind, first of all, we re-emphasize that the St$\ddot u$ckelberg-modified Lagrangian density ${\cal L}_{(S)}$
is found to be (anti-)BRST invariant (i.e. $s_{(a)b}\,{\cal L}_{(S)} = 0$). Thus, we have to focus on the gauge-fixing and FP-ghost terms
to study the anti-BRST transformation of  ${\cal L}_{B}^{(HC)} = {\cal L}_{gf}^{(B)} + {\cal L}_{FP}^{(B)}$ [cf. Eq. (43)]
and  ${\cal L}_{B}^{(GIR)} = {\cal L}_{gf}^{(B)} + {\cal L}_{FP}^{(B)}$ [cf. Eq. (65)].
Thus, the anti-BRST symmetry transformation of  ${\cal L}_{B}^{(T)}$ will lead us to obtain the {\it sum} of the r.h.s. of equations 
(61) and (68).  In other words, we shall end up with a {\it total} spacetime derivative term {\it plus}
the terms that vanish-off on the submanifold of the quantum Hilbert space of quantum fields which is defined by the 
CF-type restrictions (69). It is crucial to pin-point the fact that, in (61), {\it only} the CF-type restrictions (14)
appear which owe their origin to the HC. These are valid for the {\it massless} Abelian 3-form theory, too [17, 18].
On the other hand, on the r.h.s. of Eq. (68), we see the appearance of {\it all} the {\it six} CF-type restrictions of our theory.
As a consequence, it is clear that ${\cal L}_{B}^{(HC)} + {\cal L}_{B}^{(GIR)}$ respects the anti-BRST symmetry transformations 
only on the submanifold of the Hilbert space of the quantum fields where the CF-type restrictions (69) are satisfied.
In exactly similar fashion, when we apply the BRST symmetry transformations on ${\cal L}_{\bar B}^{(HC)} + {\cal L}_{\bar B}^{(GIR)}$,
we obtain the {\it sum } of the r.h.s. of (59) and (67) which is the {\it sum} of a {\it total}
spacetime derivative {\it plus} terms that vanish-off on the submanifold in the Hilbert space of quantum fields
that is defined by the field equations corresponding to the CF-type restrictions (69).

\section{Conclusions}

In our present investigation, we have exploited the basic tenets of the augmented version of the superfield
approach (AVSA) to BRST formalism $(i)$ to derive {\it all} the off-shell nilpotent and absolutely anticommuting 
(anti-)BRST symmetry transformations for {\it all} the fields of the {\it massive} Abelian 3-form theory in any arbitrary D-dimension of 
Minkowskian flat spacetime, and $(ii)$ to deduce {\it all} the sets of Curci-Ferrari (CF) type restrictions which are found to 
be (anti-)BRST invariant and they are responsible for the absolute anticommutativity of the (anti-)BRST symmetry transformations.
These CF-type restrictions lead to the existence of a coupled (but equivalent) set of  Lagrangian densities that respect {\it both} the 
BRST and anti-BRST symmetry transformations on the submanifold (of a quantum Hilbert space of the quantum  fields) which is defined by the field equations
corresponding to the (anti-)BRST invariant CF-type restrictions.
The CF-type restrictions are the hallmark [27, 28] of a {\it quantum} gauge theory (discussed within the ambit of BRST formalism) and they are connected
with the  geometrical objects called gerbes [27, 28]. Thus, the existence of the CF-type restrictions on our present theory is {\it crucial}.
We have {\it also} derived {\it these} restrictions from the equations of motion 
that emerge out from the (anti-)BRST invariant coupled (but equivalent)
Lagrangian densities (cf. Sec. 4) where the {\it above} EL-EoMs are derived w. r. t. the bosonic as well as  fermionic auxiliary fields of our theory.

In the whole body of our present text, we have exploited an elegant conjunction of the HC and GIR 
within the realm of AVSA. Both these restrictions have been invoked 
on the basis of geometrically- as well as physically-backed theoretical grounds. First of all, we note that {\it these} quantities, at 
the {\it classical} level {\it itself}, are interesting in the sense that they are gauge invariant and, hence, these are primarily 
{\it physical} objects. For instance, we note that 
$\delta_g\, H_{\mu \nu \lambda \zeta } = 0$ and $\delta_g\, [A_{\mu\nu\lambda} \mp\, \frac{1}{m}\, (\partial_\mu\, \Phi_{\nu\lambda} + 
\partial_\nu\, \Phi_{\lambda\mu} + \partial_\lambda\, \Phi_{\mu\nu})] = 0$ which are nothing but the 
gauge [as well as the (anti-)BRST] invariance of $(i)$ the field strength tensor 
of our theory, {\it and} $(ii)$ an elegant combination of the gauge field $A_{\mu\nu\lambda}$ and beautiful sum of a single  derivative acting on the 
St$\ddot u$ckelberg field $\Phi_{\mu\nu}$. Thus, both these quantities are invariant  at the {\it classical} as well as  {\it quantum} level
because they respect gauge as well as (anti-)BRST symmetries. 
Both these invariances can be expressed in terms of the differential-form within the framework of AVSA 
as: $\tilde d \, \tilde A^{(3)} = d\, A^{(3)} $ [cf. Eq. (8)] and $\tilde d \, \tilde A^{(3)}_{(h)} \mp\, \frac{1}{m}\, \tilde d\, \tilde \Phi^{(2)} =
d\, A^{(3)} \mp \frac{1}{m}\, d\, \Phi^{(2)}$
 [cf. Eq. (26)]. Thus, we lay emphasis on the fact that, in these couple of restrictions, the fundamental ideas of 
 the theoretical physics and differential geometry {\it both} are intertwined together in a beautiful fashion.

 In our present endeavor, we have discussed and described only the {\it complete} set of  nilpotent (anti-)BRST symmetry transformations and
 coupled (but equivalent) total Lagrangian densities that respect the {\it above} nilpotent symmetry transformations on the sumanifold
 (in the {\it total} Hilbert space of quantum fields) that is defined by the CF-type field equations whose number is {\it six} for 
 our D-dimensional massive model of Abelian 3-form theory. However, we have {\it not} devoted time on the derivation of the Noether
 currents and corresponding charges; the proof of their conservation laws by exploiting the full set of EL-EoMs that are 
 derived from the coupled (but equivalent) Lagrangian densities; deduction of the associated BRST algebra, etc. In the immediate
future, we plan to carry out research activities that will be connected with the above mentioned topics. To be precise, we shall
compute the exact expressions for the conserved, off-shell nilpotent and absolutely anticommuting (anti-)BRST charges along
with the ghost charge and demonstrate the existence of the {\it standard} BRST algebra in any arbitrary D-dimensions of spacetime.
However, for the six $(5 + 1)$-dimensional (6D) massive Abelian 3-form theory, we expect the existence of the conserved,
off-shell nilpotent and absolutely anticommuting (anti-)co-BRST charges along with the conserved 
(anti-)BRST and ghost charges. In the
very near future, we wish to derive the {\it extended} BRST algebra with all the above conserved charges for the 6D massibe Abelian 3-form
theory and plan to show that {\it this} algebra is reminiscent of the Hodge algebra [19-22] of the de Rham cohomological operators of differential geometry.

We have worked on the BRST and superfield approaches to the {\it massive} Abelian $p$-form ($p = 1, 2$) gauge theories  over the 
years where we have shown that such theories in $D= 2p$ (i.e. $D= 2, 4$) dimensions of spacetime are {\it massive}
models of Hodge theory where the fields with {\it negative} kinetic term appear (absolutely on the {\it symmetry} grounds).
These {\it latter} fields are found to be the possible candidates of dark matter because they obey Klein-Gordon equation 
with the well-defined rest mass (see, e.g. [29], [30] and references therein). In addition, such {\it exotic} fields have been christened as the ``ghost"
and/or ``phantom" fields in the realm of modern cosmology where they play a {\it crucial} role in the
cyclic, self-accelerated and bouncing models of Universe (see, e.g. [31-33] for details). 
We plan to extend our present work by introducing some {\it new} fields on symmetry grounds and
establish that our {\it massive} Abelian 3-form gauge theory in $D= 6$ is a massive model of Hodge theory
where the {\it axial} antisymmetric tensor, {\it axial} vector and pseudo-scalar fields with {\it negative} kinetic terms
({\it but} with well-defined mass) are expected to exist. These will be, once again, {\it exotic} fields which are 
very popular in the domain of research activities in modern cosmology. These exotic fields will also provide a set of possible candidates of
 dark energy\footnote{The {\it massless} version of the Abelian $p$-form ($p = 1, 2, 3$)
gauge theories have been discussed in our earlier work(s) (see, e.g. [34] and references therein) 
where the {\it massless} pseudo-scalar, axial-vector and axial antisymmetric tensor fields with {\it negative} kinetic term 
have been shown to exist on the ground of symmetries {\it alone}. These are the possible candidates for the dark energy (see, e.g. [31-33] for details).}
in their {\it massless} limits. 
It is worthwhile to mention that the St$\ddot u$ckelberg-modified (SUSY)QED has been discussed in an interesting set of papers
[35-37] where an {\it ultralight} dark matter candidate has been purposed. It will be a nice future endeavor [38] to apply the 
BRST and superfield approaches to these systems [35-37].

\vskip 0.9cm

\noindent
{\bf Acknowledgments}\\

\noindent
One of us (AKR) thankfully acknowledges the financial support as a scholarship from the {\it BHU-fellowship program} of the Banaras Hindu University (BHU),
Varanasi, under which the present investigation has been carried out.
Fruitful discussions with Mr. B. Chauhan, in the initial stages of this work, are gratefully acknowledged.

\vskip 1 cm
\begin{center}
{\bf Appendix A: On the Horizontality Condition }\\
\end{center}
\noindent
The central purpose of this Appendix is to capture a few key steps in the derivation of the complete set of (i) the 
(anti-)BRST symmetry transformations which are off-shell nilpotent and absolutely anticommuting in nature, and (ii) the 
{\it trivial} as well as {\it non-trivial} (anti-) BRST invariant CF-type restrictions from the horizontality condition: 
$\tilde d\, \tilde A^{(3)} = d\, A^{(3)}$. This condition, as is obvious, implies that all the differentials in the 
{\it super} 4-forms on the l.h.s will be set equal to zero so that the l.h.s.  
(i.e. the {\it super}  4-forms with Grassmannian differentials) and r.h.s. can be proven to be equal.
In this context, first of all, we note that {\it when} we set the coefficients of 
$(d\theta \wedge d\theta \wedge d \theta \wedge d \theta), \, (d \bar\theta \wedge d \bar\theta \wedge d \bar\theta \wedge d \bar\theta),\,
(d\theta \wedge d\theta \wedge d \theta \wedge d \bar\theta),\,(d\theta \wedge d\theta \wedge d \bar\theta \wedge d \bar\theta),\,
(d\theta \wedge d \bar\theta \wedge d \bar\theta \wedge d \bar\theta)$ equal to {\it zero}, respectively, we obtain the following  {\it five} 
restrictions on the superfields 
\[
\partial_\theta\, \bar{\cal F}_2 = 0,\qquad \partial_{\bar\theta}\, {\cal F}_2 = 0, \qquad \partial_{\bar\theta} {\cal \bar F}_2 
+ \partial_\theta {\cal \bar F}_1 = 0,
\]
\[
 \partial_{\bar\theta} {\cal \bar F}_1 +  \partial_\theta {\cal F}_1 = 0, \qquad \qquad \; 
\partial_{\bar\theta} {\cal  F}_1 + \partial_\theta {\cal F}_2 = 0,
\eqno (A.1)
\]
which lead to the derivation of secondary fields [cf. Eq. (12)] as well as the {\it trivial} (anti-) BRST invariant
CF-type restrictions as follows:
\[
\qquad \qquad s_1 \; = \;  \bar s_1\; = \;s_2 \; = \; \bar s_2 \; = \;0, \qquad  b_2^{(1)} = \bar b_2^{(2)} = 0,
\]
\[
b_1^{(2)} + \bar b_1^{(1)} = 0, \qquad b_2^{(2)} + \bar b_1^{(2)} = 0, \qquad b_1^{(1)} + \bar b_2^{(1)} = 0.
\eqno (A. 2)
\]
In the above, the last three entries are the {\it trivial} CF-type restrictions.
We make the  following (anti-)BRST invariant [$s_{(a)b}\, B_1 = s_{(a)b}\, B_2 = s_{(a)b}\, B = 0$] choices
\[
b_2^{(2)} = -\, \bar b_1^{(2)} = B_2, \qquad  \bar b_1^{(1)} = -\, b_1^{(2)} = B_1, \qquad \bar b_2^{(1)}  = -\, b_1^{(1)} = B,
\eqno (A.3)
\]
which satisfy the {\it trivial} CF-type restrictions of (A.2) [i.e. the last three entries].
As a consequence of the results and the choices in (A.2) and (A.3), respectively,  we obtain the following super expansions
of the {\it four} fermionic superfields $({\cal F}_1 , \, \bar {\cal F}_1, \,{\cal F}_2 , \, \bar {\cal F}_2 )$:
\[
\qquad {\cal F}_1^{(h)}\,(x, \theta, \bar\theta) = C_1 + \theta\, (B_1) + \bar\theta\, (-B) + \theta\bar\theta \,(0)
\]
\[
 ~~~~~~\qquad\qquad \qquad \qquad \qquad \equiv C_1 + \theta \,(s_{ab}\, C_1) + \bar\theta \,(s_b\, C_1)
 + \theta\bar\theta \,(s_b\, s_{ab}\, C_1),
\]
\[
~~\qquad \bar {\cal F}_1^{(h)}\,(x, \theta, \bar\theta) = \bar C_1 + \theta \,(- B_2) 
+ \bar\theta \,(-B_1) + \theta\bar\theta\, (0)
\]
\[
 ~~~~~~\qquad\qquad \qquad \qquad \qquad  \equiv \bar C_1 + \theta\, (s_{ab}\, \bar C_1) 
+ \bar\theta \,(s_b\, \bar C_1) + \theta\bar\theta \,(s_b\, s_{ab}\, \bar C_1),
\]
\[
{\cal F}_2^{(h)}\,(x, \theta, \bar\theta) = C_2 + \theta\, (B) + \bar\theta\, (0) + \theta\bar\theta \,(0)
\]
\[
 ~~~~~~\qquad\qquad \qquad \qquad \qquad \equiv C_2 + \theta \,(s_{ab}\, C_2) + \bar\theta \,(s_b\, C_2) 
+ \theta\bar\theta \,(s_b\, s_{ab}\, C_2),
\]
\[
~~\bar{\cal F}_2^{(h)}\,(x, \theta, \bar\theta) = \bar C_2 + \theta\, (0) + \bar\theta\, ( B_2) + \theta\bar\theta \,(0)
\]
\[
 ~  \qquad \qquad\qquad\qquad \qquad \qquad \qquad \equiv \bar C_2 + \theta \,(s_{ab}\, \bar C_2) + \bar\theta \,(s_b\, \bar C_2) 
+ \theta\bar\theta \,(s_b\, s_{ab}\, \bar C_2).
\eqno (A. 4)
\]
In the above, the superscript $(h)$ denotes that the above superfields and their super expansions along the Grassmannian directions
$(\theta, \, \bar\theta)$ of the $(D, 2)$-dimensional supermanifold have been obtained after the application of 
horizontality condition (HC). Furthermore, the coefficients of $\theta$ and $\bar\theta$ are nothing but the off-shell nilpotent (anti-)BRST symmetry 
transformations (17) and (18) of our theory.

We observe, furthermore, that the setting the coefficients of the {\it super} 4-form differentials:
$(d x^\mu \wedge d\theta \wedge d \theta \wedge d \theta), \, (d x^\mu \wedge d \bar\theta \wedge d \bar\theta \wedge d \bar\theta), \,
(d x^\mu \wedge d\theta \wedge d \theta \wedge d \bar\theta), \, (d x^\mu\wedge d\theta \wedge d \bar\theta \wedge d \bar\theta)$
equal to zero, respectively, lead to the following relationships
\[
\partial_\mu\, \bar {\cal F}_2^{(h)} - \partial_\theta\, \tilde{\bar\beta}_\mu = 0, \qquad \qquad
\partial_\mu\, {\cal F}_2^{(h)} - \partial_{\bar\theta}\, \tilde\beta_\mu = 0,
\]
\[
\partial_\mu\, \bar {\cal F}_1^{(h)} - \partial_{\bar\theta}\, \tilde{\bar\beta}_\mu - \partial_\theta\, \tilde\Phi_\mu = 0, \qquad\quad 
\partial_\mu\, {\cal F}_1^{(h)} - \partial_{\theta}\, \tilde\beta_\mu - \partial_{\bar\theta}\, \tilde\Phi_\mu = 0,
\eqno (A.5)
\]
where the superfields with superscript $(h)$ have been illustrated in (A.4). The above relationships (A.5) lead to the following:
\[
f_\mu^{(1)} = \partial_\mu\, C_2, \quad b_\mu = i\, \partial_\mu\, B, \quad \bar f_\mu^{(2)} = \partial_\mu\, \bar C_2, \quad 
\bar b_\mu = -\, i\, \partial_\mu\, B_2,
\]
\[
 b_\mu^{(3)} = i\, \partial_\mu\, B_1, \quad \bar f_\mu^{(3)} +  f_\mu^{(2)} = \partial_\mu\, \bar C_1, 
 \quad f_\mu^{(3)} + \bar f_\mu^{(1)} = \partial_\mu\, C_1. 
\eqno (A.6)
\]
In the above, we have the precise expression for the secondary fields in terms of the {\it basic} and {\it auxiliary} 
fields of our theory and we have {\it also} derived the CF-type restrictions
\[
\bar f_\mu +  { F}_\mu = \partial_\mu\, \bar C_1, \qquad  f_\mu + \bar { F}_\mu = \partial_\mu\, C_1, 
\eqno (A. 7)
\]
where we have identified $ \bar f_\mu^{(3)} = \bar {f}_\mu,\;  f_\mu^{(2)} = { F}_\mu,\; f_\mu^{(3)} = {f}_\mu,\;
\bar f_\mu^{(1)} = \bar {F}_\mu$. 
It is clear that the {\it non-trivial} CF-type restrictions (A.7) emerge out when we set  equal to {\it zero}
the coefficients of the {\it super} differentials: 
$ (d x^\mu\wedge d\theta \wedge d \bar\theta \wedge d \bar\theta)$ and $ (d x^\mu \wedge d\theta \wedge d \theta \wedge d \bar\theta) $, respectively.
The substitution of (A.6) leads to the following super expansions
\[
~~~\tilde\beta_\mu^{(h)} (x, \theta, \bar\theta)  =  \beta_\mu (x) + \theta \, (\bar F_\mu) 
+ \bar\theta \, (\partial_\mu C_2) + \theta\bar\theta \, (-\, \partial_\mu B)
\]
\[
 ~~~~~~~~~~~~ \qquad \qquad  \equiv \beta_\mu (x) + \theta \, (s_{ab}\, \beta_\mu) 
+ \bar\theta \, (s_{b}\, \beta_\mu) + \theta\bar\theta \, (s_b\,s_{ab}\, \beta_\mu),
\]
\[
~~~~\tilde{\bar\beta}_\mu^{(h)} (x, \theta, \bar\theta) = \bar\beta_\mu (x) + \theta \,(\partial_\mu \bar C_2)
+ \bar\theta \, (F_\mu) + \theta\bar\theta \, (\partial_\mu B_2)
\]
\[
 ~~~~~~~~~~~~\qquad \qquad  \equiv \bar\beta_\mu (x) + \theta \, (s_{ab}\, \bar\beta_\mu) 
+ \bar\theta \, (s_{b}\, \bar\beta_\mu) + \theta\bar\theta \, (s_b\,s_{ab}\, \bar\beta_\mu),
\]
\[
\tilde\Phi_\mu^{(h)} (x, \theta, \bar\theta) = \phi_\mu (x) + \theta \, (\bar f_\mu) 
+ \bar\theta \, (f_\mu) + \theta\bar\theta \, (-\, \partial_\mu B_1)
\]
\[
 ~\qquad \qquad\qquad \qquad  \equiv \phi_\mu (x) + \theta \, (s_{ab}\, \phi_\mu) 
+ \bar\theta \, (s_{b}\, \phi_\mu) + \theta\bar\theta \, (s_b\,s_{ab}\, \phi_\mu),
\eqno (A.8)
\]
where the superscript $(h)$ on the l.h.s., once again, denotes that the above {\it three} bosonic superfields have been 
obtained after the application of HC. It is worthwhile to point out a clarification at this stage. From (A.7), it 
is clear that the ghost numbers for pairs $(\bar F_\mu, \, F_\mu)$ are $(+1, -1)$, respectively, despite the 
notations which are a bit misleading.

At this stage, we now focus on setting the coefficients of the {\it super} 4-form differentials: 
$(d x^\mu\wedge d x^\nu \wedge d \theta \wedge d \theta), \; (d x^\mu\wedge d x^\nu \wedge d \bar\theta \wedge d \bar\theta), \; 
(d x^\mu\wedge d x^\nu \wedge d \theta \wedge d \bar\theta)$ equal to zero due to HC.
These restrictions lead to the following relationships
\[
 ~~~~\qquad \qquad \qquad \partial_\theta \bar{\cal F}_{\mu\nu} = \partial_\mu \tilde{\bar \beta}_\nu^{(h)} - \partial_\nu \tilde{\bar \beta}_\mu^{(h)}, 
\qquad
\partial_{\bar\theta} {\cal F}_{\mu\nu} = \partial_\mu \tilde{ \beta}_\nu^{(h)} - \partial_\nu \tilde{\beta}_\mu^{(h)},
\]
\[
\partial_\theta {\cal F}_{\mu\nu} + \partial_{\bar\theta} \bar{\cal F}_{\mu\nu} =
\partial_\mu \tilde{\Phi}_\nu^{(h)} - \partial_\nu \tilde{\Phi}_\mu^{(h)},
\eqno (A.9)
\]
where the super expansions of the superscript $(h)$ have been quoted in Eq. (A.8). 
The above relationships in (A.9) lead to
\[
\bar B_{\mu\nu}^{(2)} = (\partial_\mu \bar\beta_\nu - \partial_\nu \bar\beta_\mu), \quad 
\bar s_{\mu\nu} = -\, i\, (\partial_\mu F_\nu - \partial_\nu F_\mu) \equiv i\, (\partial_\mu \bar f_\nu - \partial_\nu \bar f_\mu),
\]
\[
B_{\mu\nu}^{(1)} = (\partial_\mu \beta_\nu - \partial_\nu \beta_\mu), \quad 
s_{\mu\nu} = i\, (\partial_\mu \bar F_\nu - \partial_\nu \bar F_\mu) \equiv -\, i\, (\partial_\mu f_\nu - \partial_\nu  f_\mu),
\]
\[
\bar B_{\mu\nu}^{(1)} + B_{\mu\nu}^{(2)} = (\partial_\mu \phi_\nu - \partial_\nu \phi_\mu) \Longrightarrow 
\bar B_{\mu\nu} + B_{\mu\nu} = (\partial_\mu \phi_\nu - \partial_\nu \phi_\mu).
\eqno (A.10)
\]
The last entry, in the above equation, is the {\it non-trivial} CF-type restriction where we have taken into account: 
$ \bar B_{\mu\nu}^{(1)}  = \bar B_{\mu\nu} $ and $B_{\mu\nu}^{(2)}  = B_{\mu\nu}$.
This CF-type restriction [i.e. $\bar B_{\mu\nu} + B_{\mu\nu} = (\partial_\mu \phi_\nu - \partial_\nu \phi_\mu)$]
is valid if and only if the other {\it two} fermionic CF-type restrictions (i.e. $f_\mu + \bar { F}_\mu = \partial_\mu\, C_1 , \; 
\bar f_\mu +  { F}_\mu = \partial_\mu\, \bar C_1$) are satisfied. In other words, due to the validity of the 
{\it fermionic} CF-type restrictions, we have the following
\[
\partial_\mu \bar f_\nu - \partial_\nu \bar f_\mu = -\, (\partial_\mu F_\nu - \partial_\nu F_\mu), \qquad 
\partial_\mu  f_\nu - \partial_\nu f_\mu = -\, (\partial_\mu \bar F_\nu - \partial_\nu \bar F_\mu).
\eqno (A.11)
\]
Ultimately, we have derived the following super expansions
\[
~~~~~{\cal F}_{\mu\nu}^{(h)}(x, \theta, \bar\theta) = C_{\mu\nu} + \theta\, (\bar B_{\mu\nu}) 
+ \bar\theta\, (\partial_\mu \beta_\nu - \partial_\nu \beta_\mu) + \theta\bar\theta \, [-(\partial_\mu \bar F_\nu - \partial_\nu \bar F_\mu)]
\]
\[
~~~~~~~\equiv C_{\mu\nu} + \theta\, (s_{ab}\,C_{\mu\nu}) 
+ \bar\theta\, (s_b\, C_{\mu\nu} ) + \theta\bar\theta \,(s_b\, s_{ab}\,C_{\mu\nu}),
\]
\[
\bar{\cal F}_{\mu\nu}^{(h)}(x, \theta, \bar\theta) = \bar C_{\mu\nu} + \theta\,(\partial_\mu \bar \beta_\nu - \partial_\nu \bar \beta_\mu)
+ \bar\theta\, (B_{\mu\nu})   + \theta\bar\theta \, (\partial_\mu F_\nu - \partial_\nu F_\mu)
\]
\[
~~~~~~~\equiv \bar C_{\mu\nu} + \theta\, (s_{ab}\,\bar C_{\mu\nu}) 
+ \bar\theta\, (s_b\, \bar C_{\mu\nu} ) + \theta\bar\theta \,(s_b\, s_{ab}\,\bar C_{\mu\nu}),
\eqno (A.12)
\]
where the superscript $(h)$ on the superfields carries its standard meaning.

Before we end this Appendix, we point out that setting the coefficients of 
$(d x^\mu\wedge d x^\nu \wedge d x^\lambda \wedge d \theta)$, and $(d x^\mu\wedge d x^\nu \wedge d x^\lambda \wedge d \bar\theta)$
 equal to {\it zero}, respectively, leads to  the following conditions on the superfields:
\[
\partial_\mu \bar {\cal F}_{\nu\lambda}^{(h)} + \partial_\nu \bar {\cal F}_{\lambda \mu}^{(h)} 
+ \partial_\lambda \bar {\cal F}_{\mu\nu}^{(h)} = \partial_\theta \tilde A_{\mu\nu\lambda},
\]
\[
\partial_\mu{\cal F}_{\nu\lambda}^{(h)} + \partial_\nu {\cal F}_{\lambda \mu}^{(h)} 
+ \partial_\lambda {\cal F}_{\mu\nu}^{(h)} = \partial_{\bar\theta} \tilde A_{\mu\nu\lambda}.
\eqno (A.13)
\]
The above relationships imply the following
\[
R_{\mu\nu\lambda} = \partial_\mu C_{\nu\lambda} + \partial_\nu C_{\lambda\mu} + \partial_\lambda C_{\mu\nu}, \quad
\bar R_{\mu\nu\lambda} = \partial_\mu \bar C_{\nu\lambda} + \partial_\nu \bar C_{\lambda\mu} + \partial_\lambda \bar C_{\mu\nu},
\]
\[
S_{\mu\nu\lambda} = i\, (\partial_\mu \bar B_{\nu\lambda} + \partial_\nu \bar B_{\lambda\mu} + \partial_\lambda \bar B_{\mu\nu})
 \equiv -\, i\, (\partial_\mu  B_{\nu\lambda} + \partial_\nu  B_{\lambda\mu} + \partial_\lambda B_{\mu\nu}),
\eqno (A.14)
\]
where the last entry is {\it true} only due to the CF-type restriction: $B_{\mu\nu} + \bar B_{\mu\nu} = 
\partial_\mu \phi_\nu - \partial_\nu \phi_\mu$. Finally, the equality of the coefficients of the 4-form differential
$(d x^\mu\wedge d x^\nu \wedge d x^\lambda \wedge d x^\eta )$ from the l.h.s. and r.h.s. of $\tilde d\, \tilde A^{(3)} = d\, A^{(3)}$ leads to
\[
\partial_\mu \tilde A_{\nu\lambda\eta}^{(h)} - \partial_\nu \tilde A_{\lambda\eta\mu}^{(h)} + \partial_\lambda \tilde A_{\eta\mu\nu}^{(h)}
- \partial_\eta \tilde A_{\mu\nu\lambda}^{(h)} = H_{\mu\nu\lambda\eta},
\eqno (A.15)
\]
where the third-rank superfield with superscript $(h)$ is:
\[
\tilde A_{\mu\nu\lambda}^{(h)} (x, \theta, \bar\theta)  = A_{\mu\nu\lambda} (x) 
+ \theta \,(\partial_\mu \bar C_{\nu\lambda} + \partial_\nu \bar C_{\lambda\mu} + \partial_\lambda \bar C_{\mu\nu})
+ \bar\theta\,  (\partial_\mu C_{\nu\lambda} + \partial_\nu C_{\lambda\mu} + \partial_\lambda C_{\mu\nu})
\]
\[
+ \theta\bar\theta\, (\partial_\mu  B_{\nu\lambda} + \partial_\nu  B_{\lambda\mu} + \partial_\lambda B_{\mu\nu})
\]
\[
\equiv  A_{\mu\nu\lambda} + \theta\, (s_{ab}\,  A_{\mu\nu\lambda}) + \bar\theta\, (s_b\,  A_{\mu\nu\lambda})
+ \theta \bar\theta\, (s_b\, s_{ab}\,  A_{\mu\nu\lambda}).
\eqno (A. 16)
\]
A close look at the above mentioned equations establish that we have derived the off-shell nilpotent and absolutely anticommuting
(anti-)BRST symmetry transformations [cf. Eqs (17), (18)] and the {\it non-trivial} (anti-)BRST invariant CF-type restrictions.
We lay emphasis on the fact that (A.15) is satisfied due to the precise expression for the secondary fields 
in Eq. (A.14). It is interesting to mention, in passing, that the (anti-)BRST invariance of the CF-type restrictions, off-shell nilpotency
and absolute anticommutativity properties lead to the {\it complete} derivations of the precise forms of the 
(anti-)BRST symmetry transformations (17) and (18) for the {\it massless} version of our {\it massive} Abelian 3-form theory.

\vskip 1cm 
\begin{center}
{\bf Appendix B: On the Complete Set of (Anti-)BRST Symmetries and Total BRST and Anti-BRST Invariant Lagrangian Densities}\\
\end{center}
\noindent
The purpose of this Appendix is to collect {\it all} the off-shell nilpotent $[s_{(a)b}^2 = 0]$ and absolutely anticommuting
$(s_b\, s_{ab} + s_{ab}\, s_b = 0)$ (anti-)BRST symmetry transformations  [$s_{(a)b}$] of our theory that have been derived 
from the HC (8) and GIR (26) in the main body of our text. We have also used the {\it requirements} of the off-sell nilpotency as well
as the absolute anticommutativity properties and the  
invariance of the CF-type restrictions under the (anti-)BRST symmetry transformations to obtain the {\it complete} list of the 
(anti-)BRST symmetry transformations [$s_{(a)b}$].
The infinitesimal, continuous  and off-shell nilpotent  $(s_b^2 = 0)$ BRST transformations $(s_b)$ are as follows:
\[
~~~~~~s_b A_{\mu\nu\lambda} = \partial_\mu C_{\nu\lambda} + \partial_\nu C_{\lambda\mu} + \partial_\lambda C_{\mu\nu}, \quad
s_b C_{\mu\nu} = \partial_\mu \beta_\nu - \partial_\nu \beta_\mu,
\quad 
\]
\[
 ~~~~~~~s_b \bar B_{\mu\nu} = -\, (\partial_\mu \bar F_\nu - \partial_\nu \bar F_\mu) \equiv (\partial_\mu\, f_\nu - \partial_\nu\, f_\mu), 
 \quad s_b \bar C_{\mu\nu} = B_{\mu\nu}, \quad
\]
\[
~~~~~s_b\, \Phi_{\mu\nu} = \pm\, m\, C_{\mu\nu} - \, (\partial_\mu\, C_\nu - \partial_\nu\, C_\mu), \qquad \quad s_b \bar F_\mu = - \partial_\mu  B, 
\]
\[ 
~~~s_b\, C_\mu = \pm\, m\, \beta_\mu - \partial_\mu\, \beta, \qquad  s_b \phi_\mu = f_\mu, \qquad  s_b \beta_\mu = \partial_\mu C_2, 
\]
\[
 ~~~s_b\, \bar B_\mu = \pm\, m\, f_\mu - \partial_\mu f,  \qquad s_b \bar f_\mu = -\, \partial_\mu B_1,  \quad s_b \bar \beta_\mu = F_\mu,
\]
\[
s_b\, \bar C_\mu = B_\mu, \qquad  s_b \bar C_2 = B_2, \qquad  s_b \bar C_1 = -\, B_1, ~~~~~~~~~
\]
\[
s_b\, \phi = f, \qquad   s_b\, \beta = \pm \, m\, C_2, \qquad s_b\, \bar\beta = F, ~~~~~~~~~~~~
\]
\[
s_b\, \bar F = \mp\, m\, B, \qquad s_b\, \bar f = \mp\, m\, B_1, \qquad  s_b \, C_1 = - B, 
\]
\[
s_b\, [H_{\mu\nu\lambda\zeta}, \, B_{\mu\nu}, \, B_\mu, \, f_\mu, \, F_\mu, \, F, \, f,\,   B, \, B_1,\, B_2, \, C_2] = 0.
\eqno (B.1)
\]
On top of the above transformations, we have the corresponding off-shell nilpotent $(s_{ab}^2 = 0)$, infinitesimal and continuous 
 anti-BRST symmetry transformations  $(s_{ab})$ as:
\[
~~~~~~~~~~~~ s_{ab} A_{\mu\nu\lambda} = \partial_\mu \bar C_{\nu\lambda} + \partial_\nu \bar C_{\lambda\mu} + \partial_\lambda \bar C_{\mu\nu}, \quad
s_{ab} \bar C_{\mu\nu} = \partial_\mu \bar \beta_\nu - \partial_\nu \bar\beta_\mu, \quad
\]
\[
~~~~~~~~~~ s_{ab} B_{\mu\nu} = -\, (\partial_\mu  F_\nu - \partial_\nu  F_\mu) \equiv (\partial_\mu  \bar f_\nu - \partial_\nu  \bar f_\mu),
 \quad s_{ab}  C_{\mu\nu} = \bar B_{\mu\nu},
\]
\[
~~~~~~s_{ab}\, \Phi_{\mu\nu} = \pm\, m\, \bar C_{\mu\nu} - \, (\partial_\mu\, \bar C_\nu - \partial_\nu\, \bar C_\mu), \quad s_{ab}\,  F_\mu = 
-\, \partial_\mu\, B_2
\]
\[
~~~~~~~ s_{ab}\, \bar C_\mu = \pm\, m\, \bar\beta_\mu
 - \partial_\mu\, \bar\beta, \qquad s_{ab} \phi_\mu = \bar f_\mu, \quad  s_{ab}  \bar \beta_\mu = \partial_\mu\, \bar C_2,~
\]
\[
~~~~~~~~~~~ s_{ab}\, B_\mu = \pm \, m\, \bar f_\mu - \, \partial_\mu \, \bar f, \quad  
 \quad s_{ab}  \beta_\mu = \bar F_\mu,  \quad s_{ab}  f_\mu =  \partial_\mu B_1,~~~~
\]
\[
s_{ab}\, C_\mu = \bar B_\mu, \qquad  s_{ab}\, f = \pm\, m\, B_1,  \qquad s_{ab}\,\phi = \bar f, ~~~~~
\]
\[
~~ s_{ab}\,\bar\beta = \pm\, m\, \bar C_2, \qquad 
s_{ab}\,\beta = \bar F, \qquad s_{ab}\, F = \mp\, m\, B_2~~
\]
\[
 \qquad s_{ab} \bar C_1 = - B_2, \qquad  s_{ab} C_1 =  B_1, \qquad s_{ab} C_2 =  B, ~~~~~~~~~~~
\]
\[
~~~~s_{ab}\, [H_{\mu\nu\lambda\zeta}, \, \bar B_{\mu\nu}, \, \bar B_\mu, \;  \bar F_\mu,\,\bar f_\mu, \, 
 \bar F,\,\bar f, \,  B, \, B_1, \, B_2, \, \bar C_2] = 0.
\eqno (B.2)
\]
The above off-shell nilpotent [$s_{(a)b}^2 =0$] symmetry transformations leave the Lagrangian densities 
${\cal L}_B^{(HC, \, GIR)}$ and ${\cal L}_{\bar B}^{(HC, \, GIR)}$ {\it perfectly} invariant because we observe that 
$s_{ab}\, {\cal L}_{\bar B}^{(HC, \, GIR)}$ and  $s_{b}\, {\cal L}_{ B}^{(HC, \, GIR)}$ are total spacetime derivatives
 [cf. Eqs. (57), (58), (64), (66)]. In addition, we have already mentioned that the St$\ddot u$ckelberg-modified Lagrangian 
 density (4) remains invariant $[s_{(a)b}\, {\cal L}_{S} = 0]$ under the (anti-)BRST symmetry transformations, too.

The above complete set of (anti-)BRST symmetry transformations of (B.2) and (B.1) are {\it also} anticommuting in nature.
As pointed out earlier in the main body of the text [cf. Eq. (19)] that
$\{s_b, s_{ab} \} \;A_{\mu\nu\lambda} = 0, \;
\{s_b, s_{ab} \} \; C_{\mu\nu} = 0,$ and $
\{s_b, s_{ab} \}\; \bar C_{\mu\nu} = 0$
are {\it satisfied} only when the CF-type conditions:
$ B_{\mu\nu} + \bar B_{\mu\nu} = \partial_\mu \phi_\nu - \partial_\nu \phi_\mu,\, 
f_\mu +  \bar F_\mu = \partial_\mu C_1,$ and $
\bar f_\mu +  F_\mu = \partial_\mu \bar C_1$  are imposed, respectively,
from {\it outside}\footnote{It should be pointed out that $(i)$ the proper (anti-)BRST symmetry 
transformations for $A_{\mu\nu\lambda}, \, C_{\mu\nu}$ and $\bar C_{\mu\nu}$, 
and $(ii)$ the (anti-)BRST invariant CF-type restrictions: $ B_{\mu\nu} + \bar B_{\mu\nu} = \partial_\mu \phi_\nu - \partial_\nu \phi_\mu,\, 
f_\mu +  \bar F_\mu = \partial_\mu C_1,$ and $\bar f_\mu +  F_\mu = \partial_\mu \bar C_1$ have been derived from the HC (cf. Appendix A below for details).}.
It is very interesting to observe that the following absolute anticommuting properties, namely;
\[
\{s_b, s_{ab} \} \; \Phi_{\mu\nu} = 0, \quad  \quad  \{s_b, s_{ab} \} \; C_{\mu} = 0, \quad \quad  \{s_b, s_{ab} \} \; \bar C_{\mu} = 0,
\eqno (B.3)
\]
are satisfied provided we take into account the following {\it pair} of (anti-)BRST invariant CF-type restrictions
that have been derived using HC and GIR, namely; 
\[
\qquad B_{\mu\nu} + \bar B_{\mu\nu} = \partial_\mu \phi_\nu - \partial_\nu \phi_\mu, \qquad  
B_\mu + \bar B_\mu = \pm m\, \phi_\mu - \partial_\mu\ \phi,
\]
\[
f_\mu +  \bar F_\mu = \partial_\mu C_1, \qquad \qquad f + \bar F = \pm\, m\, C_1,
\]
\[
\bar f_\mu +  F_\mu = \partial_\mu \bar C_1, \qquad \qquad \bar f + F = \pm\, m\, \bar C_1.
\eqno (B.4)
\]
In other words, the absolute anticommutativity property for the St$\ddot u$ckelberg-compensating field and associated 
fermionic (anti-)ghost fields is satisfied if and only if we take into account the {\it pair} of CF-type restrictions (B.4)
{\it together}. This happens due to the fact that the GIR [cf. Eq. (26)] incorporates HC as well as gauge invariance {\it together} 
for its validity.

Before we end this Appendix, we write down the {\it perfectly} BRST-invariant and {\it perfectly} anti-BRST 
invariant {\it coupled} Lagrangian densities. First of all, we express the {\it perfectly} BRST-invariant {\it total}
Lagrangian density ${\cal L}_B^{(T)}$ as follows
\[
{\cal L}_B^{(T)} = {\cal L}_{(S)} + {\cal L}_{gf}^{(B)} + {\cal L}_{FP}^{(B)} ,
\eqno (B.5)
\]
where the above symbols stand for the following:
\[
{\cal L}_{(S)} = \frac {1}{24}\,H^{\mu \nu \lambda \zeta }\,H_{\mu \nu \lambda \zeta} - \frac {m^2}{6} 
\, A^{\mu\nu\lambda }\,A_{\mu\nu\lambda } - \frac{1}{6}\, \Sigma^{\mu \nu \lambda }\, \Sigma_{\mu \nu \lambda }
\pm \frac{m}{3}\,A^{\mu\nu\lambda} \, \Sigma_{\mu \nu \lambda }, ~~
\]
\[
 {\cal L}_{gf}^{(B)}  + {\cal L}_{FP}^{(B)} 
  \equiv {\cal L}_{B}^{(HC)} + {\cal L}_{B}^{(GIR)}, 
\]
\[
\equiv (\partial_\mu A^{\mu\nu\lambda}) B_{\nu\lambda}  - \frac{1}{2} B_{\mu\nu}\, B^{\mu\nu}
+ \frac{1}{2}\, B^{\mu\nu}\, \Big[ \partial_\mu \, \phi_\nu - \partial_\nu\, \phi_\mu \mp \, m\, \Phi_{\mu\nu}  \Big]
\]
\[
-\, (\partial_\mu\, \Phi^{\mu\nu})\, B_\nu - \frac{1}{2}\, B^\mu\, B_\mu + \frac{1}{2}\, B^{\mu}\, \Big[\pm \, m\phi_\mu - \partial_\mu\, \phi\Big] 
 + \frac{m^2}{2}\, \bar C_{\mu\nu}\, C^{\mu\nu}  ~~~
\]
\[
~~~~~~~~~+ (\partial_\mu \bar C_{\nu\lambda} + \partial_\nu \bar C_{\lambda\mu} + 
\partial_\lambda \bar C_{\mu\nu}) (\partial^\mu C^{\nu\lambda}) \pm m\, (\partial_\mu\, \bar C^{\mu\nu})\, C_\nu \pm \, m\, \bar C^\nu\, (\partial^\mu\, C_{\mu\nu})
\]
\[ 
~~~ + (\partial_\mu\, \bar C_\nu - \partial_\nu\, \bar C_\mu)\, (\partial^\mu \ C^\nu)  
- \frac{1}{2}\, \Big[\pm m\, \bar \beta^\mu 
- \partial^\mu\, \bar\beta \Big]\, \Big[\pm m\,  \beta_\mu - \partial_\mu\, \beta \Big]
\]
\[
~~~~~~~-\, (\partial_\mu \, \bar\beta_\nu - \partial_\nu\, \bar\beta_\mu)\, (\partial^\mu\,\beta^\nu) - \partial_\mu \bar C_2 \partial^\mu  C_2
 - \, m^2\, \bar C_2\, C_2 + [(\partial \cdot \bar\beta) \mp m\, \bar \beta]\, B 
\]
\[
~~~~~~~~~- [(\partial \cdot \phi) \mp m\, \phi]\, B_1 - [(\partial \cdot \beta) \mp m\, \beta]\, B_2
+ \Big[\partial_\nu \bar C^{\nu\mu}  + \partial^\mu\, \bar C_1 \mp \frac {m}{2}\, \bar C^\mu\Big]\, f_\mu
\]
\[
~~~~~~~~~~~~~~~- \, 2\, F^\mu\,f_\mu - 2 \, F\, f - \,\Big[\partial_\nu  C^{\nu\mu}  + \partial^\mu\,  C_1 \mp \frac {m}{2}\,  C^\mu\Big]\, F_\mu
 + F\, \Big[\pm m\, C_1 - \frac{1}{2}\, (\partial \cdot C)\Big]
\]
\[
 - \Big[\frac{1}{2}\, (\partial \cdot \bar C) \mp m\, \bar C_1\Big]\, f-\, B\, B_2 -\, \frac{1}{2}\, B_1^2.~~~~~~~~~~~~~~~~~~~~~~~~~~~~~~~~~
 \eqno (B.6)
\]
It is worth pointing out that the above {\it perfectly} BRST-invariant {\it total}
Lagrangian density is function of the auxiliary fields $(B_{\mu\nu}, \, B_\mu, \,  F_\mu, \, f_\mu, \, F, \,  f)$ and their
counterparts $(\bar B_{\mu\nu}, \, \bar B_\mu, \, \bar F_\mu, \, \bar f_\mu, \, \bar F, \, \bar f)$ do {\it not} appear in the theory due to our exercise in (49).
To be precise, the {\it total} Lagrangian density (B.6) is an exact sum of (43) and the {\it modified} form of (47) and/or (63)
that is re-expressed  as:
\[
{\cal L}_B^{(GIR)} =  
-\, (\partial_\mu \Phi^{\mu\nu})\, B_\nu -\, \frac{1}{2}\, B^\mu B_\mu   + \frac{1}{2}\,  B^{\mu}\, \big(\pm \, m\phi_\mu - \partial_\mu\, \phi \big)
\mp \frac{m}{2}\, B^{\mu\nu}\, \Phi_{\mu\nu} ~~~~~~~
\]
\[
~~~~+ \frac{m^2}{2}\, \bar C_{\mu\nu}\, C^{\mu\nu} 
+ (\partial_\mu \bar C_\nu - \partial_\nu \bar C_\mu)\, \partial^\mu C^\nu 
\pm m\, (\partial_\mu \bar C^{\mu\nu})\, C_\nu \mp\, \frac{m}{2}\, F^\mu\, C_\mu  
\pm\, m\, \bar C^\nu\, (\partial^\mu C_{\mu\nu}) 
\]
\[
 - \, \frac{1}{2}\, \big[\pm \, m\, \bar\beta^\mu - \partial^\mu \bar\beta \big]\, \big[ \pm \, m\, \beta_\mu - \partial_\mu \beta \big]
 \mp \frac{m}{2}\, \bar C_\mu f^\mu 
+ F\, \big[ \pm m\, C_1 -\,  \frac{1}{2}\,\partial \cdot C\big]~~~~~~~
\]
\[
-  \big[ \frac{1}{2}\, \partial \cdot \bar C \mp m\, \bar C_1 \big]\, f \pm m\, B_1\, \phi 
 -\, 2\, F\, f  \,
-\, m^2\, \bar C_2\, C_2 \mp\, m\, B \, 
\bar\beta \pm m\, B_2\, \beta.~~~~~~
\eqno (B.7)
\]
The analogues of (B.5) and (B.6) can be written as the {\it perfectly} anti-BRST {\it total} Lagrangian density $[{\cal L}_{\bar B}^{(T)}]$ as follows
\[
{\cal L}_{\bar B}^{(T)} = {\cal L}_{(S)} + {\cal L}_{gf}^{(\bar B)}  + {\cal L}_{FP}^{(\bar B)},
\eqno (B.8)
\]
where the above symbols stand for the following:
\[
{\cal L}_{(S)} \; = \; \frac {1}{24}\,H^{\mu \nu \lambda \zeta }\,H_{\mu \nu \lambda \zeta} - \frac {m^2}{6} 
\, A^{\mu\nu\lambda }\,A_{\mu\nu\lambda } - \frac{1}{6}\, \Sigma^{\mu \nu \lambda }\, \Sigma_{\mu \nu \lambda }
\pm \frac{m}{3}\,A^{\mu\nu\lambda} \, \Sigma_{\mu \nu \lambda }, ~~
\]
\[
 {\cal L}_{gf}^{(\bar B)} + {\cal L}_{FP}^{(\bar B)} 
 \equiv {\cal L}_{\bar B}^{(HC)} + {\cal L}_{\bar B}^{(GIR)}, 
\]
\[
 ~~\equiv -\, (\partial_\mu A^{\mu\nu\lambda}) \bar B_{\nu\lambda}  - \frac{1}{2} \bar B_{\mu\nu}\, \bar B^{\mu\nu}
+ \frac{1}{2}\, \bar B^{\mu\nu}\, \Big[ \partial_\mu \, \phi_\nu - \partial_\mu\, \phi_\nu \pm \, m\, \Phi_{\mu\nu}  \Big]
\]
\[
~~~~+\, (\partial_\mu\, \Phi^{\mu\nu})\, \bar B_\nu - \frac{1}{2}\,  \bar B^\mu\, \bar B_\mu +
  \frac{1}{2}\,  \bar B^{\mu}\, \Big[\pm \, m\phi_\mu - \partial_\mu\, \phi \Big] 
+ \frac{m^2}{2}\, \bar C_{\mu\nu}\, C^{\mu\nu}
\]
\[
~~~~~~~~~~~ + (\partial_\mu \bar C_{\nu\lambda} + \partial_\nu \bar C_{\lambda\mu} + 
\partial_\lambda \bar C_{\mu\nu}) (\partial^\mu C^{\nu\lambda}) 
\pm m\, (\partial_\mu\, \bar C^{\mu\nu})\, C_\nu
\pm \, m\, \bar C^\nu\, (\partial^\mu\, C_{\mu\nu})
\]
\[
~~~~~~ + 
(\partial_\mu\, \bar C_\nu - \partial_\nu\, \bar C_\mu)\, (\partial^\mu \ C^\nu) 
 - \frac{1}{2}\, \Big[\pm m\, \bar \beta^\mu 
- \partial^\mu\, \bar\beta \Big]\, \Big[\pm m\,  \beta_\mu - \partial_\mu\, \beta \Big]
\]
\[
~~~~~~~~~~~-\, (\partial_\mu \, \bar\beta_\nu - \partial_\nu\, \bar\beta_\mu)\, (\partial^\mu\,\beta^\nu) - \partial_\mu \bar C_2 \partial^\mu  C_2
 - \, m^2\, \bar C_2\, C_2 + [(\partial \cdot \bar\beta) \mp m\, \bar \beta]\, B 
\]
\[
~~~~~~~~~~~~- [(\partial \cdot \phi) \mp m\, \phi]\, B_1 -  [(\partial \cdot \beta) \mp m\, \beta]\, B_2
+ \Big[\partial_\nu C^{\nu\mu}  - \partial^\mu\,  C_1 \mp \frac {m}{2}\,  C^\mu\Big]\, \bar f_\mu
\]
\[
~~~~~~~~~~~~~~~+ \, 2\, \bar F^\mu\, \bar f_\mu + 2 \, \bar F\, \bar f - \,\Big[ \partial_\nu  \bar C^{\nu\mu}  - \partial^\mu\,  
\bar C_1 \mp \frac {m}{2}\,  \bar C^\mu\Big]\, \bar F_\mu
 + \Big[ \frac{1}{2}\, (\partial \cdot \bar C) \pm m\, \bar C_1 \Big]\, \bar F 
\]
\[
~~~~- \Big[\frac{1}{2}\, (\partial \cdot C) \pm m\, C_1 \Big]\, \bar f  -\, B\, B_2 -\, \frac{1}{2}\, B_1^2. ~~~~~~~~~~~~~~~~~~~~~~~~~~~~~~~~~
 \eqno (B.9)
\]
It is worthwhile to point out the fact that, in the total {\it perfectly} anti-BRST {\it total} Lagrangian density $[{\cal L}_{\bar B}^{(T)}]$,
 we have the presence of the auxiliary fields $(\bar B_{\mu\nu}, \, \bar B_\mu, \, \bar F_\mu, \, \bar f_\mu, \, \bar F, \, \bar f)$
and the set of auxiliary fields $(B_{\mu\nu}, \, B_\mu, \, F_\mu, \, f_\mu, \, F,\,  f)$ does {\it not} appear anywhere [cf. Eq. (49)].
In other words, the above equation (B.9) is the {\it exact} sum of the Lagrangian density (44) and the {\it modified} version of (48)
and/or (65) that is re-expressed as:
\[
{\cal L}_{\bar B}^{(GIR)} = 
(\partial_\mu \Phi^{\mu\nu})\,  \bar B_\nu \pm \frac{m}{2}\, \bar B^{\mu\nu}\, \Phi_{\mu\nu} + \frac{m^2}{2}\, \bar C_{\mu\nu}\, C^{\mu\nu} 
+ (\partial_\mu \bar C_\nu - \partial_\nu \bar C_\mu)\, \partial^\mu C^\nu 
\]
\[
\pm m\, (\partial_\mu \bar C^{\mu\nu})\, C_\nu
\pm\, m\, \bar C^\nu\, (\partial^\mu C_{\mu\nu}) 
 - \, \frac{1}{2}\, \big(\pm \, m\, \bar\beta^\mu - \partial^\mu \bar\beta \big)\, \big( \pm m\, \beta_\mu - \partial_\mu \beta \big)
 \]
 \[
~~~~~~- \frac{1}{2}\,  \bar B^\mu\, \bar B_\mu + \frac{1}{2}\,  \bar B^{\mu}\, \big(\pm \, m\phi_\mu - \partial_\mu\, \phi \big) 
+ \bar f\, \big[ \frac{1}{2}\, \partial \cdot C  \pm m\,  C_1\big]
\pm \frac{m}{2}\, \bar C_\mu \bar F^\mu  \pm\, \frac{m}{2}\, \bar f^\mu\, C_\mu
\]
\[
~~~~+ \big[\frac{1}{2}\,\partial \cdot \bar C \mp\, m\,  \bar C_1\big]\, \bar F \pm m\, B_1\, \phi 
+ 2\, \bar F\, \bar f -\, m^2\, \bar C_2\, C_2 \mp\, m\, B \bar\beta \pm m\, B_2\, \beta, 
\eqno (B.10)
\]
where we have used the modifications of Eq. (49).

We end this Appendix with the concluding remarks that the {\it total} Lagrangian densities ${\cal L}_{ B}^{(T)}$ and 
${\cal L}_{\bar B}^{(T)}$ are {\it perfectly} BRST and anti-BRST invariant [cf. Eqs. (71), (72)].
Furthermore, we observe that $(i)$ ${\cal L}_{ B}^{(T)}$ respects anti-BRST symmetry, {\it and} (ii) ${\cal L}_{\bar B}^{(T)}$ remains invariant under the 
BRST symmetry transformations {\it provided} we use all the CF-type restrictions [cf. Eqs. (14), (69)].
Hence, the Lagrangian density  ${\cal L}_{ B}^{(T)}$ and ${\cal L}_{ \bar B}^{(T)}$ are coupled (but equivalent) w.r.t. the nilpotent
(anti-)BRST symmetry transformations.\\

\vskip 1cm 
\begin{center}
{\bf Appendix C: On the Existence and Emergence of CF-Type Restrictions}\\
\end{center}
\noindent
Within the framework of BRST formalism, the existence of the CF-type restriction(s) is a key and 
decisive feature for a BRST-quantized theory. Even in the case of a D-dimensional BRST-quantized 
Abelian 1-form theory, a CF-type restriction exists but it is {\it trivial} 
in the sense that it is the limiting case of the CF-condition [12] which is a {\it non-trivial} restriction
in the case of  a D-dimensional non-Abelian 1-form theory.
All the higher $p$-form $(p = 2, 3, ...)$ (non-)Abelian gauge theories are endowed with the {\it non-trivial} CF-type restriction(s) because 
their existence for a BRST-quantized theory is {\it as} crucial and fundamental {\it as} the existence of the 
{\it first-class} constraints [in the terminology of the Dirac's prescription for the classification scheme (see, e.g. [23-26] for details) of constraints]
for a given {\it classical} gauge theory. In our present Appendix, we shed some light on the existence and emergence of the {\it six}
(anti-)BRST invariant CF-type restrictions 
on our arbitrary D-dimensional BRST-quantized modified  {\it massive} Abelian 3-form gauge theory.

Towards the above goal in mind, 
we dwell a bit on the Fig. 1 where we have shown {\it all} the fields (with their ghost numbers)
which appear in the BRST and anti-BRST symmetry transformations (B.1) and (B.2) [cf. Appendix B].
Every individual field has been shown by a circle at an appropriate ghost number that is associated with it.
The BRST symmetry transformations that {\it increase} the ghost number by {\it one} (for the field with a {\it specific} ghost number
on which they operate) have been shown by the {\it red} arrows. On the other hand, the anti-BRST transformations
that {\it decrease} the ghost number by one (for the field with a {\it specific} ghost number
on which they operate) have been displayed by the {\it green} arrows. All the BRST and anti-BRST symmetry transformations
of (B.1) and (B.2) have been taken into account in our Fig. 1 and these have been displayed with appropriate colours.

\begin{figure*}[ht]
\begin{center}
\includegraphics[scale=0.80]{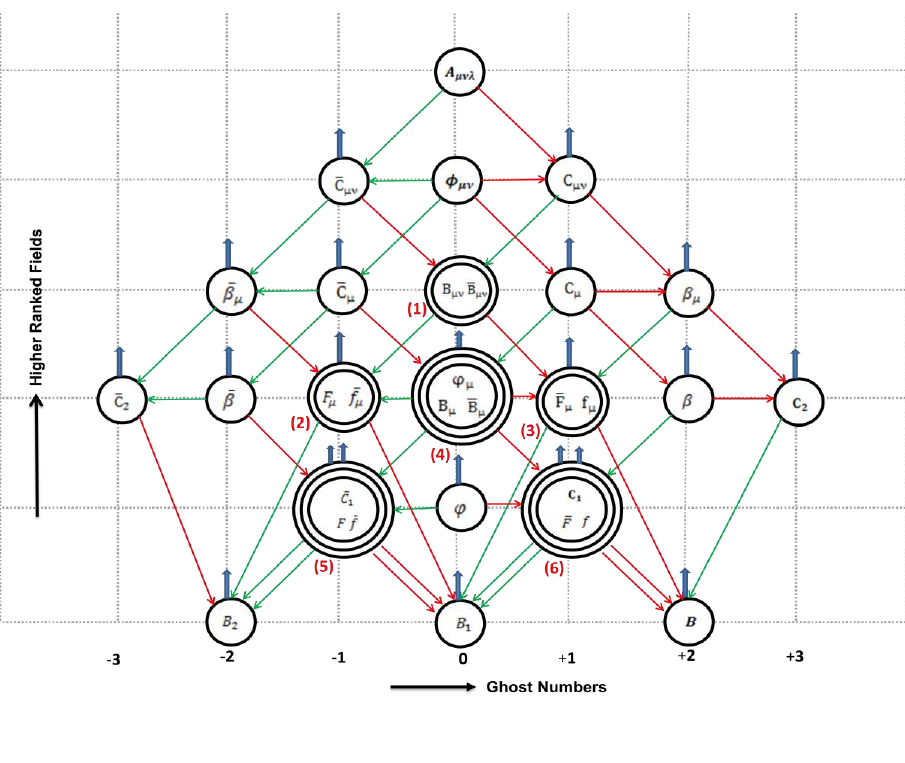} 
\caption{Emergence of CF-type Restrictions} \label{Fig.1}
\end{center} 
\end{figure*}

Except for the fields $A_{\mu \nu \lambda }, \, B_{\mu\nu}, \, \bar B_{\mu\nu}$ and $ \Phi_{\mu\nu}$, we have an exterior derivative $d$
that has been shown by the blue-bold arrow in the upward direction. This carries the informations about
the {\it full} (anti-)BRST symmetry transformations for our {\it massive} D-dimensional St$\ddot u$ckelberg-modified Abelian 3-form gauge 
theory. For instance, the exterior derivatives 
$d$ that have been shown on the fields $C_{\mu\nu}$ and $\bar C_{\mu\nu}$ lead to the following: 
\[
s_b\, A^{(3)} = d\, C^{(2)}, \qquad \qquad \qquad   s_{ab}\, A^{(3)} = d\, \bar C^{(2)}.
\eqno (C.1)
\]
In other words, the operation of $d$ on the {\it fermionic}  2-forms $C^{(2)} = [(d\, x^\mu \wedge d\, x^\nu)/ 2!]\, C_{\mu\nu}$
and $\bar C^{(2)} = [(d\, x^\mu \wedge d\, x^\nu)/ 2!]\, \bar C_{\mu\nu}$
lifts these 2-forms  fields to the layer of the 3-form fields ($A_{\mu\nu\lambda}$) in the sense that we have the following:
\[
s_b\, A_{\mu\nu\lambda} = \partial_\mu\, C_{\nu\lambda} + \partial_\nu\, C_{\lambda \mu} + \partial_\lambda\, C_{\mu\nu}, 
\]
\[
s_{ab}\, A_{\mu\nu\lambda} = \partial_\mu\, \bar C_{\nu\lambda} + \partial_\nu\, \bar C_{\lambda \mu} + \partial_\lambda\, \bar C_{\mu\nu}. 
\eqno (C.2)
\]
In exactly similar fashion, we note that
\[
s_b\, \Phi^{(2)} = C^{(2)} - d\, C^{(1)}, \qquad s_{ab}\, \Phi^{(2)} = \bar C^{(2)} - d\,\bar C^{(1)},
\eqno (C.3)
\]
where $\Phi^{(2)} = [(d\, x^\mu \wedge d\, x^\nu)/ 2!]\, \Phi_{\mu\nu}$, $C^{(1)} = d\, x^\mu\, C_\mu$, 
$\bar C^{(1)} = d\, x^\mu\, \bar C_\mu$ are the set of a {\it single} bosonic 2-form and a {\it couple} of fermionic 1-form fields.
This makes it clear that $\Phi^{(2)} $ and $ C^{(2)}$ as well as $\bar C^{(2)}$ are in the {\it same} layer but 
$C^{(1)}$ and $\bar C^{(1)}$ are one-step {\it lower} so that when $d$ operates, it lifts {\it them} to the level of 
2-forms. In other words, we have
\[
s_b\, \Phi_{\mu\nu} = \pm\, m\, C_{\mu\nu} - (\partial_\mu\, C_\nu - \partial_\nu\, C_\mu),  \qquad 
s_{ab}\, \Phi_{\mu\nu} = \pm\, m\, \bar C_{\mu\nu} - (\partial_\mu\, \bar C_\nu - \partial_\nu\, \bar C_\mu).
\eqno (C.4)
\]
Similarly, all the other such kinds of (anti-)BRST symmetry transformations 
(e.g. $s_b\, \bar B_{\mu}  = \pm m\, f_\mu - \partial_\mu\, f, \, s_b\, C_\mu = \pm \, m\, \beta_\mu - \partial_\mu\, \beta_\mu$, etc.) 
for our {\it massive} D-dimensional St$\ddot u$ckelberg-modified Abelian 3-form gauge theory can be explained.

It is worthwhile to pinpoint the fact that we have {\it two} exterior derivative operators on the {\it triplets} of concentric circles at (5) and (6)
in our Fig. 1 because of the following reason
\[
s_b\, \bar B_\mu =  \pm m\, f_\mu - \partial_\mu\, f,  \qquad f_\mu + \bar F_\mu = \partial_\mu\, C_1, 
\]
\[
s_{ab}\,  B_\mu =  \pm m\, \bar f_\mu - \partial_\mu\, \bar f,  \qquad \bar f_\mu +  F_\mu = \partial_\mu\, \bar C_1,
\eqno (C.5)
\]
which can be explained in the language of differential-forms as
\[
s_b\, \bar B^{(1)} = \pm \, m\, f^{(1)} - d\, f^{(0)}, \qquad f^{(1)} + \bar F^{(1)} = d\, C_1^{(0)}, 
\]
\[
s_{ab}\, B^{(1)} = \pm \, m\,\bar  f^{(1)} - d\, \bar f^{(0)}, \qquad \bar f^{(1)} +  F^{(1)} = d\, \bar C_1^{(0)},
\eqno (C.6)
\]
where the 1-forms and 0-forms are:
$\bar B^{(1)} = d\, x^\mu\, \bar B_\mu, \, B^{(1)} = d\, x^\mu\,  B_\mu, \, f^{(1)} = d\, x^\mu\, f_\mu, \,
 \bar f^{(1)} = d\, x^\mu\, \bar f_\mu, \, C_1^{(0)} = C_1,\, \bar C^{(0)} = \bar C_1, \, f^{(0)} = f, \, \bar f^{(0)} = \bar f, \, 
 d\, C_1^{(0)} = d\, x^\mu\, (\partial_\mu C_1), \,
d\, \bar C_1^{(0)} = d\, x^\mu\, (\partial_\mu \bar C_1 )$. In other words, we have {\it two} CF-type restrictions as well as {\it two}
(anti-)BRST transformations (i.e. $s_b\, \bar B_\mu$ and $s_{ab}\, B_\mu$) such that the exterior derivatives appear {\it twice}
which lift the 0-forms $f^{(0)},\, \bar f^{(0)}, \, C_1^{(0)}$ and $\bar C_1^{(0)}$ to their appropriate counterparts 1-forms.

As far as the existence and emergence of CF-type restrictions are concerned, we note that whenever there is clustering of the fields at a 
given ghost number, there is a  relationship amongst these fields [see, e.g. the bracketed numbers (1), (2), (3), (4), (5) and (6)
in our Fig. 1]. Sometime the {\it clustering} of the fields is in the {\it same} layer [e.g. (5) and (6) where we have: $ f + \bar F = \pm\, m\, C_1$ 
and $\bar f + F = \pm \, m\, \bar C_1$]. On the other hand, at other time, the fields are in different layers {but} they are connected by the 
exterior derivative $d = d\, x^\mu\, \partial_\mu$ (with $d^2 = 0$). It is very interesting to point out that the fields:
$\phi, \, \phi_\mu$ and $B_{\mu\nu}$ (and/or $\bar B{\mu\nu}$) are located in {\it three} different layers.
The lowest layer is occupied by $\phi$ (which is a 0-form).  However, the CF-type restriction: 
$B_\mu + \bar B_\mu = \pm m\, \phi_\mu - \partial_\mu\, \phi$ shows that the 0-form field $\phi$
is lifted to the layer corresponding to the 1-forms by an exterior derivative. In other words, 
we have this CF-type restriction in the differential-form language as
\[
B^{(1)} + \bar B^{(1)} = \pm\, m\, \Phi^{(1)} - d\, \Phi^{(0)},
\eqno (C.7)
\]
where $\Phi^{(0)} = \phi, \, \Phi^{(1)} = d\, x^\mu \, \phi_\mu, \, B^{(1)} = d\, x^\mu\, B_\mu$ and $\bar B^{(1)} = d\, x^\mu\, \bar B_\mu$
This observation  establishes  that the fields $\phi_\mu, \, B_\mu$ and $\bar B_\mu$ are in the {\it same} layer of the fields but $\phi$
is one-step down of {\it this} layer. In exactly similar fashion, we note that the CF-type restriction: $B_{\mu\nu} + \bar B_{\mu\nu} 
= \partial_\mu\, \phi_\nu - \partial_\nu\, \phi_\mu$ shows that the field $\phi_\mu$ is located one-step {\it lower} than 
the fields $B_{\mu\nu}$ and/or $\bar B_{\mu\nu}$. This statement becomes transparent when we express the CF-type 
restrictions in the form-language as 
\[
B^{(2)} + \bar B^{(2)} =  d\, \Phi^{(1)}, \qquad  \Phi^{(1)} =  d\, x^\mu\, \phi_\mu,
\eqno (C.8)
\]
where $B^{(2)} = [(d\, x^\mu \wedge d\, x^\nu)/ 2!]\, B_{\mu\nu}$ and $\bar B^{(2)} = [(d\, x^\mu \wedge d\, x^\nu)/ 2!]\, \bar B_{\mu\nu}$.
All the {\it six} (anti-)BRST invariant CF-type restrictions can be explained, in exactly similar fashion, 
provided we express them in the form-language of differential geometry [19-22].
We end this Appendix with the {\it final} remark that the CF-type restrictions are connected with the geometrical objects 
called gerbes [27,28] which, primarily, provide the mathematical basis for the independent identity of the 
off-shell nilpotent (anti-)BRST symmetry transformations and corresponding conserved Noether charges.

\end{document}